\documentclass[tradiabstract,longauth]{aa}
\pdfoutput=1

\usepackage[varg]{txfonts}
\usepackage{graphicx}
\usepackage{lscape}
\usepackage{url}
\usepackage{hyperref}
\usepackage{natbib}
\usepackage{multirow}
\usepackage{color}
\usepackage{amssymb}
\usepackage{color}
\usepackage[dvipsnames]{xcolor}
\usepackage{enumerate}
\usepackage{subcaption}

\newcommand{\pf}{JPAS-{\it Pathfinder}}
\newcommand{\pfs}{JPAS-PF}
\newcommand{\jplus}{J-PLUS}

\newcommand{\jp}{J-PAS}
\newcommand{\js}{J-spectra}
\newcommand{\mjp}{miniJPAS}
\newcommand{\ang}{\normalfont\AA}

\newcommand{\degree}{$^{\circ}$}

\newcommand{\photoz}{photo-$z$}

\newcommand{\hsc}{HSC-SSP}

\def\sext{\texttt{SExtractor}}

\def\psfex{\texttt{PSFEx}}

\def\mainurl{http://archive.cefca.es/catalogues/minijpas-pdr201912/}
\newcommand{\weburl}[2]{%
\href{{\mainurl}/#1}{#2}%
}

\def\flags{\texttt{FLAGS}}
\def\mflags{\texttt{MASK\_FLAGS}}
\def\magauto{\texttt{MAG\_AUTO}}
\def\magiso{\texttt{MAG\_ISO}}
\def\magpetro{\texttt{MAG\_PETRO}}
\def\magpsfcor{\texttt{MAG\_PSFCOR}}
\newcommand\magaper[2]{\texttt{MAG\_APER\_#1\_#2}}

\bibpunct{(}{)}{;}{a}{}{,}
\begin{document}
\author{ \small
 S.~Bonoli\inst{1, 2, 3}\thanks{\url{silvia.bonoli@dipc.org}},
A.~Mar\'in-Franch\inst{4},
J.~Varela\inst{4},
H.~V\'azquez~Rami\'o\inst{4},
L.~R.~Abramo\inst{5},
A.~J.~Cenarro\inst{4},
R.~A.~Dupke\inst{6,31,32}\thanks{\url{rdupke@on.br}},
J.~M.~V\'ilchez\inst{7},
D.~Crist\'obal-Hornillos\inst{1},
R.~M.~Gonz\'alez Delgado\inst{7}
% ...
}
\institute{\small
Centro de Estudios de F\'isica del Cosmos de Arag\'on (CEFCA),  Plaza San Juan, 1, E-44001, Teruel, Spain \goodbreak
\and
Donostia International Physics Center (DIPC),  Manuel Lardizabal Ibilbidea, 4, San Sebasti\'an, Spain \goodbreak
\and
Ikerbasque, Basque Foundation for Science, E-48013 Bilbao, Spain\goodbreak
\and
Centro de Estudios de F\'isica del Cosmos de Arag\'on (CEFCA), Unidad Asociada al CSIC, Plaza San Juan, 1, E-44001, Teruel, Spain \goodbreak
\and
Instituto de F\'isica, Universidade de S\~ao Paulo, Rua do Mat\~ao 1371, CEP 05508-090,  S\~ao Paulo, Brazil \goodbreak
\and
Observat\'orio Nacional, Minist\'erio da Ciencia, Tecnologia, Inovaç\~ao e Comunicaç\~oes, Rua General Jos\'e Cristino, 77, S\~ao Crist\'ov\~ao, 20921-400, Rio de Janeiro, Brazil \goodbreak
\and
Instituto de Astrof\'isica de Andaluc\'ia - CSIC, Apdo 3004, E-18080, Granada, Spain \goodbreak
\and
Núcleo de Astrofísica e Cosmologia, PPGCosmo \& Dep.~de Física, Universidade Federal do Espírito Santo, 29075-910, ES, Brazil  \goodbreak
\and
Departamento de F\'isica, Universidade Federal de Sergipe, Av. Marechal Rondon, S/N, 49000-000 S\~ao Crist\'ov\~ao, SE, Brazil \goodbreak 
\and
Instituto de F\'isica, Universidade Federal da Bahia, 40210-340, Salvador, BA, Brazil \goodbreak
\and
Academia Sinica Institute of Astronomy \& Astrophysics (ASIAA), 11F of Astronomy-Mathematics Building, AS/NTU, No. 1, Section4, Roosevelt Road, Taipei 10617, Taiwan\goodbreak
\and
Instituto de F\'isica de Cantabria (CSIC-UC). Avda. Los Castros s/n. 39005, Santander, Spain\goodbreak
\and
Departamento de Astronomia, Instituto de Astronomia, Geofísica e Ciências Atmosf\'ericas, Universidade de São Paulo, São Paulo, Brazil\goodbreak
\and
Instituto de Matematica Estatistica e Fisica, Universidade Federal do Rio Grande (IMEF--FURG), Rio Grande, RS, Brazil\goodbreak
\and
Campus Duque de Caxias, Universidade Federal do Rio de Janeiro, 25265-970, Duque de Caxias, RJ, Brazil \goodbreak
\and
Department of Physics, Ben-Gurion University of the Negev, Be'er-Sheva 84105, Israel \goodbreak
\and
Department of Physics, University of Notre Dame, Notre Dame, IN 46556, USA\goodbreak
\and
JINA Center for the Evolution of the Elements, USA\goodbreak
\and
Instituto de F\'isica, Universidade Federal do Rio de Janeiro, 21941-972, Rio de Janeiro, RJ, Brazil\goodbreak
\and
Observat\'orio do Valongo, Universidade Federal do Rio de Janeiro, 20080-090, Rio de Janeiro, RJ, Brazil\goodbreak
\and
Department of Astronomy, Beijing Normal University, Beijing 100875, China\goodbreak
\and
Instituto Universitario de F\'isica Aplicada a las Ciencias y las Tecnolog\'ias, Universidad de Alicante, San Vicent del Raspeig, E03080, Alicante, Spain\goodbreak
\and
Departamento de F\'isica Te\'orica and Instituto de F\'isica de Part\'iculas y del Cosmos, IPARCOS, Universidad Complutense de Madrid, 28040, Madrid, Spain \goodbreak
\and
Departamento de F\'{\i}sica - CFM - Universidade Federal de Santa Catarina, Florian\'opolis, SC, Brazil \goodbreak
\and
Dipartimento di Fisica, Sezione di Astronomia, Università di Trieste, Via Tiepolo 11, I-34143 Trieste, Italy \goodbreak
\and
INAF -- Osservatorio Astronomico di Trieste, via Tiepolo 11, I-34131 Trieste, Italy\goodbreak
\and
IFPU -- Institute for Fundamental Physics of the Universe, via Beirut 2, 34151, Trieste, Italy \goodbreak
\and
INFN -- Sezione di Trieste, I-34100 Trieste, Italy \goodbreak
\and
Tartu Observatory, University of Tartu, Observatooriumi~1, 61602 T\~oravere, Estonia \goodbreak
\and
European Southern Observatory (ESO), Alonso de C\'ordova 3107, Vitacura, Santiago, Chile\goodbreak
\and
Department of Astronomy, University of Michigan, 311West Hall, 1085 South University Ave., Ann Arbor, USA\goodbreak
\and
Department of Physics and Astronomy, University of Alabama, Box 870324, Tuscaloosa, AL, USA \goodbreak
% ...
}

\title{The miniJPAS survey:  a preview of the Universe in 56 colours}
\date{\today}
\abstract{ 
The {\it Javalambre-Physics of the Accelerating Universe Astrophysical Survey} (\jp) will soon start to scan thousands of square degrees of the northern extragalactic sky with a unique set of $56$ optical filters from a dedicated $2.55$m telescope, JST, at the  {\it Javalambre Astrophysical Observatory}. Before the arrival of the final instrument, JPCam (a 1.2 Gpixels, $4.2\deg^2$ field-of-view camera), the JST was equipped with an interim camera (\pf), used to test the telescope performance and execute the first scientific operations. The \pf\ camera is composed of one  $9 \rm{k} \, \times \,9 \rm{k}$ CCD, with a  $0.3\deg^2$ field-of-view and same pixel size as JPCam, 0.23\,arcsec\,pixel$^{-1}$.  To demonstrate the scientific potential of \jp, with the \pf\ camera we carried out a survey on the AEGIS field (along the Extended Groth Strip), dubbed \textit{miniJPAS}. 
We observed  a total of $\sim 1 \deg^2$, with the $56$ \jp\ filters, which include $54$ narrow band (NB, $\rm{FWHM} \sim 145$~\AA) and  two broader filters extending to the UV and the near-infrared, complemented by the $u,g,r,i$ SDSS broad band (BB) filters. In this paper we present the \mjp\ data set, the details of the catalogues and data access,  and illustrate the scientific potential of our multi-band data.
The data  surpass the target depths originally planned for \jp, reaching $\rm{mag}_{\rm {AB}}$ between $\sim 22$ and $23.5$  for the NB filters and up to $24$ for the BB filters ($5\sigma$ in a $3$~arcsec aperture).
 The \mjp\ primary catalogue contains more than $64,000$ sources extracted in the $r$ detection band with forced photometry in all other band. We estimate the catalogue  to be complete up to $r=23.6$~AB for point-like sources and up to $r=22.7$~AB for extended sources. 
 Photometric redshifts reach subpercent precision for all sources up to $r=22.5$~AB , and a precision of $\sim 0.3$\% for about half of the sample. 
 On this basis, we outline several scientific applications for our data,
 including the study of spatially resolved stellar populations of nearby galaxies, the analysis of the large scale structure  to $z\sim  0.9$ and the detection of large numbers of clusters and groups.  We further show that sub-percent redshift precision can be reached also for quasars within a broad range of redshifts, allowing to push the study of the large scale structure to $z>2$.
 \mjp\ demonstrates the capability of the \jp\ filter system to accurately  characterize a broad variety of sources and paves the way for the  upcoming arrival of \jp, which is expected to multiply this data by three orders of magnitude.
The \mjp\ data and associated value added catalogues are publicly accessible via this url: \url{http://archive.cefca.es/catalogues/minijpas-pdr201912}. 
  }

\keywords{surveys -- astronomical databases: miscellaneous -- techniques: photometric --  stars:general -- galaxies:general -- cosmology:observations}
\titlerunning{miniJPAS: paving the way for J-PAS}
%
\authorrunning{S.~Bonoli et al.}
\maketitle
%
%
%
%

\section{Introduction}
\label{sec:intro}

From the pioneering Sloan Digital Sky Survey \citep[SDSS,][]{york2000}, to the most recent Dark Energy Survey \citep[DES,][]{des2005}, Panoramic Survey Telescope and Rapid Response System 1 \citep[Pan-STARRS1,][]{chambers2016} and Hyper Suprime-Cam Subaru Strategic Program \citep[\hsc,][]{aihara2018}, wide photometric surveys have been providing in the last few decades a huge wealth of data to study all kinds of objects in our Universe. From stellar population studies to the analysis of the large scale structure, from the discovery of dwarf galaxies to gravitational lensing, all fields in astronomy have benefited from these surveys, and a leap forward is expected with the arrival of the Legacy Survey of Space and Time \citep[LSST,][]{ivezic2019}. 
However, typically featuring information in only a few broad bands, the characterization of the nature and redshift of the detected objects from these photometric surveys has large intrinsic errors, and for the precise analysis of the sources and their use for precision cosmology, large follow-up spectroscopic campaigns have been needed.  

COMBO-17 \citep{wolf2003} was the first experiment using a significantly larger number of bands, narrower in full-width-half-maximum (FWHM), allowing a more precise characterization of the sources using photometric data alone \citep[see also the seminal work of][]{hickson1994}. The approach has been then followed by other projects, such as COSMOS \citep{scoville2007}, ALHAMBRA \citep{moles2008} and SHARDS \citep{perezgonzalez2013}. All these experiments focused on the observation of small, but deep, fields, and opened the way for more extensive   ``spectrophotometric'' observations. 
 The PAU survey  \citep[e.g.,][]{eriksen2019}  is currently targeting tens of  $\mathrm{deg}^2$ over a few known fields  (e.g., CFHTLS, KiDS )   with 40 narrow band  filters  on the PAU Camera \citep{padilla2019}, a guest instrument at the $4.2$~m William Herschel Telescope.
The \jplus\ \citep{cenarro2019} and S-PLUS \citep{mendes2019} surveys are scanning very wide areas with $12$ filters, including $7$ narrow bands positioned at strategic wavelengths to be able to properly classify and characterize spectral features of stars and local galaxies. Initially conceived to provide an alternative way to calibrate \jp\ data, \jplus\ started in $2017$ and is being carried out from a dedicated $80$~cm telescope, the JAST/T80, at the {\it Observatorio Astrof\`isico de Javalambre} \citep[OAJ, ][]{cenarro2014},  located in the province of Teruel, in an exceptionally dark region in continental Spain. \jplus\  has already observed thousands of square degrees and its data are being used for a broad variety of science studies, e.g., the search for metal-poor stars \citep{whitten2019b}, the 2D study of nearby galaxies \citep{sanroman2019} and star forming regions \citep{logrono2019}. 
S-PLUS is observing using the T80-South, a replica of the JAST/T80, located at  Cerro Tololo, Chile. S-PLUS is imaging the southern sky with different survey strategies to optimize a variety of science goals \citep[see][]{mendes2019}.




The  project which aims at pushing the spectrophotometric approach to the largest scales is the {\it Javalambre-Physics of the Accelerating Universe Astrophysical Survey} \citep[\jp,][]{benitez2009, benitez2014}. \jp\ is about to start observing thousands of square degrees with a unique set of $56$ filters, $54$ overlapping narrow bands (NB, FWHM~$\sim145$~\ang)  complemented with two broader filters in the blue and red extremes of the optical spectrum.
Designed and optimized to achieve extremely accurate redshift precision to perform baryonic acoustic oscillation (BAO) measurements and other cosmological experiments across a wide range of cosmic epochs, the \jp\ filter system will effectively offer a low-resolution spectrum for every pixel of the sky observed, with no issues such as target selection or fiber collision incompleteness, making \jp\ a very versatile project to carry out a broad range of astrophysical studies.
From asteroids to stellar physics, from galaxy evolution to cosmology, \jp\ will provide a unique data-set with a powerful legacy value. The survey will be carried out with 
the primary telescope at the OAJ, the $2.55$~m {\it Javalambre Survey Telescope} (JST/T250), characterized by a very large Field of View ($3$~deg~\diameter).
The main instrument of the JST/T250 is the {\it Javalambre Panoramic Camera} (JPCam), a 1.2~Gpixel camera that will cover an area of  $\sim 4.2 \deg^2$ with its 14 CCDs.
While waiting for the final steps of the assembly and testing of JPCam, the JST/T250 has been equipped with the \pf\ camera, a single CCD camera located at the center of the focal plane, with a field of view of $\sim 0.3 \deg^2$. 
This camera has been used to perform the first scientific operations of the JST/T250 and to deliver the first J-PAS-like data.
A variety of observations have been performed, but most of the effort has been devoted in completing $\sim 1 \deg^2$ on the famous Extended Groth Stip (EGS) field, an area covered by many experiments, from the AEGIS \citep{davis2007} to the ALHAMBRA \citep{moles2008}  and the \hsc\ \citep{aihara2018, aihara2019} surveys.
The \pf\ observations on the EGS field, dubbed {\it miniJPAS}, have been used to test the performance of the telescope and to prove the potential of the \jp\ filter system. In this paper we present the \mjp\ data set and the data products, which are part of the first  public release of the \jp\ collaboration. We also provide a flavour of the diverse science cases that can be studied with future \jp\ data. 

The paper is organized as follows. Section~\ref{sec:survey} provides details on the OAJ,  the instruments and filter system, and describes \mjp\, from the observational strategy to basic data statistics. The data reduction and catalogues construction pipelines are described in Sect.~\ref{sec:upad}. Section~\ref{sec:data} offers an overview of the data products, including photometric redshifts and other value-added catalogues. In Sect.~\ref{sec:dava} we discuss the quality of the data and in Sect.~\ref{sec:science} we give a brief overview of the science that can be achieved with the \jp\ filter system. Section~\ref{sec:summary} summarizes our results.

Throughout this work, we adopt a Lambda Cold Dark Matter ($\Lambda$CDM) cosmology with $h = 0.674$, $\Omega_M = 0.315$, $\Omega_{\Lambda} = 0.685$, as in the latest Planck results \citep[][]{planck2018}. Magnitudes are quoted in the AB-system \citep{oke1983}.

\section{Survey definition and data acquisition} \label{sec:survey}

In this section we first summarize the technical aspects of the survey, and then focus on the data acquisition. 

\subsection{The Observatorio Astrof\'{\i}sico de Javalambre (OAJ)}
\label{sec:oaj}

The OAJ\footnote{\url{http://www.oaj.cefca.es}} \citep{cenarro2014} is an astronomical infrastructure conceived to carry out large sky photometric  surveys from the Northern hemisphere with dedicated telescopes of large field-of-view (FoV). The definition, design, construction, exploitation and management of the observatory and the data produced at the OAJ are responsibility of the Centro de Estudios de F\'{\i}sica del Cosmos de Arag\'on (CEFCA\footnote{\url{http://www.cefca.es}}), and since  2014 the OAJ is a Spanish Singular Scientific and Technical Infrastructure (known with the Spanish acronym ICTS for Infraestructura Cient\'{\i}fico T\'ecnica Singular). The observatory is located at the Pico del Buitre of the Sierra de Javalambre, Teruel, Spain, and is organized around two telescopes, the Javalambre Survey Telescope \citep[JST/T250, ][]{cenarro2018b} and the Javalambre Auxiliary Survey Telescope (JAST/T80), both manufactured by AMOS\footnote{\url{https://www.amos.be/}}.  The site, at an altitude of $1957$\,m, has excellent median seeing (0.71\,arcsec in V band, with a mode of 0.58\,arcsec) and darkness (typical sky surface brightness of $V\sim 22$\,mag\,arcsec$^{-1}$), a feature quite exceptional in continental Europe. Full details about the site testing of the OAJ can be found in \cite{moles2010}.

\begin{figure}[t!] \centering
\includegraphics[width=0.49\textwidth]{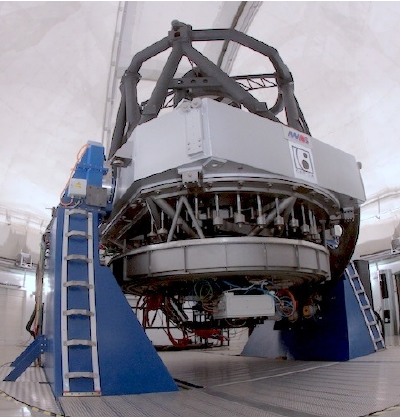}
\caption{View of the JST/T250 telescope, the primary telescope of the OAJ. It is the telescope used for the \mjp\  survey, with the  JPAS-{\it Pathfinder} camera  integrated at the Cassegrain focus.} 
\label{fig:JST}
\end{figure}

\jp\ will be conducted from the JST/T250 (see Fig.~ \ref{fig:JST}), an innovative Ritchey-Chr\' etien-like, alt-azimuthal, large-\'etendue telescope, with an aperture of 2.55\,m and a 3\,deg diameter FoV. The focal plane of JST/T250 is flat and corresponds to a Cassegrain layout.  The effective collecting area of JST/T250 is 3.75\,m$^{2}$, yielding an \'etendue of 26.5 m$^{2}$ deg$^{2}$. Motivated by the need of optimizing the etendue, JST/T250 is a very fast optics telescope (F$\#$3.5) with a plate scale of 22.67~arcsec~mm$^{-1}$. The optical design has been optimized to provide a good image quality in the optical spectral range (330-1100\,nm) all over the 48 cm diameter focal plane (7\,deg$^{2}$).
 Table~\ref{tab:JST_technical} illustrates a summary of the main technical characteristics of JST/T250 \citep[see also][]{cenarro2018b}.

\begin{table} \caption{Main technical characteristics of the JST/T250 telescope} \label{tab:JST_technical} \centering \begin{tabular}{r l} 
\hline\hline 
Mount: & Altazimuthal  \\ 
Optical configuration: & Ritchey Chr\'etien modified, \\
  & equipped with a field corrector \\
  & and rotator\\ 
M1 diameter: & 2.55\,m \\ 
Field corrector: & 3 aspherical lenses \\ 
Effective collecting area: & 3.75\,m$^2$\\
Focus: & Cassegrain\\ 
F$\#$: & 3.5\\ 
Focal length:& 9098\,mm\\ 
Plate scale: & 22.67\,arcsec\,mm$^{-1}$\\ 
FoV (diameter) & 3.0\,deg \\
\'Etendue: & 26.5\,m$^2$deg$^2$\\ 
EE50 (diameter, as-built) & $12\,\mu{\rm m}$\\
EE80 (diameter, as-built) & $25\,\mu{\rm m}$ \\ 
\hline \end{tabular}
\end{table}

The main scientific instrument of the JST/T250, which will be used to carry out \jp, is the Javalambre Panoramic Camera \citep[JPCam,][]{taylor2014, marin2017}, a 1.2\,Gpixel camera with an effective FoV of 4.2\,deg$^{2}$ and a plate scale of 0.23\,arcsec\,pix$^{-1}$. 
JPCam has completed the assembly, integration and verification phases and is currently being installed and commissioned at the Cassegrain focus of the JST/T250 telescope. Before the arrival of JPCam, the JST/T250 telescope has been equipped with a first light instrument, single CCD camera,  called \pf, specifically built to test the system optical performance and carry out the first scientific operations.

\subsection{The JPAS-{\it Pathfinder} camera} \label{sec:pf}

The JPAS-{\it Pathfinder} camera (\pfs\ hereafter) is the first scientific, interim instrument installed at the JST/T250 telescope. This single CCD, direct imager had been developed with two primary goals: (i) to minimize the time and risk of JPCam commissioning at the telescope, as it was used to perform the JPCam Actuator System\footnote{Because of the large FoV and fast optics, the JST/T250 secondary mirror and the focal plane are actively controlled with two hexapod actuators to ensure optimum image quality.} commissioning at the telescope before the JPCam instrument was assembled in the OAJ clean room, and (ii) to start the scientific operation of the JST/T250.

The \pfs\ camera is equipped with one large format, 9.2k x 9.2k, 10\,$\mu$m pixel, low noise detector from Teledyne-e2V located at the center of the JST/T250 FoV. The detector is read out from 16 ports simultaneously. It has an image area of 92.16\,mm $\times$ 92.32\,mm and a broadband anti-reflective coating for optimized performance from 380\,nm to 850\,nm. The \pfs\ consists of two main subsystems, the filter and shutter unit (FSU) and the cryogenic camera. 

The cryogenic camera is a 1110S CCD camera manufactured by Spectral Instruments\footnote{\url{www.specinst.com}} (USA). It comprises the scientific detector and its associated controllers, the cooling and vacuum systems and the image acquisition electronics and control software. The proximity drive electronics provide 16 different operational modes. The \mjp\ survey has been observed with a read mode that achieves total system level noise performance of 3.4\,${\rm e}^{-}$ (RMS), allowing read out times of 12\,s (full frame) and 4.3\,s (2x2 binning). The FSU includes a two-curtain shutter provided by Bonn-Shutter UG\footnote{\url{www.bonn-shutter.de}} (Germany) that allows taking integration times as short as 0.1\,s with an illumination uniformity better that 1$\%$ over the whole \pfs\ FoV. The FSU incorporates a single filter wheel designed to integrate 7 different \jp\ filters simultaneously. As these filters have a physical dimension slightly smaller than the size of the CCD, the CCD is vignetted on its periphery (see Sect.~\ref{sec:single_frames}). As a result, \pfs\ provides an effective FoV of 0.27\,deg$^2$ with a pixel scale of 0.23\,arcsec\,pixel$^{-1}$. A summary of the technical characteristics of the camera is presented in Table~\ref{tab:PF_technical}. 

The \pfs\ camera was dismounted at the end of 2019, leaving room for JPCam. 

\begin{table} \caption{Main technical performances of the \pf\ camera} \label{tab:PF_technical}
\centering
\begin{tabular}{r l } \hline\hline 
CCD format	     & $9216\times9232$\,pix \\ 
                              & $10\,\mu{\rm m}\,{\rm pix}^{-1}$ \\ 
Pixel scale	     & $0.23$\,arcsec\,pix$^{-1}$ \\ 
Effective FoV         & $0.54\deg$ $\times$ $0.50\deg$ $=$ $0.27\deg^2$ \\ 
Read out time	     & 12\,s (full frame) \\ 
                  	     & 4.3\,s (2x2 binning) \\ 
Read out noise	     & $3.4\,{\rm e}^{-}\,{\rm pix}^{-1}$ \\ 
Full well                 & $123\,000\,{\rm e}^{-}$ \\ 
CTE                       & 0.99995 \\ 
Dark current	     & $0.0008\,{\rm e}^{-}\,{\rm pix}^{-1}\,{\rm s}^{-1}$ \\ 
Quantum Efficiency  & $40\%$ $@$ 350nm \\
 " & $86\%$  $@$ 400nm \\
 " & $93\%$ $@$ 500nm \\
 " & $93\%$  $@$ 650nm \\
 " & $61\%$ $@$ 900nm  \\
Filters per filter wheel    & 7 \\
\hline \end{tabular} 
\end{table}

\subsection{The J-PAS filter system}
\label{sec:filters}

\begin{figure*}[t]
\centering
\includegraphics[width=0.8\textwidth]{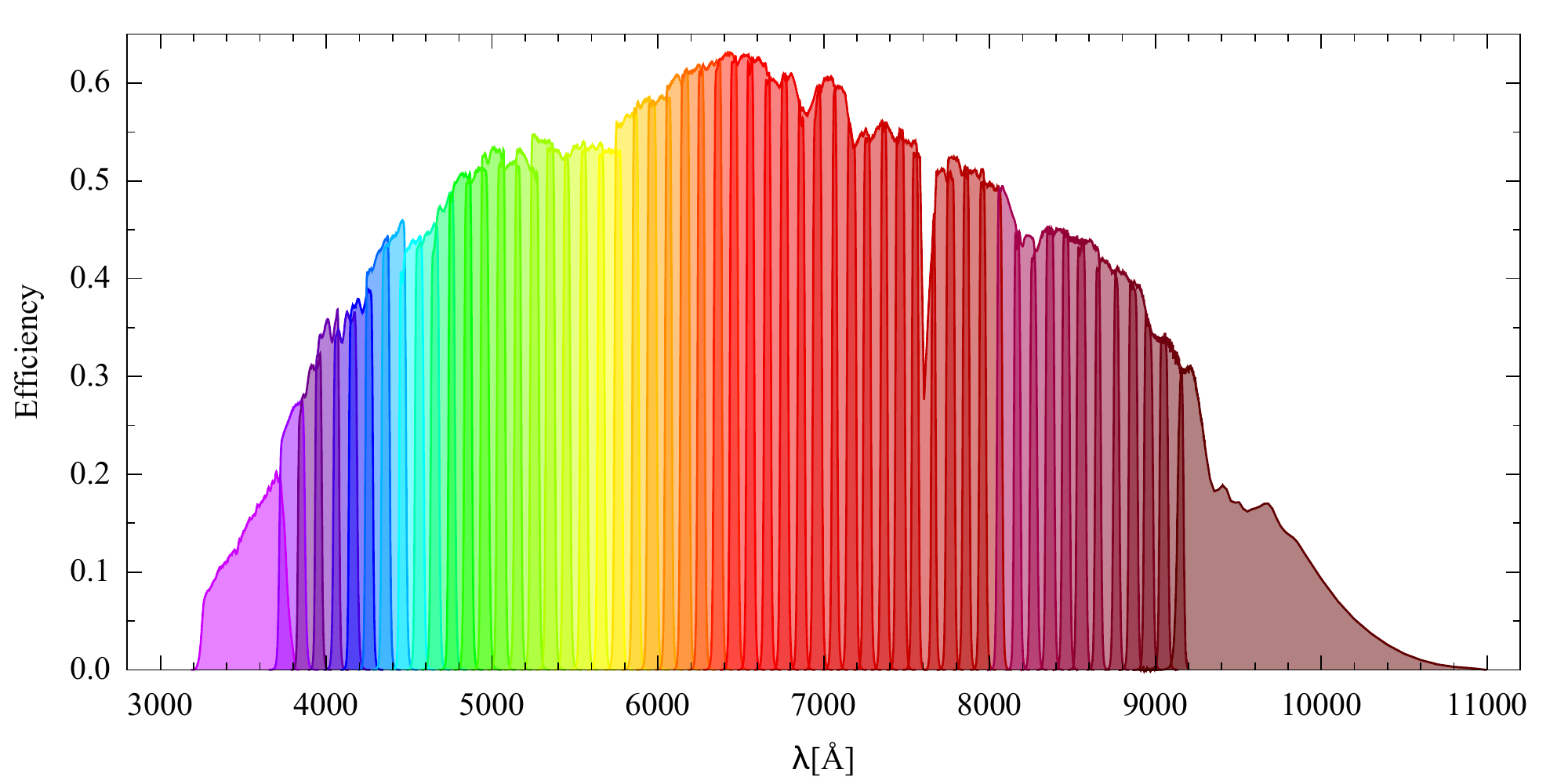}
\caption{The measured transmission curves of the \jp\ filters.  Effects of the CCD quantum efficiency, the entire optical system of the JST/T250 telescope and sky absorption are included.  The HTML color representation of each filter is provided in the \mjp\ database in the table \texttt{minijpas.Filter}.}
\label{fig:filters}
\end{figure*}

The novel and unique aspect of \jp\ lies on its filter system: $54$ narrow band filters ranging from $3780$~\ang\ to $9100$~\ang, complemented with two broader filters in the blue and red wings. The NB filters have a FWHM of $145$~\ang\ and are spaced by about $100$~\ang\ (except for the filter $J0378$), thus creating a continuous spectral coverage through the entire optical range. The two additional filters are one medium band covering the UV edge ($J0348$, called $u_{\rm JAVA}$) and a broad filter red-wards of $9100$~\ang\ ($J1007$). Table~\ref{tab:filters} lists the main characteristics of the \jp\ filter set, while Fig.~\ref{fig:filters} shows the transmission curves of the 56 filters described above, where the overall system efficiency has been taken into account (including the atmospheric transmission, the CCD efficiency, and  the telescope optics). This filter system effectively delivers a low-resolution ($R \sim 60$)\footnote{The wavelength resolution $R_{\lambda}$ is defined as $R_{\lambda}=\lambda/\Delta \lambda$, so $R \sim 60$ is the approximate value in the intermediate wavelength range in the \jp\ filter system.} spectra (\js\, hereafter) for every object observed and has been particularly designed and optimized to achieve a photometric redshift (\photoz) accuracy sufficient to carry out cosmological experiments using a variety of tracers at different redshift ranges \citep[see][and Sect.~\ref{sec:photoz}]{benitez2009, benitez2014}. The observations in the $56$ filters discussed above are complemented with broad-band observations. The \mjp\ field has been observed with the SDSS-like broad band filters $u_{\rm JPAS}$\footnote{the $u_{\rm JPAS}$ filter has a redder cut-off than the SDSS $u$}, $g_{\rm SDSS}$, $r_{\rm SDSS}$, and $i_{\rm SDSS}$. The $r_{\rm SDSS}$ filter has been chosen as the detection band for the \mjp\ ``dual-mode'' catalogues, as explained in Sect.~\ref{sec:photometry}. From now on, unless otherwise stated, we will refer to these filters simply with $u$, $g$, $r$ and $i$. The information on FWHM and central wavelengths of these filters is also available, as for the other filters, in the \mjp\ data release ADQL table \texttt{minijpas.Filter} (see Sect.\ref{sec:data_access}). The filter transmission curves are publicly available at the Spanish Virtual Observatory page\footnote{\url{http://svo2.cab.inta-csic.es/theory/fps/index.php?mode=browse&gname=OAJ}}.

Beyond the specified theoretical filter transmission curves, whose definition is exclusively driven by the main scientific goals, additional functional requirements are influenced by the telescope and instrument opto-mechanical designs.  Some of the most demanding requirements are: filter physical dimension (101.7\,mm $\times$ 96.5\,mm), central wavelength (CW) uniformity across the filter usable area (CW varies less than $\pm$ 0.2$\%$), high band-pass transmission and flatness (higher than 90$\%$, except for the bluest filters, with a flatness better than 5$\%$ peak-to-valley), out of band blocking (OD5 from 250 to 1050\,nm) and filter-to-filter continuity (overlap at transmissions higher that 75$\%$). 

The \jp\ filters have been designed by CEFCA and SCHOTT Suisse SA (Switzerland) and manufactured by SCHOTT\footnote{\url{www.schott.com}}. A detailed technical description of the filter requirements, design, measurements and characterization can be found in \citet{marin-franch2012}, \cite{brauneck2018a} and \cite{brauneck2018b}.  


\begin{table}[ht] 
\caption{Filter system main characteristics. The full table is available in the \mjp\ database in the ADQL table \texttt{minijpas.Filter}.} 
\label{tab:filters}
\centering 
	\begin{tabular}{c c c l } 
	\hline\hline 
                     &       & Central         &   		 \\ 
       Filter $\#$ & Filter name  & Wavelength  & FWHM    \\ 
	             &	  & [\AA]             & [\AA]        \\ 
	\hline
	1 & $uJAVA$ 	& 3497 	& 495	  \\ 
	2 & $J0378$ 	& 3782 	& 155	  \\ 
	3 & $J0390$ 	& 3904 	& 145	  \\ 
	4 & $J0400$ 	& 3996 	& 145	  \\ 
	5 & $J0410$ 	& 4110 	& 145	  \\ 
	$...$ & $...$ 	         & $...$ 	& $...$	  \\ 
	54 & $J0900$ 	& 9000 	& 145	  \\ 
	55 & $J0910$ 	& 9107 	& 145	  \\ 
	56 & $J1007$ 	& 9316 	& {\it High-pass filter}	  \\ 

	\hline 
\end{tabular}
\end{table}

\begin{figure*}[ht]
\centering
\includegraphics[width=0.48\textwidth]{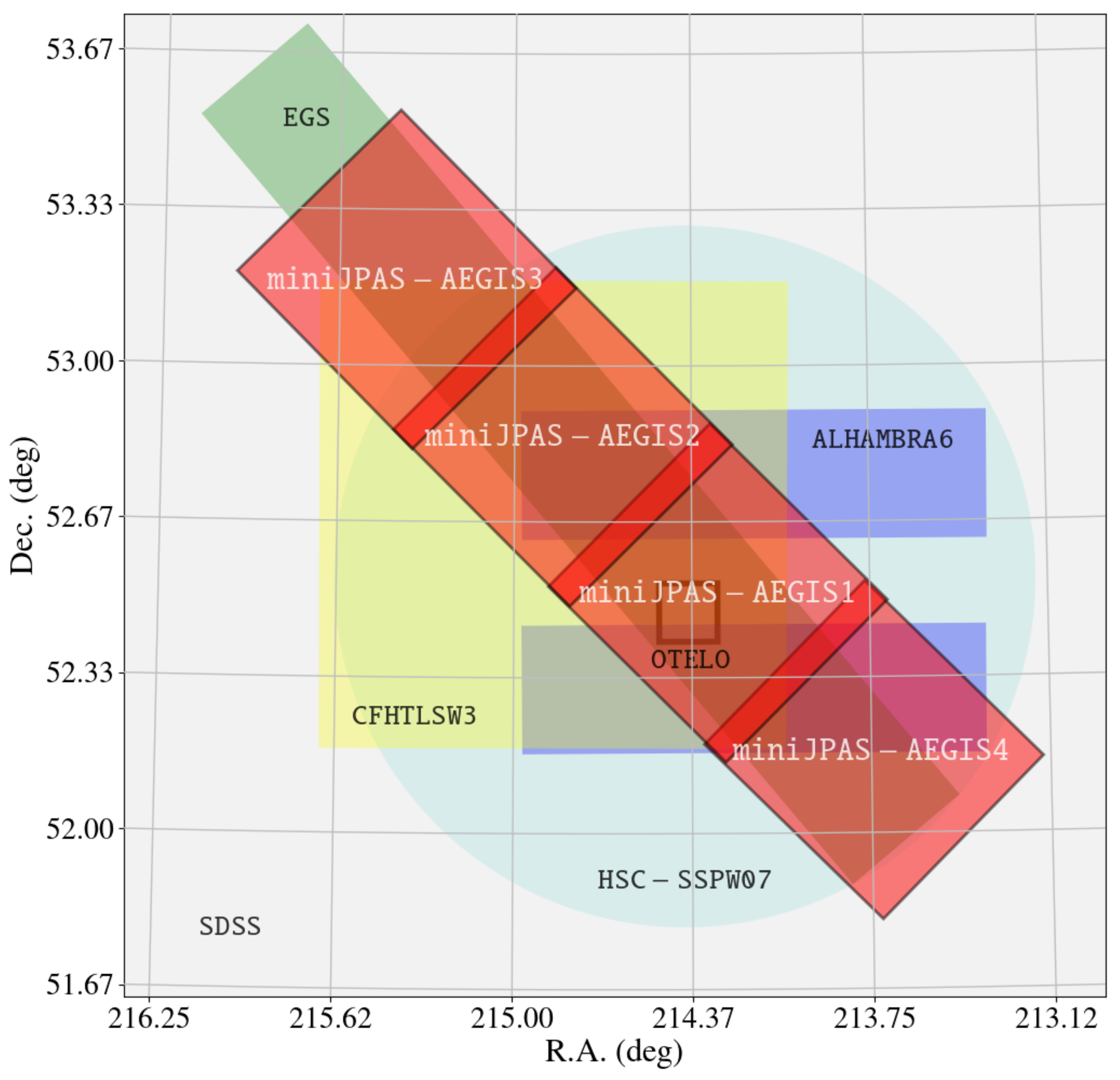}
\includegraphics[width=0.48\textwidth]{./Mini_jpas_figures/fig3b.pdf}
\caption{Left: Footprint of the \mjp{} field, with pointings shown as red squares (see their coordinates in Table~\ref{tab:pointings}). It overlaps with the footprints of other projects, also shown for reference: EGS (in green), pointing \#6 of the ALHAMBRA Survey (in violet), the W07 wide field of the \hsc\ (large circle in pale blue), field W3 of the CFHTLS (in yellow), OSIRIS Tunable Emission Line Object survey (OTELO\protect\footnote{http://research.iac.es/proyecto/otelo/pages/otelo.php}) (small square close to the center of the figure) and SDSS (in light gray occupying the whole area). We refer to Table~\ref{tab:xmatch} for the references to these overlapping surveys. Right: $g,r,i$ color image of \mjp\, with zoom in three selected areas.}
\label{fig:EGS}
\end{figure*}

\subsection{\mjp\ observations} \label{sec:observations}

This section is devoted to the description of the \mjp\ observations, from the definition of the 
footprint to the observational strategy and  the primary statistics of the collected data.

\subsubsection{Footprint and survey strategy} \label{sec:footprint} 

The  area targeted for the \mjp{} observations is the well-known
EGS field, in the north galactic hemisphere. The field
has been chosen for two main reasons: (i) its sky location ($+215.00$\degree{},$
+53.00$\degree{}), which makes it optimally observable at  altitudes $>30$\degree{} from the
OAJ from February to July and (ii) the wealth of multi-wavelength data available in the field, from the AEGIS project, to SDSS and the \hsc\ wide field. The size of the \pfs{} camera FoV allowed to cover the EGS field 
almost entirely with only $4$ pointings, with observations carried out with the instrument rotated at $45$\degree{} with respect to the celestial North. The $4$ pointings composing the \mjp{} field are listed in Table~\ref{tab:pointings} and shown in Fig.~\ref{fig:EGS}, together with the
footprints of other projects.

 The overlap between the pointings is
of $3.6$\arcmin{}. Each tile was covered with a minimum of $4$ exposures, with a
dithering of $10$\,arcsec along the horizontal and vertical direction of the
CCD. The total area covered is
$\sim 1 \deg^2$, while the total area with overlapping tiles  is  $\sim 0.09 \deg^2$,
$9\%$ of the total area. After taking the mask into account, the effective area
 is $0.895 \deg^2$ (see Sect.~\ref{sec:masks}).

\begin{figure*}[t]
\centering
\includegraphics[width=0.49\textwidth]{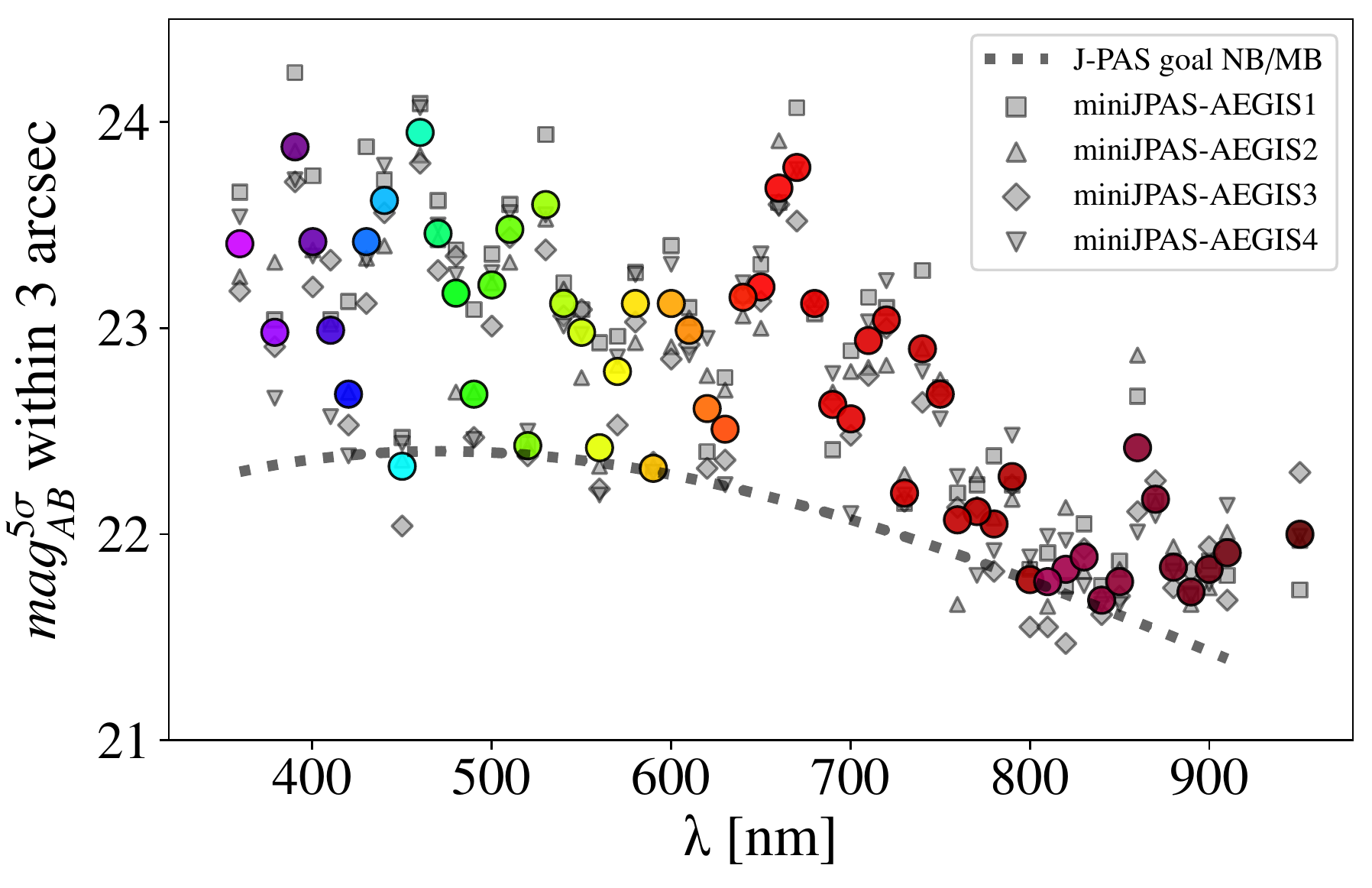}
\includegraphics[width=0.49\textwidth]{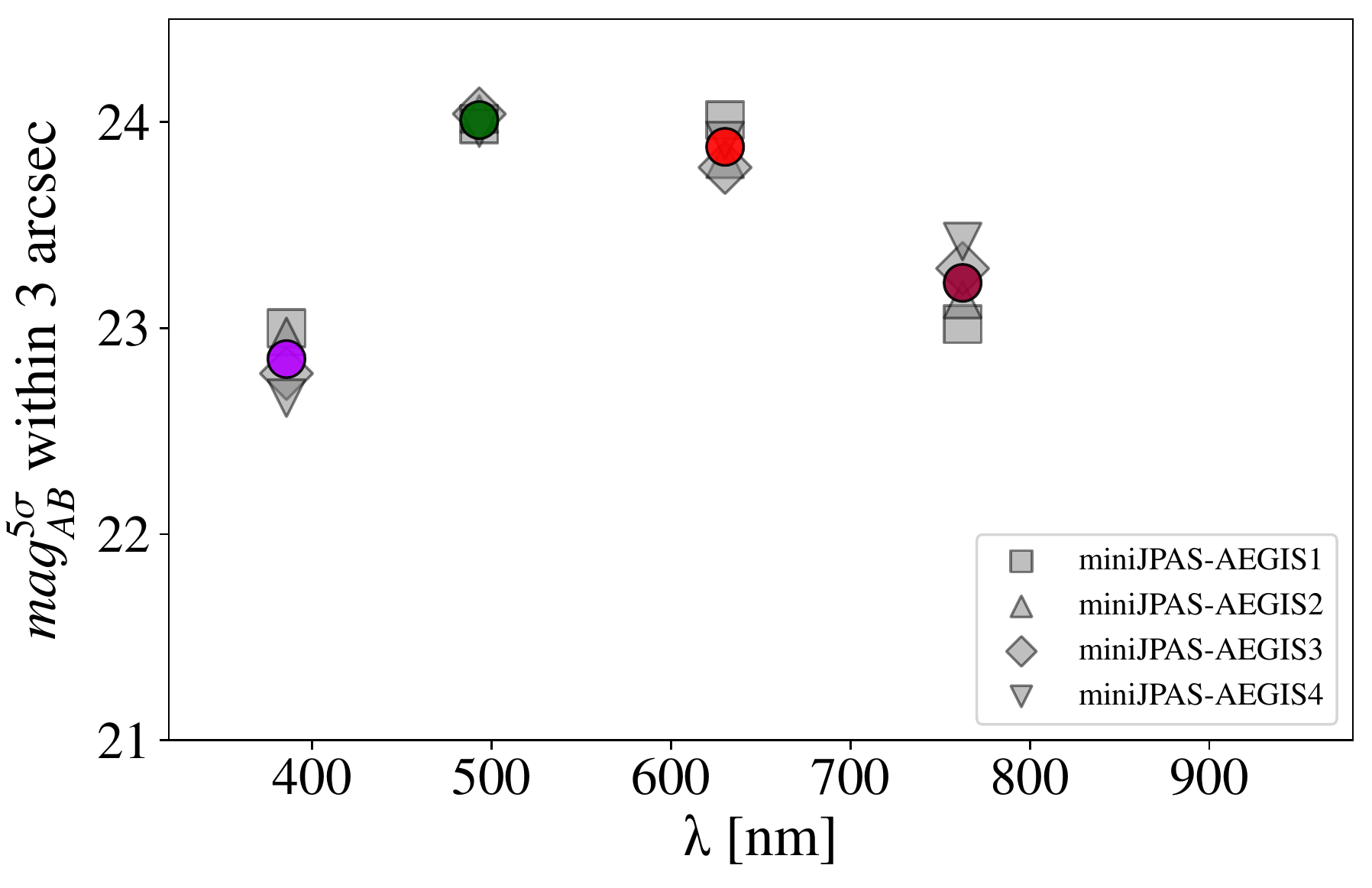}
\caption{Estimated depths ($5\sigma$ at $3$~arcsec aperture), computed from the noise in each tile, for the narrow bands (left) and broad bands (right). The coloured symbols show the average values for each filter, while the gray ones are the values for the co-added images of each pointing. For the narrow bands, the dashed gray line indicates the approximate targeted   minimum depth, as defined in \citet{benitez2014}.} 
\label{fig:depths}
\end{figure*}

The exposure times for the \mjp\ observations have been scaled up with respect to the ones quoted in \citet{benitez2014} to account for the degraded reflectivity of M1\footnote{The first re-aluminization of M1 has been performed at the beginning of 2020 together with the integration of JPCam, hence optimizing the telescope performance for the beginning of \jp\ observations.} during observations. 
For the $56$ \jp\ filters and  the $u$ filter, each independent 
exposure was of $120 \sec$, while each exposure for the broad bands $g$, $r$ and $i$ was of $30
\sec$, to avoid saturation. The basic strategy for the narrow bands and $u$ required $4$ exposures, one per dithered
position (except for the reddest filters, where a minimum of $8$ exposures, two per dithered position, was planned), while the strategy for the  broad bands was of  $4$ exposures per
dithered position. Therefore, the total minimum exposure time per filter was set to $480
\sec$, $4 \times 120 \sec$ for the NB  and $u$ ($960\sec$ for the reddest filters )  and $4 \times 4 \times 30 \sec$ for the
BB. 
However, for several filters more than this minimum number of  images is currently available, as multiple observations have been carried out with different sky conditions to test the system.  
Concerning the readout modes, the same setup expected for \jp\ was adopted:  full frame mode for $u$, $g$, $r$, $i$ and 2x2 binning for the remaining filters. The readout noise is the same in the two cases (see Table~\ref{tab:PF_technical}). 
All good-quality images (see Sect.~\ref{sec:upad}) have been included in the coadded images used to generate the \mjp{} catalogues. The number of exposures available per filters and per tile are provided in Appendix~\ref{sec:log_obs}.
The resulting depths are shown in Fig.~\ref{fig:depths}. The  minimum target depths are reached in all the filters, with most actually reaching fainter magnitudes. 
The differences in depth from band to band depend both on the net effect of sky brightness when the observations were acquired and on the final number of combined images. This inhomogeneity is expected to be minimized for \jp\ data, as the exposure times for each pointing will be modulated according to the sky brightness of the night, following a similar procedure than the one applied in \jplus\ \citep[][]{cenarro2019}.

\begin{table}
    \centering
    \caption{Central coordinates of each pointing. }
    \begin{tabular}{c c c}
    \hline
    \hline
    Tile &  RA J2000 (deg)  & DEC J2000 (deg) \\ 
    \hline
    \mjp{}-AEGIS1 & 214.2825 & 52.5143 \\
    \mjp{}-AEGIS2 & 214.8285 & 52.8487 \\
    \mjp{}-AEGIS3 & 215.3879 & 53.1832 \\
    \mjp{}-AEGIS4 & 213.7417 & 52.1770 \\
    \hline
      \end{tabular}\vspace{1mm}
        \label{tab:pointings}
\end{table}

\subsubsection {Details on data acquisition}

Data were acquired between May and September 2018, although for few filters further images were taken in the next available season in 2019 (see Appendix~\ref{sec:log_obs}).  Because of the nature of the filter wheel available for the \pf{} camera (see Sect.~\ref{sec:pf}), observations were performed in groups of six filters.  The filter ordering was chosen to maximize the data validation and scientific analysis as the survey was progressing, and to ensure a broad enough spectral coverage in case some filters could not be observed for unforeseen reasons. Even though the nominal desired minimum elevation of the telescope during observations is $40^{\circ}$, for this first data set this restriction was relaxed to allow observations for a longer period of time. The quality of the data in terms of PSF and depth suffered only mildly from this choice, as images with FWHM below $2$~arcsec can be taken down to an elevation of around $30^{\circ}$. 

\begin{figure}[t]
\centering
\includegraphics[width=0.49\textwidth]{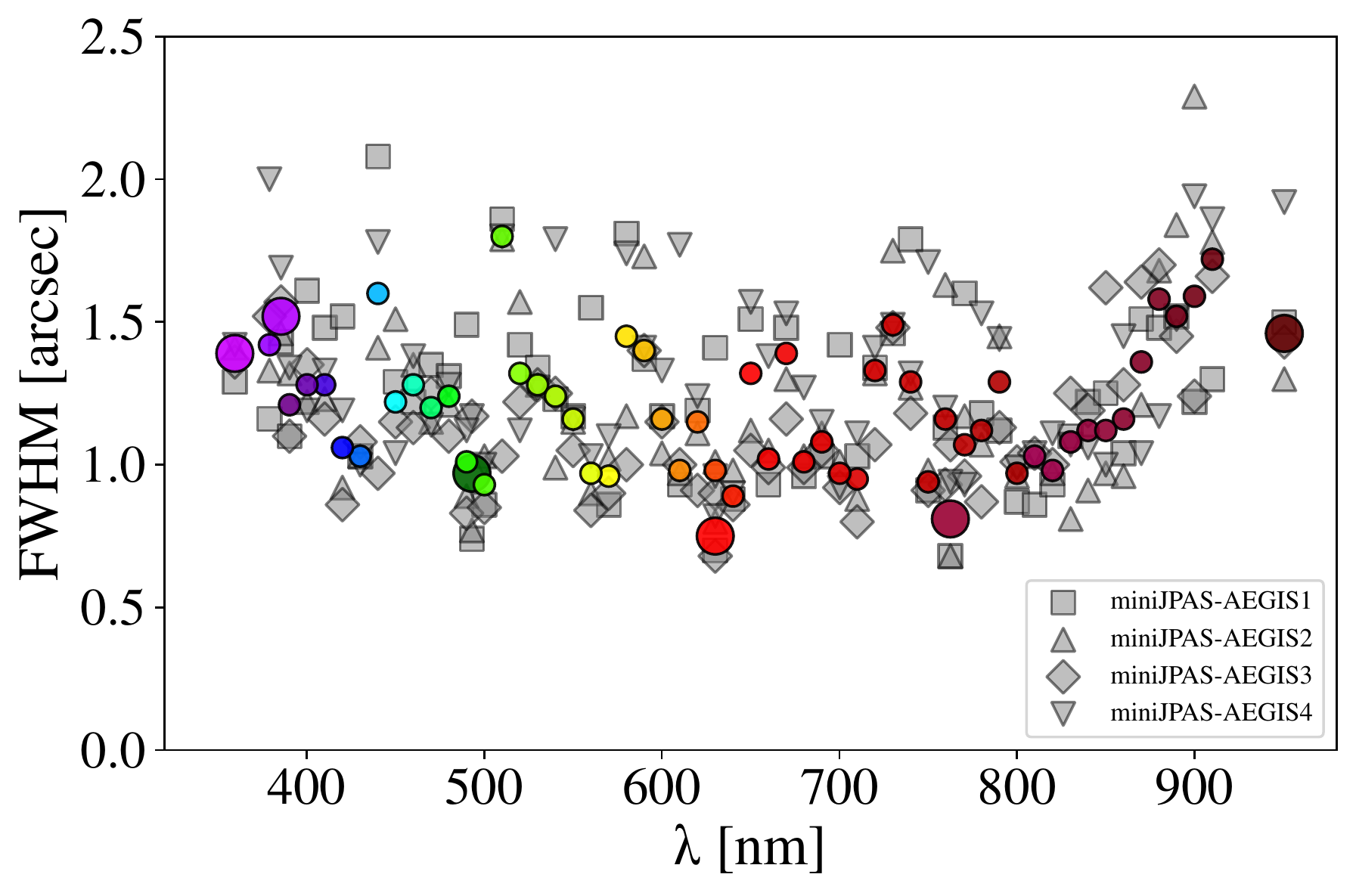}
\caption{Statistics of the PSF FWHM. The coloured symbols represent the average values for each filter, while the gray ones are the value for each pointing. The larger symbols indicate the FWHM of the the broad bands. } 
\label{fig:fwhm}
\end{figure}

The average FWHM per tile and per filter are shown in Fig.~\ref{fig:fwhm}. Most of the bands have FWHM below $1.5$~arcsec. A slight systematic increase in FWHM can be observed for the reddest bands, from $800$~nm up and, especially, from $860$~nm to higher wavelengths. The main reason for this behaviour is that the reddest filters were scheduled to be observed last, and towards  the end of the observing campaign the EGS/AEGIS field reached the lowest elevations. As a consequence, the FWHMs are larger than the ones we would have obtained at lower zenith distances with the same atmospheric conditions. Nonetheless, very few tiles have PSFs with FWHM above $2$~arcsec. Regarding the 2D stability of the PSF across the images, the median rms FWHM relative variation is, accounting for all filters and tiles, $3.7\%$, with a normalized median absolute deviation, $\sigma_{NMAD}$,  of 2.3\%.

\section{The \jp\ pipeline and data management}\label{sec:upad}

All data collected by the OAJ observatory  are handled and processed  by the {\it Data Processing and Archiving Unit} (hereafter UPAD) group at CEFCA. The UPAD data center \citep[][]{cristobal2014}
has the capacity to provide reduced and calibrated data and to archive and allow external access to the whole scientific community.
 In this section we go over the several steps of data managing, 
from the construction of coadded images to the development of the source catalogues.

\subsection{Processing of single frames} \label{sec:single_frames}

The detrending of the single frames follows the standard steps of
bias subtraction, pre/overscan subtraction, trimming, flat field
correction, illumination correction and, if needed, fringing
corrections. We refer to ~\citet{cenarro2019} for details on how these steps have been implemented, as the same procedure as for \jplus\ has been followed.
In what follows we focus on few particular aspects that needed specific treatment for \mjp\  data (e.g., reflections, background patterns, etc.), provided that \pfs\ was not specifically designed for JST/T250 and its FoV is significantly smaller than the one of the telescope. For the same reason, we do not expect many of these issues in the final \jp\ images from JPCam, since both camera and the telescope were designed together as a unique optical system.

 The main issues that needed special treatment were: 

\begin{description}

\item [{\bf Vignetting.}] The first detected issue on
  \pfs\ images is a strong vignetting in the outer parts of the
  CCD. As a result, the images show a strong gradient of
  efficiency. To minimize the impact on the final measurements, the
  images need to be trimmed to exclude regions with low efficiency.
  The preliminary tests required a reduction of the effective area of
  the CCD from 9216 $\times$ 9232 pixels to 7777 $\times$ 8473. The
  resulting effective FoV  of \pf\ is thus 0.27\,deg$^2$ (see Table~\ref{tab:PF_technical}). \\

\item [{\bf Background patterns.}] The other important issue detected in \pfs\ images 
  are background patterns with
  strong gradients and variations in time scales of few minutes, which
  affect in different ways filters at different wavelengths.  We were
  able to identify two kinds of reflections with two different
  origins. The first one is characterized by straight patterns and is
  likely due to the optics of the camera. The second one, instead,
  features circular patterns affecting many of the reddest filters,
  likely due to the small variation of the central wavelength of the
  transmission curves of the filters. In both cases, the net effect is
  an irregular distribution of the sky light which affects the images
  in two ways:  affecting the flat field images via spurious
  changes of efficiency and creating an irregular background
  pattern. The first needs to be corrected through the illumination
  correction while the second needs a careful subtraction to not
  affect the photometry of the objects.

  The technical details of the procedures followed to remove the
  background patterns can be found in
  Appendix~\ref{sec:app_bgk}. The left panels of Fig.~\ref{fig:processed} show an example
  of an image before and after the background subtraction. \\

\item[{\bf Fringing.}] For filters redder than J0740, the presence of fringing requires  an extra  step to remove 
the pattern. In this case, we follow
the standard implementation already  used for the $z$ filter of
\jplus\ and  constructed
master fringing images using all available images taken with \pfs\ in each filter. This is possible as the fringing
pattern is very stable across nights. However, some small residuals of the fringing pattern could still be visible in the final images of a few filters for which the pattern can be  particularly strong (special
mention here for the J1007 filter), and for which the number of available images is small. 

\end{description}

The right panels of Fig.~\ref{fig:processed} show an example of an image of the J0880 filter before and
after full processing. In the raw image one can see vignetting in the edges, present in all the images, and the circular pattern and fringing, which are common in the red filters.

\begin{figure*}[t]
  \begin{subfigure}{0.5\textwidth}
    \centering
  \includegraphics[height=1.8in]{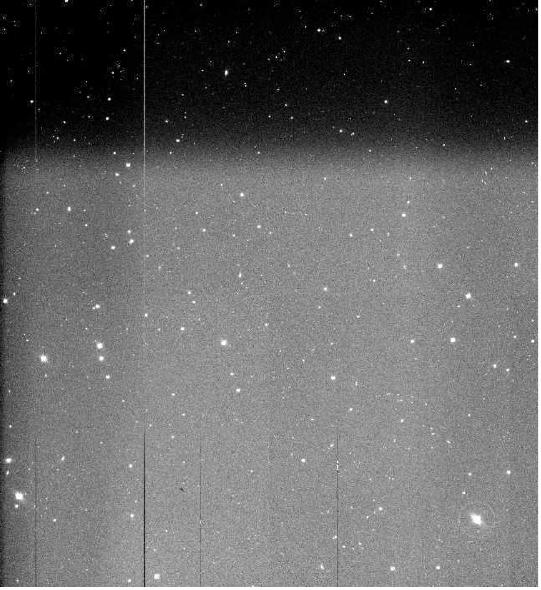}
  \includegraphics[height=1.8in]{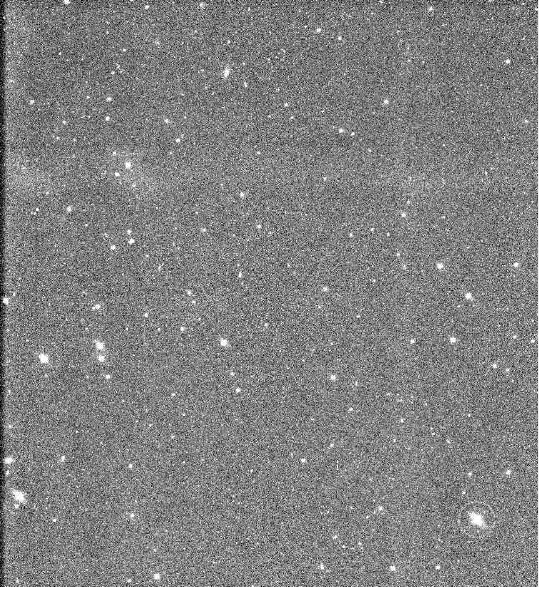}
  \caption{Example of an image before and after background subtraction (miniJPAS-AEGIS2, in the filter J0690).} 
  \end{subfigure}
  ~
   \begin{subfigure}{0.5\textwidth}
    \centering
    \includegraphics[height=1.8in]{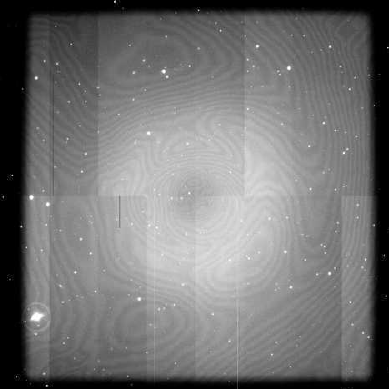}
    \includegraphics[height=1.8in]{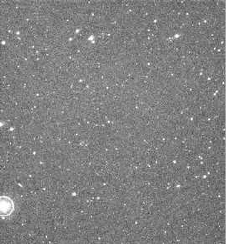}
   \caption{Example of an image before and after the full processing  (miniJPAS-AEGIS1, in the filter J0880).}
  \end{subfigure}
  \caption{Examples of image processing. In the left panels, an example of the background subtraction. In the right panels, an example of the full processing, from a raw image to the final one.}
  \label{fig:processed}
\end{figure*}

Astrometric calibration is the last step needed for the proper combination of the images. 
This is carried out again as a part of the
standard procedure for OAJ images using the software
\texttt{Scamp} \citep{bertin2010scamp} and the Gaia DR2 \citep{gaia2018}
as reference catalogue.  The typical uncertainty of the final astrometry of the
single frames, with respect to Gaia, is $\sim0.035$~arcsec. That is a small fraction ($\sim 7-15\%$) of the
pixel size, which is $\sim0.23$~arcsec for not binned images and $\sim0.46$~arcsec for the binned ones.


\subsection{Final coadded images} \label{sec:coadded}

Once the single frames are cleaned of instrumental effects, they
are combined with the
Astromatic\footnote{\url{https://www.astromatic.net}} software
\texttt{Swarp}\ \citep{bertin2010}. All the images are sampled to the
fiducial pixel size of the camera (i.e., $0.23$~arcsec~pixel$^{-1}$), including all
the images of the narrow bands that were observed with a binning of
2$\times$2. We opted for keeping the orientation as the one 
of the original frames instead of forcing to have the coadded images
oriented with North towards up (increasing Y axis) and East toward
left (decreasing X axis). In addition to the combined scientific
images, \texttt{Swarp} also constructs a normalized weight image that
keeps track, for example, of areas were the number of combined images
is lower (for instance, in the edges due to the dithering pattern but
also in regions were spurious detection were previously masked in one
or more single frames). Therefore, the weight image can be considered
as a map of the effective exposure time of the different parts of the
image. This is taken into account, for example, in the masking
process explained below, in Sect.~\ref{sec:masks}.




\subsection{PSF treatment} \label{sec:psf}

One of the main challenges of large field surveys is to provide
homogeneous photometry for a large number of objects. Each \mjp\ 
pointing is composed of 60 coadded images from multiple
exposures. Variations in the PSF from filter to filter 
 can produce
inhomogeneous photometry, thus artificial color terms, since the light coming from the same area of the
sky will be redistributed differently in the final image depending on
the filter used.

This discontinuous PSF variations can cause problems in the photometry
for objects with a significantly different PSF between the tiles. The
homogenization process allows PSF-Matched aperture-corrected
photometry measurements to be more consistent, at the expense of degrading
some images (see Sect.~\ref{sec:photometry}).  This method uses
\texttt{PSFEx}~\citep{bertin2011} to calculate a PSF homogenization kernel, to
convert variable instrumental PSFs to constant round Moffat profiles
for practical purposes. The
homogenization process has consequences in the image noise and
generates variable noise correlations over the image pixels.  For that
reason, the algorithm recalculates the noise model of the images that
later is used to compute photometric errors \citep{molino2014}.  We
  only apply the homogenization kernel to the corrected PSF apertures
  (\texttt{PSFCOR} and \texttt{WORST\_PSF}, see
  Sect.~\ref{sec:photometry}) in the dual-mode catalogues, while for all
  other apertures no convolution takes place.

As product of the PSF analysis of the images, we provide the overall PSF model produced by \texttt{PSFEx} for each image, as well as a service that generates on-demand an actual image of
the PSF model in any position of any image (see Sect.~\ref{sec:psf_models}).


\subsection{Photometry} \label{sec:photometry}

The detection of sources in the images is done with the widely used
software \sext\ \citep{bertin1996}. To cover as much as possible the
needs of the astronomical community we have run \sext\ in two
different complementary ways:

\begin{description}

\item [{\bf Dual-mode}.] To construct catalogues in which the photometry in
  the different bands for each object is done consistently, we have
  run \sext\ in the so called ``dual-mode''. In dual-mode, \sext\ is
  first run on a reference image for source detection and for the definition
  of the position and sizes of the apertures.  Afterwards, it is run on the other 
  images, where the photometry is performed within the apertures defined
  by the reference image (forced photometry). In our case, we chose the $r$ band co-added images as
  reference images for constructing the dual-mode
  catalogues\footnote{For very faint objects, with fluxes at the
    noise level in bands different than the detection one, it is possible
    to obtain negative fluxes. This is not a problem and is
    consistent with the fact of measuring within the noise.}. \\
    
\item [{\bf Single-mode}.] In the so called ``single-mode'', both the detection
  and the photometry are performed for each filter separately.
  This has the advantage that
  objects not detected in the reference band (e.g., faint objects with
  strong emission lines out of the $r$ band) can be identified. The
  drawback is that, for the same object, the photometry can be done in
  slightly different positions in the different bands, increasing the
  chance for inconsistent colours. 
\end{description}

The detection parameters were set to particular values to compromise
between large detectability (i.e., completeness), while avoiding
spurious detections (i.e., purity). Due to higher noise in all the
bands different to the $r$ band, the detection threshold (in units of
the $\sigma$ of the background) for the generation of single-mode
catalogues were set to a higher value (\texttt{DETECT\_THRESH=2}) than
for the dual-mode
catalogues~(\texttt{DETECT\_THRESH=0.9})\footnote{The value of the other parameter controlling the source detection is 
  \texttt{DETECT\_MINAREA}, which was set to 5.}. The full list of \sext\ parameters can be found in the Appendix~\ref{sec:sext_params}.


\subsection{Photometric calibration} \label{sec:calibration}

The final step to generate the final catalogues is the photometric
calibration of the fluxes. 
The photometric calibration of \mjp\ data was adapted from the methodology presented in \citep{clsj2019cal}, developed to provide the calibration of \jplus. We applied the following steps:

\begin{enumerate}[i.]
\item Definition of a high-quality sample of stars for calibration. We selected those \mjp +{\it Gaia} sources with $S/N > 10$ in all the photometric bands and with $S/N > 3$ in {\it Gaia} parallax. We constructed the dust de-reddened absolute $G$ magnitude vs. $G_{BP} - G_{RP}$ colour using the information from the 3D dust maps in \citet{green2018} and selected those sources belonging to the main sequence. This provides 334 calibration stars.

\item Calibration of the $gri$ broad-band filters to the Pan-STARRS photometry. 
We compared our $6^{\prime\prime}$ aperture magnitudes corrected to total
magnitudes with the PSF magnitudes in Pan-STARRS. The aperture
corrections were computed from the light growth curves of unsaturated,
bright stars in each tile and are stored in the ADQL table
\texttt{minijpas.TileImage}. This step provides the zero points of the $gri$ broad band filters. We computed also the zero points by comparison with \jplus\ photometry, obtaining differences below $0.01$ mag.

\item Homogenization of the narrow bands with the stellar locus. For each NB, we compute the dust de-reddened $(\mathcal{X}_{\rm ins}-r)_0$ vs. $(g-i)_0$ colour-colour diagrams  of the calibration stars, where $\mathcal{X}_{\rm ins}$ is the instrumental magnitude of the selected NB. From this, we computed the offsets that lead to a consistent stellar locus among the pointings. This provides a homogeneous instrumental photometry in \mjp.

\item Absolute colour calibration. The last step is to translate the instrumental magnitudes on top of the atmosphere. This step was done in \jplus\ with the white dwarf (WD) locus, but there is no high-quality WD in the \mjp\ area. We therefore used BOSS stellar spectra as reference to provide the absolute calibration of \mjp\  colours. After a visual inspection of the available BOSS spectra, we ended up with 40 stars. We compared the synthetic $(\mathcal{X}-r)$ colours from BOSS spectra with the instrumental magnitudes from \mjp\ to obtain one offset per passband except for $r$, which is used as reference and anchored to the Pan-STARRS calibration.

\item The BOSS spectra do not cover the $u_{\rm JAVA}$ and $u_{\rm JPAS}$ filters. For these bands, we obtained the colour offset by direct comparison with \jplus\ photometry. 
\end{enumerate}

Summarising, the flux calibration of \mjp\ photometry is
referred to the Pan-STARRS $r$ band and the colours are anchored
to the BOSS spectra except for the $u_{\rm JAVA}$ and $u_{\rm JPAS}$, which are anchored to the \jplus\ photometry.

We estimate the zero points also by direct comparison with \jplus\ photometry in the shared or similar passbands. Such comparison provides consistent zero points at the 4\% level. Thus, we conclude that the current photometric calibration has an absolute error of $\sim 0.04$~mag.
We set this as a safe upper limit in the calibration performance because of the limited statistics to provide an accurate estimation.
The consistency in the calibration among tiles is tested in Sect.~\ref{sec:overlap} by comparing the
photometry in overlapping areas. Note that this comparison only reflects the errors
in the \mjp\ homogenization (step iii) and is not sensitive to the
uncertainties in
the absolute colour calibration (steps iv and v).
 We expect to reach a 1-2\% accuracy in the \jp\ photometric calibration when $\sim500$ deg$^2$ are gathered. This will provide a few hundred WDs to derive a consistent colour calibration with the WD locus and a robust estimation of the uncertainties thanks to a large number of
overlapping areas with duplicated
measurements of the same sources.


\subsection{Masks} \label{sec:masks}

In order to help the identification of problematic areas in the images, we
also computed masks, provided in the \texttt{MANGLE}
format\footnote{\url{https://space.mit.edu/~molly/mangle/}}\citep{swanson2008}. 
In addition, we flag the objects falling in
those masked areas storing this information in the
\texttt{FLAGS\_MASK} column of the catalogues (see Sect.~\ref{sec:flags}).

The problematic areas that we are currently identifying and masking
are the following:
\begin{description}

\item [\textbf{Window frame mask.}] This mask identifies regions where
  the normalized weight map values are less than 85\%. This is
  determined from the tiling weight-map image, in order to homogenize
  the effective exposure times for the same coadd. The threshold value
  has been selected to be a compromise to maximize the valid
  observation area and minimize regions with less effective exposure
  time where, usually, a lower number of images have been
  combined. Nevertheless, for each object we compute the value of the
  normalized weight map at its position, which is stored in the
  parameter \texttt{NORM\_WMAP\_VAL}. This parameter takes a value of
  1 in the area with highest effective exposure. \\
  
\item [\textbf{Mask of bright stars.}] The mask of bright stars discards 
  regions around bright stars found in the Bright Star
  Catalog\footnote{\url{https://heasarc.gsfc.nasa.gov/W3Browse/star-catalog/bsc5p.html}}
  and the Tycho-2 catalog \citep{hoog2000}. The radius of each masked
  region is a function of the magnitude of the star. \\
  
\item [\textbf{Mask of artefacts.}] This mask identifies obvious
 artefacts in the images, usually due to light reflections in the
  telescope or its optical elements. We developed specific algorithms able to automatically
   detect, analyse and mask artefacts or patterns. In addition, artefacts 
  not detected automatically are manually masked.
\end{description}

We provide the masks for each tile or coadded image as well as a
combined mask of the whole \mjp\ footprint\footnote{The masks are
  available for download in the ``Image Search'' service of the
  Catalogues Web Portal.}.

\section{The \mjp\ data set and data access} \label{sec:data}

In this section we  describe the data products available and  how to access them.

\subsection{Data products\label{sec:data_products}}%

This data release includes  images, basic catalogue data
(parameters measured from images, such as photometry or morphology
data), as well as value-added information.

For images we provide not only the final coadded image but also the
following ancillary data\footnote{All this information is available
  through the ``Image Search'' service of the Catalogues Web Portal (see Sect.~\ref{sec:data_access}).}:
\begin{itemize}
\item [$\bullet$] {Basic parameters of the observations, like total exposure time, photometric zero
  points, estimations of the photometric depth, etc. \vspace{1.5mm}}
\item [$\bullet$] {Information on the single frames used for generating the co-added image, including the details of the sky conditions during  the observations. \vspace{1.5mm}}
\item [$\bullet$] {The weight image resulting from the \texttt{Swarp}\ co-adding procedure (see Sect.~\ref{sec:coadded}). \vspace{1.5mm}}
\item [$\bullet$] {Mask (see Sect.~\ref{sec:masks}).\vspace{1.5mm}}
\item [$\bullet$] {PSF model as resulting from \texttt{PSFEx} (see Sect.~\ref{sec:psf})}.
\end{itemize}

The information of individual objects or detections is placed in different tables in a relational database.
We provide two kinds of tables storing the data coming from running
\sext\ in both, ``dual-mode'' and ``single-mode''\footnote{In the
  database, these tables have in their names the tag \texttt{Dual} and
  \texttt{Single}, respectively.}, as explained in
Sect~\ref{sec:photometry}.
For the dual-mode, the photometry of each detection in all the
bands is unique by construction. In the catalogues, for
each detection in the reference band, we provide the list of 
geometrical parameters (position of the barycentre, shape, FWHM,...)
as well as different types of photometric measurements in all the
bands.
For the single-mode, the relation between
measurements in different bands is not always straightforward and the
actual identification of measurements in different bands will depend
on the strategy for the cross-matching between catalogues. Therefore,
we opted for providing a catalogue with all the detections treated
independently leaving to the user the freedom to perform the
cross-match in the most convenient way. 

In summary, each entry in the dual-mode catalogue corresponds to one object detected in the $r$ band and its photometry in all the \jp\ bands while each entry in the single-mode catalogue is a detection of one object in one band (and only if it is detected in that band) and, hence, each detection  in each band has its own entry in the single-mode catalogue. 
Table~\ref{tab:stats} shows a brief summary of the basic number counts
for each pointing and for dual- and single-mode catalogues.

To cover the needs of different kinds of analyses, we provide the
photometric catalogues in three different units:
\begin{itemize}
\item [$\bullet$] {AB magnitudes (names of the tables tagged with \texttt{MagAB}). \vspace{1.5mm}}
\item [$\bullet$] {Fluxes as a function of wavelength in units of
  $10^{-19}$erg$\,$s$^{-1}\,$cm$^{-2}\,\AA^{-1}$ (names of the tables tagged with
  \texttt{FLambda}). \vspace{1.5mm}}
\item [$\bullet$] {Fluxes as a function of frequency in units of
  $10^{-30}$erg$\,$s$^{-1}\,$cm$^{-2}\,$Hz$^{-1}$ (names of the tables tagged with \texttt{Fnu}.).}
\end{itemize}


\begin{table}[h]
  \centering
   \caption{Detected objects for each \mjp\ pointing. N$_{\rm Dual}$ refers to the total number of detections in
  the dual-mode catalogues as derived by \sext\ (see Sect.~\ref{sec:photometry} and \sext\ parameters in Appendix~\ref{sec:sext_params}). N$_{\rm Single}$ refers to the total number of detections in the
  single-mode catalogues. Note that for the single-mode catalogues a higher detection threshold for \sext\ has been applied (see Sect.~\ref{sec:photometry}).  }
  \begin{tabular}{c c c}
  \hline
  \hline 
 Pointing & N$_{\rm Dual}$ & N$_{\rm Single}$ \\
\hline
miniJPAS-AEGIS1  & 20016 & 167150 \\
miniJPAS-AEGIS2  & 13836 & 142481 \\
miniJPAS-AEGIS3  & 15792 & 142496 \\
miniJPAS-AEGIS4  & 14649 & 152443 \\
\hline
Total & 64293 & 604570 \\
\hline
  \end{tabular}
  \label{tab:stats}
\end{table}

We detail the different types of photometry that are provided in
the database:\footnote{For clarity, we use in this description the
  names of the columns in the tables storing magnitudes. For tables
  storing fluxes, the names are equivalent, exchanging \texttt{MAG} with
  \texttt{FLUX}.}
\begin{description}
\item[{\bf \magauto, \magiso, \magpetro .}] {\ \\
These are different types of
  estimations of total magnitudes. The reader is referred to \sext\
  User's Manual for a detailed description. \vspace{1.5mm} }
\item[{\bf \texttt{MAG\_APER\_...}.}] {\ \\
These correspond to aperture
  photometry in apertures of different sizes. The numbers in the names
  refer to the sizes of the aperture in arcseconds,
  e.g. \magaper{1}{5} corresponds to the photometry in an aperture of
  1.5~arcsec of diameter. \vspace{1.5mm}} 
\item[{\bf {\magiso}\texttt{\_WORSTPSF}, \texttt{MAG\_APER3\_WORSTPSF}.}] {\ \\
  These correspond to \magiso\ and \magaper{3}{0} with the
  particularity of being measured after transforming all the images of
  a given pointing to present a PSF size equal to the worst PSF
  among all the images. This is a straightforward procedure to remove
  the effect of the variation of the PSF among different filters on
  the photometry and to have more robust measurements of the
  colours. The disadvantage is the loss of
  information from the images with better PSF. \vspace{1.5mm}}
\item[{\bf \magpsfcor.}] {\ \\
This photometry is performed following the approach 
  of \citet{molino2019} \citep[see also][]{molino2014, molino2017} with the aim of applying
  corrections object by object in the images with worst PSF to correct
  for the differences in PSF among different bands. To
  increase the robustness of the color determination, instead of total
  magnitudes estimators like \magauto\ or \magiso, the fluxes are
  measured within an aperture with the same shape as the Kron aperture
  and a semimajor axis equal to 1 \texttt{KRON\_RADIUS}, also called
  restricted magnitudes \citep{molino2017, molino2019}. Being smaller
  than the aperture of \magauto, the signal-to-noise ratio is higher
  and, therefore, color measurements, which are key in spectral
  analysis like \photoz\ determination, are more robust. We stress, though, that these are not total magnitudes.}
\end{description}

\begin{table}
\caption{List of individual \mflags\ values. The final value of the
  \mflags\ parameter of a given object will be the sum   of all the
  individual \mflags\ flags that apply to that
  object.\label{tab:mask_flags}}
\begin{tabular}{ c l l }
\hline
\hline
Value &  Name & Description \\
\hline
0 & not mask & Not inside a mask \\
1 & window & Object is outside the window frame \\
2 & bright star & Object is bright star or near one \\
4 & artefact & Object masked due to nearby artefact \\
\hline
\end{tabular}
\end{table}

\subsubsection{Flags\label{sec:flags}}

In order to easily identify objects with known specific issues, for
each detected object two types of flags are provided to help in that
identification. 

\begin{description}

\item[{\bf \sext\ \flags.}]{ This parameter, inherited from \sext , contains 10 flag
bits with basic warnings about the source extraction process (see
Table~\ref{tab:flags}). To the original first 8 flags from \sext, the
values 1024 and 2048 have been added to identify, respectively, objects with
detections in more than one image (in the area where tiles overlap) and known variable objects from the cross-match with other surveys. Like in \sext, the final value of the \flags\ parameter
of a given object is the sum of the individual values of the \flags\
that affect that object. \vspace{1.5mm}}

\begin{table*}[h]
\centering
\caption{List of individual \flags\ values. The final value of the
  \flags\ of a given object will be the sum   of all the individual
  \flags\ values that apply to that  object.\label{tab:flags}}
\begin{tabular}{c l p{9cm}}
\hline
\hline
Value & Name & Description \\
\hline
1 & close neighbours & The object has neighbours, bright and close enough to significantly bias the photometry, or bad pixels (more than 10\% of the integrated area affected) \\
2 & blending & The object was originally blended with another one \\
4 & saturation & At least one pixel of the object is saturated (or very close to saturation) \\
8 & truncated & The object is truncated (too close to an image boundary) \\
16 & aperture incomplete & Object's aperture data are incomplete or corrupted \\
32 & isophotal incomplete & Object's isophotal data are incomplete or corrupted \\
64 & memory overflow deblending & A memory overflow occurred during deblending \\
128 & memory overflow extraction & A memory overflow occurred during extraction \\
1024 & duplicated & The object has been marked as duplicated \\
2048 & known variable & The object could be a known variable \\
\hline
\end{tabular}
\end{table*}

\item[{\bf Mask \flags.}] {Besides the issues affecting individual objects, there are large areas
of the images affected by different problems. Those areas are
identified and masked (see Sect.~\ref{sec:masks}) and the objects
falling inside those masked areas are flagged using the \mflags\
parameter. In Table~\ref{tab:mask_flags} are shown the individual
values that the \mflags\ parameter can have and the corresponding
image issues. Like in the case of the \flags\ parameter, the final
value of the \mflags\ parameter for each object is the sum of all the
individual \mflags\ that applied to that object. \vspace{1.5mm}}
\end{description}

\subsubsection{PSF models\label{sec:psf_models}}

In Sect.~\ref{sec:psf} we described how we treated variations of the PSF from filter to filter to achieve 
a homogeneous photometry. In the database the PSF models are provided in two different ways:
\begin{itemize}
\item[$\bullet$] {As the direct output of \psfex. This can be downloaded for each
  image through the ``Image Search'' service of the Web Portal. \vspace{1.5mm}}
  
\item[$\bullet$] As FITS images of the actual model of the PSF for a given
  position in a given image. This is an ``on-demand'' service accessible via HTTP request to allow programmed access.\footnote{See
    \url{http://archive.cefca.es/catalogues/minijpas-pdr201912/download_services.html\#link_get_psf_by_position}.}
\end{itemize}

\begin{figure}
\centering
\includegraphics[width=\columnwidth]{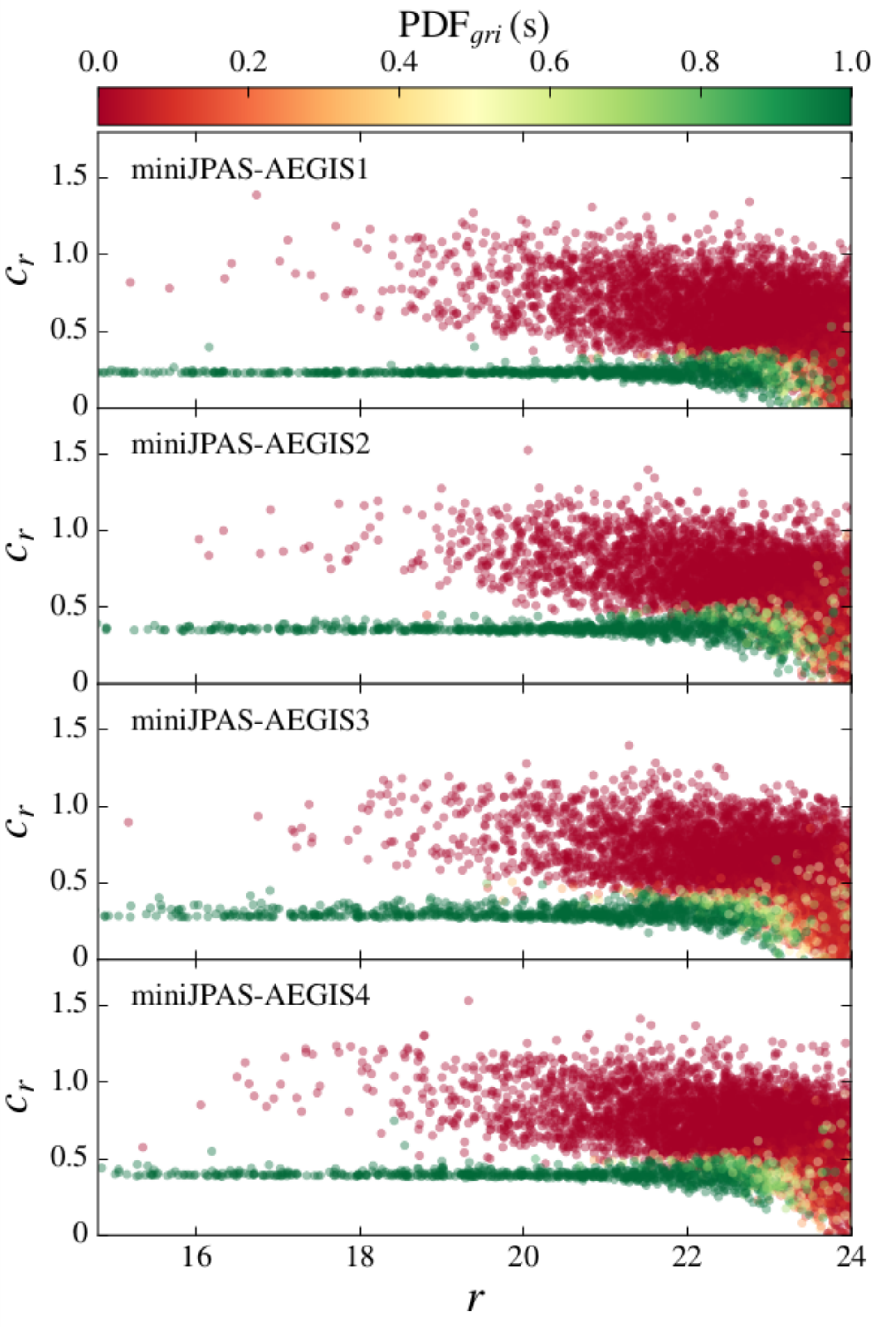}
\caption{Concentration-magnitude relation (\magauto) for sources in each individual \mjp\ pointing. Symbols are color-coded  according to the probability of being compact or extended, based on the SGLC of \citet{clsj2019mor}, adapted to \mjp\ data. }
\label{fig:conc_mag}
\end{figure}

\subsection{Value-added catalogues} \label{sec:value_added}

On top of the basic photometric information described above, the database contains a wealth of ancillary information to facilitate the scientific analysis of the data. We describe the most relevant ones below.

\subsubsection{Stellarity index} \label{sec:star_galaxy}

\begin{figure}
\centering
\includegraphics[width=\columnwidth]{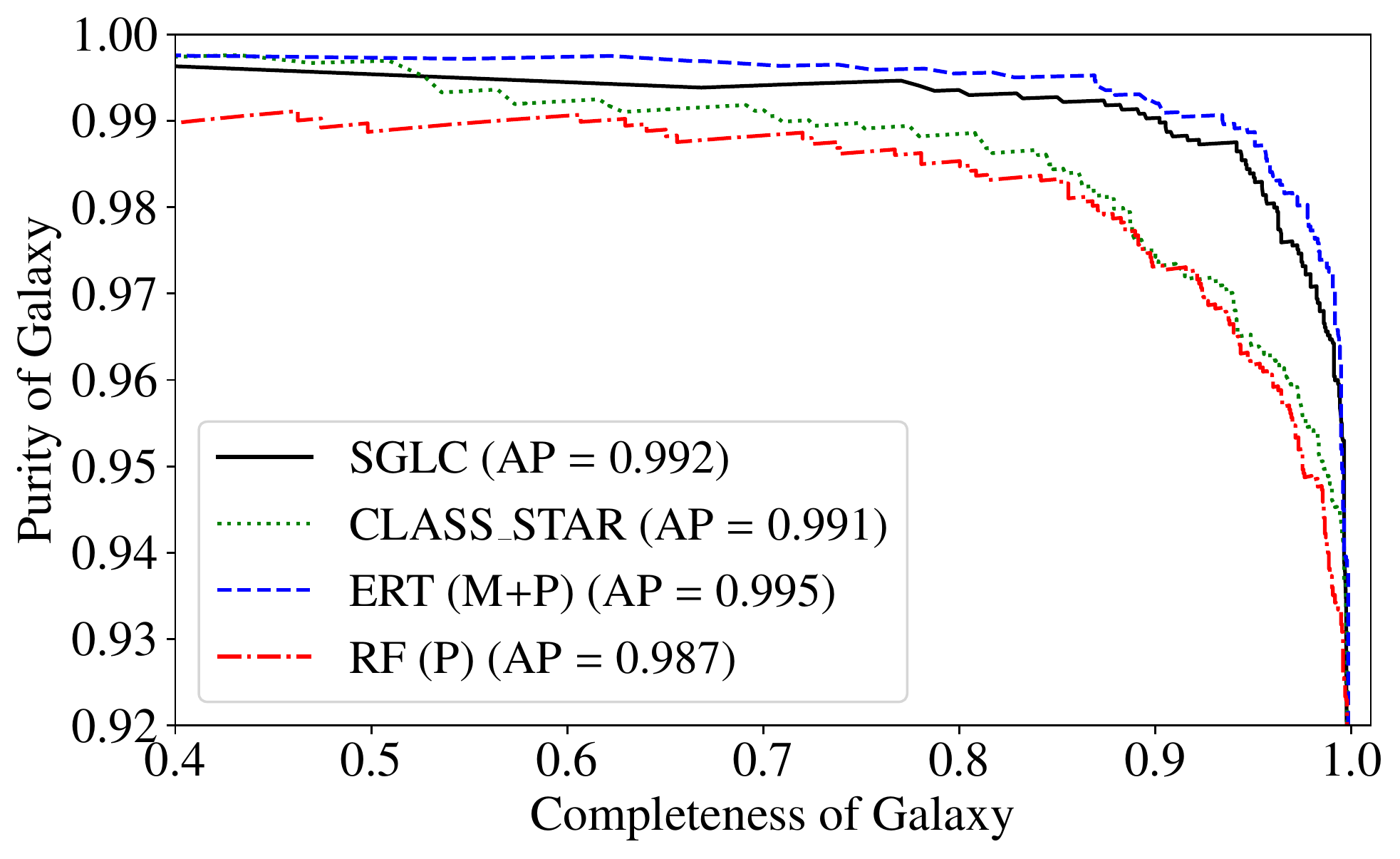}
\caption{Purity-completeness curves for galaxies in the magnitude range $15 \leq r \leq 23.5$, based on the morphological classification of SExtractor (CLASS\_STAR, green dotted line), SGLC (black solid line), Extremely Randomized Trees (blue dashed line) and Random Forest (red dot-dashed line). ERT uses both the photometric (P) and morphological (M) information of the sources, while the RF uses photometric information only.}
\label{fig:MLclass}
\end{figure}

The database provides the results of different complementary methods for the star/galaxy classification of the sources, where ``stars'' are considered all compact/point-like objects (thus including quasars and very compact galaxies)  and ``galaxies'' all resolved ones. We provide here a brief description of the three approaches:

\begin{description} 
	\item[{\bf SExtractor classification}] {\sext\ automatically provides a morphological classification (the CLASS\_STAR parameter) for all detected objects. We refer to the manual of \sext\ for details on this classification. \vspace{1.5mm}}
	
	\item[{\bf Stellar-Galaxy Locus Classification (SGLC)}] {We applied the Bayesian star/galaxy morphological classifier developed in	\citet{clsj2019mor} for \jplus\ data. The concentration vs.
		magnitude diagram presents two distinct populations,
		corresponding to compact and extended
		 sources (see Fig.~\ref{fig:conc_mag}). We
		modelled both populations to obtain for each source a probability for being
		 compact or extended, as indicated in Fig.~\ref{fig:conc_mag} by the color of the symbols. 
		 The most relevant update with
		respect to the \jplus\ methodology is the modification of the
		galaxy population model. The galaxy locus is assumed to be 
		constant with magnitude in the \jplus\ analysis, but such assumption does not
		hold at the fainter magnitudes probed by \mjp\ data. Galaxies
		become apparently smaller at larger magnitudes, with the galaxy
		locus approaching asymptotically to the stellar locus. We
		modelled such trend with an error function calibrated with 
		\mjp\ data. The morphological information in the $g$, $r$ and $i$ broad band filters
 		 was combined. A prior in $r$ magnitude, accounting for the larger
		number of galaxies at fainter magnitudes and estimated in each
		pointing independently, was applied. The modelling of the
		stellar and galaxy populations was done pointing-by-pointing as
		in \citet{clsj2019mor}. 
		As we will show in Sect.~\ref{sec:number_counts}, the number
		counts derived from this Bayesian classification agree with the
		expectations from the literature up to $r \sim 23.5$. The derived
		probabilities for the morphology of each source are publicly
		available in the ADQL table \texttt{minijpas.StarGalClass}.\vspace{1.5mm}}

	\item[{\bf Machine learning classification}] {
We used machine learning (ML) to classify sources of \mjp\ as stars or galaxies in the magnitude interval $15 \leq r \leq 23.5$. In order to train and test our classifiers, we cross-matched the \mjp\ dataset with SDSS and \hsc\ data, whose classification we assume to be trustworthy within the intervals $15 \leq r \leq 18.5$ and $18.5 < r \leq 23.5$, respectively. The best ML classifiers are Extremely Randomized Trees (ERT) and Random Forest (RF), whose performance is shown in Fig.~\ref{fig:MLclass} as compared to \sext\ and SGLC, described above. We can see that, when using morphological parameters, ERT outperforms SGLC. For the case in which we use only photometric bands, RF is the best classifier. For a more detailed analysis with other methods and metrics see Baqui et al. (2020). A value added catalogue is available in the \mjp\ database via the ADQL table \texttt{minijpas.StarGalClassML}}
	
\end{description}

\begin{figure}
\includegraphics[width=\columnwidth ]{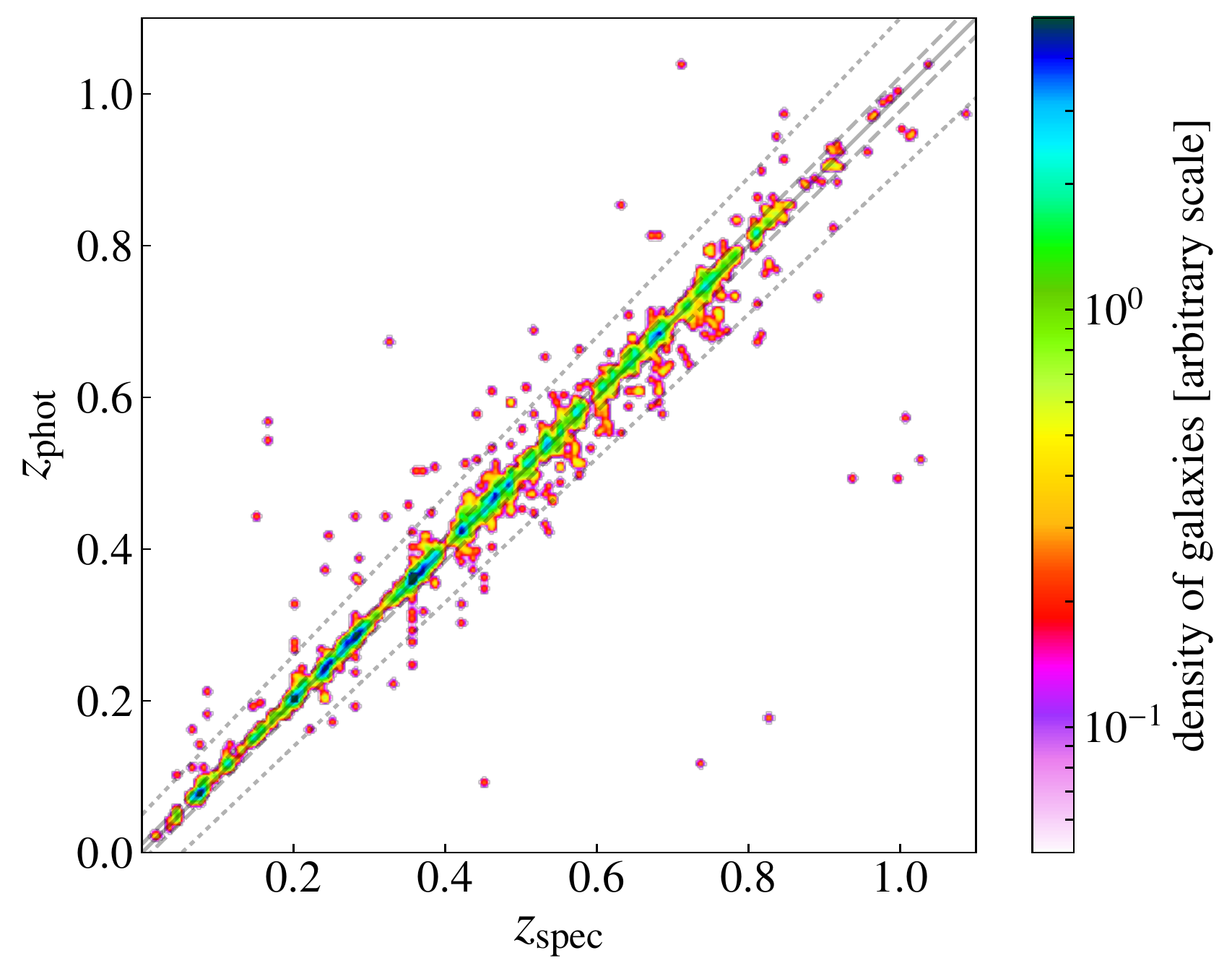}
\caption{
Comparison between photometric and spectroscopic redshifts for
  \mjp\ galaxies  at $r<22.5$ with ODDS>0.5. In order to better highlight the density of points, we have applied a 2-D Gaussian smoothing to the data. 
The solid line marks the 1:1 relation. The two dashed lines enclose a
region containing $80\%$ of the sources, while the dotted lines indicate
the $\lvert\Delta z\rvert = 0.05$  threshold used to define redshift outliers. Note that each outlier smoothed regions correspond to individual sources.}
\label{fig:zphot_zspec}
\end{figure}

\begin{figure}
\includegraphics[width=\columnwidth ]{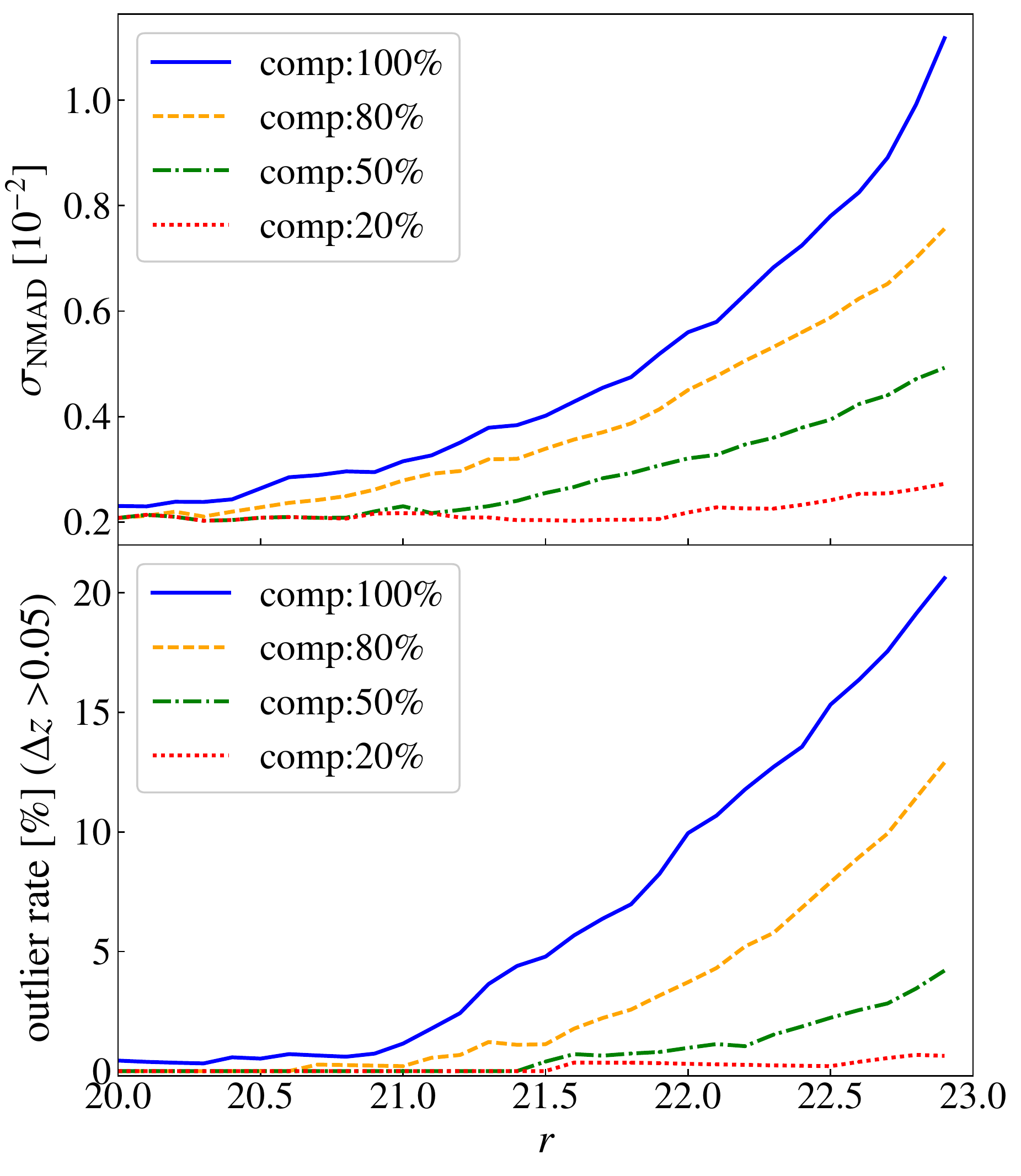}
\caption{Normalized median absolute deviation of \photoz\ errors (top panel, in units of $[10^{-2}]$) and outlier rate (bottom panel) as a function of the limiting magnitude in the $r$-band (\magauto).  Outliers are defined as having redshift errors $\lvert\Delta z\rvert$ $>$ 0.05.  The colors refer to different cuts in the ODDS parameter, resulting in the selection of 100\%, 80\%, 50\% or 20\% of the sources in the spectroscopic sample. The $\sigma_{\rm{NMAD}}$  values and the associated errors are provided in  Table~\ref{tab:photoz}.}
\label{fig:snmad_mag_odds}
\end{figure}

\subsubsection{Photometric redshifts} \label{sec:photoz}


The photometric redshifts (photo-$z$) for \mjp\ were obtained with the \textsc{jphotoz} package, developed at CEFCA as part of the reduction pipeline for the \jplus\ and \jp\ surveys. \textsc{jphotoz} is a set of python scripts that acts as interface between the database and the actual \photoz\ computing code, which is a custom version of \textsc{LePhare} \citep{arnouts2011} modified to work with a larger number of filters and higher resolution in redshift than typically required for broadband photometry. 

\textsc{LePhare} computes \photoz\ by fitting the observed photometry of each source with a set of templates. We used 50 galaxy templates specifically tailored for \mjp\ data. The templates are stellar population synthesis models generated with \texttt{CIGALE}\footnote{\url{http://cigale.lam.fr}} that match the \js\ of individual \mjp\ galaxies. The process of generation and selection of the most suitable set of templates is described in detail in Hern\'an-Caballero et al. (in prep.). 


To test the accuracy of \mjp\ \photoz\ we use a subsample of galaxies with spectroscopic redshifts taken from SDSS DR12 and the DR4\footnote{http://deep.ps.uci.edu/DR4/home.html} of the DEEP2 Galaxy Redshift Survey \citep[][]{newman2013}. The later covers the
footprint of the EGS and includes $12,051$ reliable galaxy redshifts down to magnitude $r=24.1$, with no preselection in magnitude or colour. We
matched sources in the \mjp\ catalogue with those in SDSS and DEEP2 using a search radius of 1.5~arcsec. To ensure a proper evaluation of the
\photoz\ accuracy in galaxies, we considered only sources with a reliable redshift determination (Q >= 3 in DEEP2 or zwarning = 0 in SDSS) and
spectroscopic classification as galaxy. In addition, we excluded sources with FLAGS>0 in the \mjp\
photometric catalogue. Table~\ref{tab:photoz} summarizes the total number of \mjp\ sources and the number of those used for evaluation of the \photoz\ accuracy in bins of magnitude.

The error in the \photoz\ for a given source is expressed by the quantity $\Delta z =( z_{\rm{phot}} - z_{\rm{spec}})/(1+z_{\rm{spec}})$. The distribution of $\Delta z$ is approximately Gaussian but with heavier wings far from the core due to an almost flat distribution of outliers (defined as those galaxies with catastrophic redshift errors $\lvert\Delta z\rvert$ $>$ 0.05) in the redshift search range.
A common statistic used to represent the width of the distribution is the normalized median absolute deviation $\sigma_{\rm{NMAD}} = 1.4826 \times {\rm median} (\lvert\Delta z - {\rm median} (\Delta z) \rvert)$, which equals the standard deviation for a purely Gaussian distribution but is less sensitive to the outliers. 

It is possible to select samples with more accurate \photoz\ (both in terms of $\sigma_{\rm{NMAD}}$ and outlier rate) by sacrificing the sources with the least reliable estimates. Our confidence in the \photoz\ determination of individual sources depends on the shape of their redshift probability distribution function (PDF). A common parametrization of this confidence is the ODDS parameter \citep{benitez2000}, defined as the integral of the PDF in a window of fixed width centred at the mode of the PDF.  For \mjp, we
choose a half-width of $0.03(1+z)$ for the integration window. The value
of the ODDS ranges from ~0 to 1, with higher values implying higher
confidence in the \photoz. 

In Fig.~\ref{fig:zphot_zspec} we show the photometric vs. spectroscopic redshift plane for the sources at $r<22.5$ with ODDS>0.5, which qualitatively indicates the high level of precision of our \photoz\ solutions.  The redshift precision for this sample is $\sigma_{\rm NMAD}=0.004$ and the outlier rate $\sim 4\%$.

\begin{figure}
\includegraphics[width=\columnwidth ]{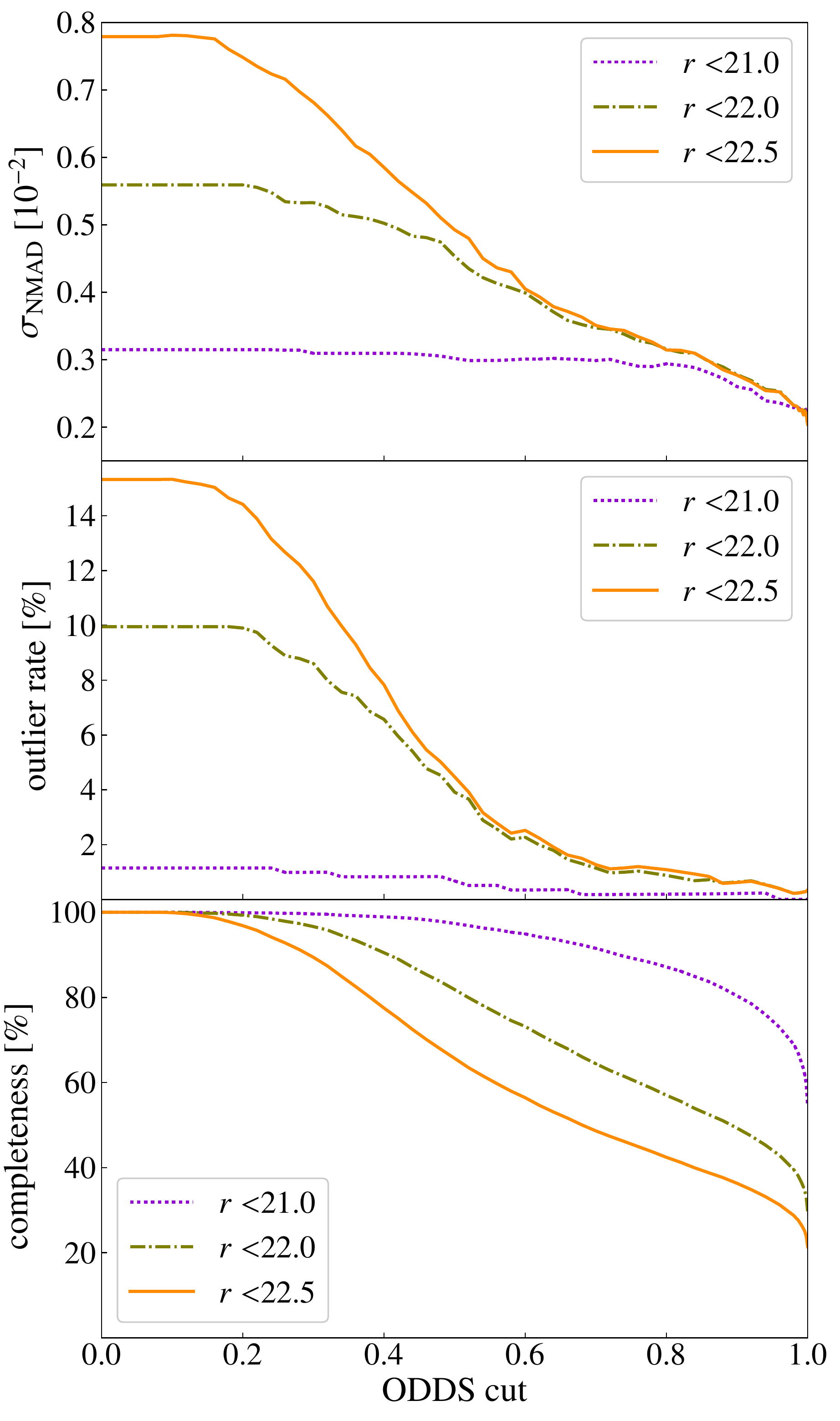}
\caption{Dependence of the Normalized median absolute deviation of \photoz\ errors (top panel) outlier rate (middle panel) and completeness (bottom panel) on the ODDS parameter for different $r$-band cuts.}
\label{fig:accuracy_odds}
\end{figure}

In Fig.~\ref{fig:snmad_mag_odds} we show the $\sigma_{\rm{NMAD}}$ and outlier rate for the sample with spectroscopic counterparts as a function of the limiting magnitude in the $r$ band, after applying cuts in ODDS corresponding to completeness of 20, 50, 80 and 100\%. As expected, both $\sigma_{\rm{NMAD}}$ and the outlier rate increase for fainter magnitude cuts due to the decrease in the average S/N of the \js. For the entire sample of 2421 galaxies at $r<22.5$ with reliable spectroscopic redshifts, we obtain $\sigma_{\rm{NMAD}} = 0.0078 \pm 0.0004$ and an outlier rate of $15.3\pm8\%$ (for $\Delta z=0.05$), where the uncertainties are calculated using bootstrap resampling. However, it is possible to select subsamples with $\sigma_{\rm{NMAD}}$ as low as $\sim 0.002$ and an outlier rate of $\sim 1-2\%$ just by imposing a more restrictive cut in ODDS. We  note that further refinement of the photometry might lead to even better \photoz\ depth.
The threshold value of the ODDS parameter required to achieve the desired redshift accuracy or completeness is shown in Fig.~\ref{fig:accuracy_odds}. In Sect.~\ref{sec:lss} we show how different cuts in ODDS translate to cuts in number densities and redshift precision for sources at different redshifts.

\begin{table*}[tb]
  \centering
  \caption{Redshift precision, in units of $[10^{-2}]$ for different magnitude cuts, differential and cumulative, and for different completeness cuts ($100\%$, $80\%$, $50\%$ and $20\%$) \label{tab:photoz}}
  \begin{tabular}{l l l l l l l}
  \hline
  \hline
  $r$ (\magauto) & Ntot\tablefootmark{a} & Nused\tablefootmark{b} &   $ \sigma_{\rm{NMAD}} \, [10^{-2}]$ ({\tiny 100\%})     &   $\sigma_{\rm{NMAD}}  \, [10^{-2}]$ ({\tiny 80\%})    &   $\sigma_{\rm{NMAD}} \, [10^{-2}]$ ({\tiny 50\%})     &   $\sigma_{\rm{NMAD}}  \, [10^{-2}]$ ({\tiny 20\%})  \\
 \hline
20.0--20.5 &  1103 &  155   & 0.34$\pm$0.04 & 0.28$\pm$0.03 & 0.25$\pm$0.04 & 0.20$\pm$0.04\\
20.5--21.0 &  1633 &  226   & 0.43$\pm$0.04 & 0.39$\pm$0.04 & 0.34$\pm$0.05 & 0.26$\pm$0.04\\
21.0--21.5 &  2360 &  394   & 0.68$\pm$0.06 & 0.58$\pm$0.07 & 0.40$\pm$0.04 & 0.33$\pm$0.04\\
21.5--22.0 &  3404 &  645   & 1.21$\pm$0.10 & 0.90$\pm$0.10 & 0.67$\pm$0.06 & 0.41$\pm$0.08\\
22.0--22.5 &  4795 &  773   & 2.30$\pm$0.29 & 1.72$\pm$0.20 & 1.11$\pm$0.12 & 0.58$\pm$0.07\\
22.5--23.0 &  6972 &  935   & 4.56$\pm$0.34 & 3.74$\pm$0.37 & 2.39$\pm$0.38 & 1.30$\pm$0.17\\
\hline
\hline
  $r$ (\magauto) & Ntot\tablefootmark{a} & Nused\tablefootmark{b} & $ \sigma_{\rm{NMAD}} \, [10^{-2}]$ ({\tiny 100\%})     &   $\sigma_{\rm{NMAD}}  \, [10^{-2}]$ ({\tiny 80\%})    &   $\sigma_{\rm{NMAD}} \, [10^{-2}]$ ({\tiny 50\%})     &   $\sigma_{\rm{NMAD}}  \, [10^{-2}]$ ({\tiny 20\%})  \\    
  \hline
  <20.5   &  4016  &  383 &  0.26$\pm$0.02 & 0.23$\pm$0.02 & 0.21$\pm$0.02 &  0.21$\pm$0.02  \\
  <21.0   &  5649  &  609 &  0.31$\pm$0.02 & 0.28$\pm$0.02 & 0.23$\pm$0.02 &  0.22$\pm$0.02  \\
  <21.5   &  8009  & 1003 &  0.40$\pm$0.02 & 0.34$\pm$0.02 & 0.25$\pm$0.02 &  0.20$\pm$0.02  \\
  <22.0   & 11413  & 1648 &  0.56$\pm$0.02 & 0.45$\pm$0.03 & 0.32$\pm$0.02 &  0.22$\pm$0.02  \\
  <22.5   & 16208  & 2421 &  0.78$\pm$0.04 & 0.59$\pm$0.02 & 0.39$\pm$0.02 &  0.24$\pm$0.02  \\
  <23.0   & 23180  & 3356 &  1.21$\pm$0.05 & 0.83$\pm$0.04 & 0.51$\pm$0.02 &  0.28$\pm$0.02  \\
\hline
  \end{tabular}
    \tablefoot{ 
 \tablefoottext{a}{Total number of objects (including compact sources) in \mjp\ for each magnitude bin/cut.}
 \tablefoottext{b}{Number of objects with spectroscopic counterpart used to calculate the  $\sigma_{\rm{NMAD}}$ statistics for  each magnitude bin/cut. The sub-sample is selected imposing no flags in \mjp\ (flags=0) in all bands, $0 < z_{\rm spec} <1.5$ and flags $zQ \geq 3$ for DEEP2 and \texttt{zwarning}=0 for SDSS.}
  } 
\end{table*}

\begin{figure}
\includegraphics[width=\columnwidth]{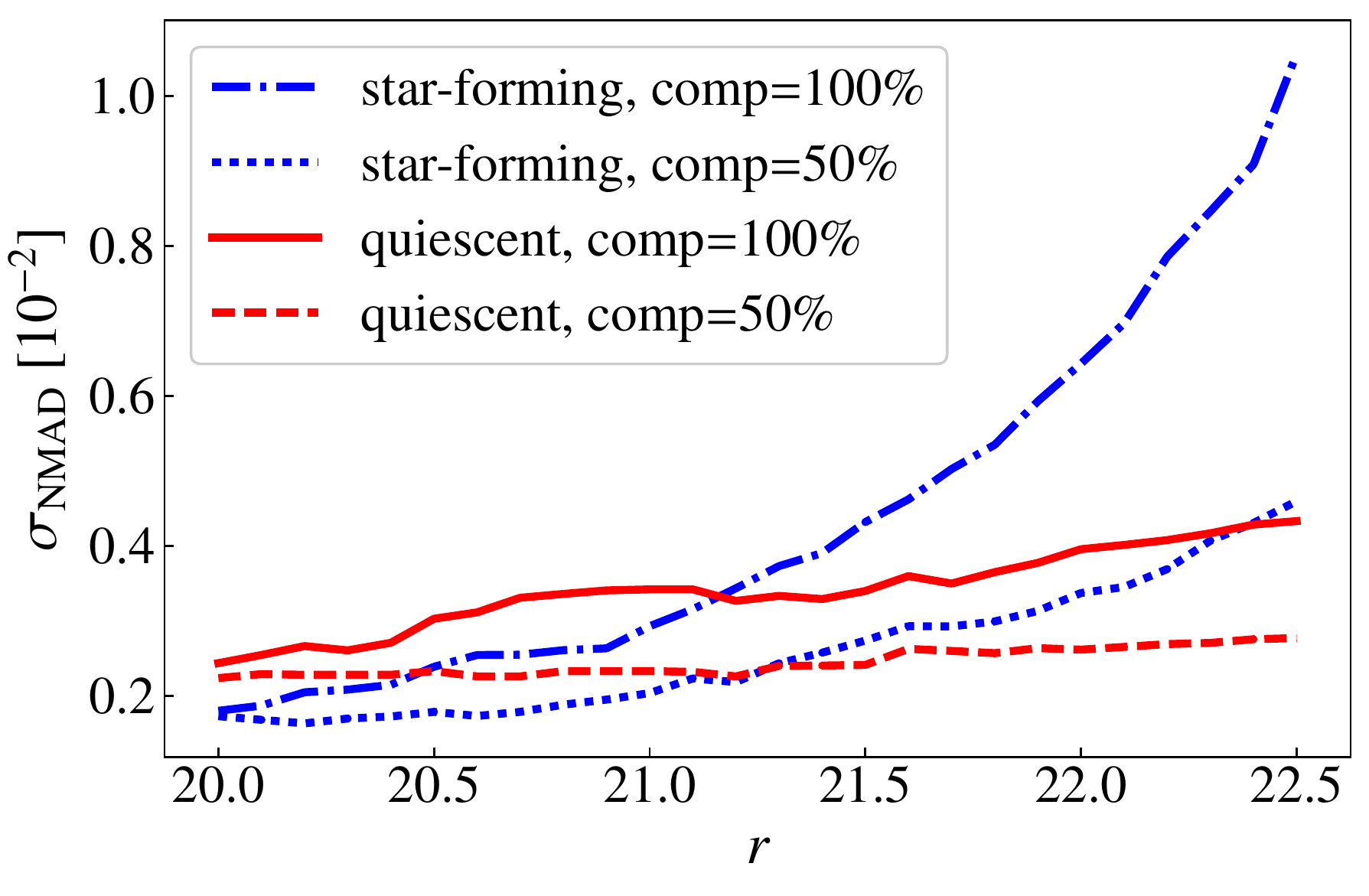}
\caption{Dependence of \photoz\ errors on the limiting magnitude in the $r$-band for galaxies in the ``passive'' subsample (red line) compared to other galaxies (main sequence and starburst galaxies, blue line).
Solid lines indicate values for 100\% completeness (no cut in ODDS) while dashed lines correspond to the 50\% with highest ODDS in each magnitude bin.}
\label{fig:snmad_redblue}
\end{figure}

The accuracy of the \photoz\ for galaxies also depends on their spectral class. In particular, LRGs are expected to have more accurate \photoz\ compared to the general population at the same magnitude thanks to a strong 4000~\AA\ spectral break. To estimate the \photoz\ accuracy for LRGs we split \mjp\ galaxies in two samples according to 
 the starforming/quiescent classification based on SED-fitting, discussed in Sect.~\ref{sec:gal_evol}, which has been performed for galaxies with $r < 22.5$. This classification is broadly consistent with a red/blue classification based on the Dn(4000) index, which measures the strength of the $4000$~\AA\  break \citep[see ][]{balogh1999}. 
Figure~\ref{fig:snmad_redblue} shows the redshift accuracy as a function of the limiting magnitude for the subsamples of passively evolving and main sequence/star-forming galaxies. Note that the redshift precision of blue galaxies depends more strongly with magnitude. This is because at fainter magnitude the emission lines typical of star-forming galaxies become weaker and we have to rely on the continuum to estimate the \photoz. For passive galaxies, instead, the precision of the \photoz\ estimate decreases only weakly with magnitude.
This same classification and the results on redshift accuracy are used in Sect.~\ref{sec:lss} to discuss the derived number densities of red and blue galaxies, in view of clustering studies of the large scale structure.  

\subsubsection{Extinction and cross-matches} \label{sec:xmatch}

We provide an estimation of the integrated $E(B-V)$ colour excess due to the 
Milky Way extinction along the line of sight of each detected source in
 the database table \texttt{minijpas.MWExtinction}. This was derived from
the \texttt{Bayestar17} \citep{bayestar2017} dust maps and coupled with the extinction coefficients $k_{\lambda}$ gathered in
table \texttt{minijpas.Filter}
and computed from the  \citet{schlafly2016} extinction curve
following the prescription in \citet{whitten2019b}. In addition to the
$E(B - V)$ value, we also provide a precomputed extinction correction
$k_{\lambda}\,E(B-V)$ for each source.

Finally, we provide the cross-match with the surveys listed in
Table~\ref{tab:xmatch}. For catalogues with a Vizier ID, the cross-match 
has been performed using the CDS X-Match
Service\footnote{\url{http://cdsxmatch.u-strasbg.fr/}}. We warn the reader that we report, for each \mjp\ source, all the objects within the search radius. Therefore, a \mjp\ source may have multiple counterparts in the cross-match catalogues.

\begin{table*}[t]
  \centering
  \caption{List of external catalogues for which a precomputed
    cross-match with \mjp\ is provided.\label{tab:xmatch}}
  \begin{tabular}{lllp{1cm}l}
  \hline
  \hline
    Catalogue Name  & Reference & Vizier ID
    & Search radius & Data base table \\
\hline\hline
J-PLUS DR1 &
\citet{cenarro2019}&
--- &
3" &
\texttt{xmatch\_jplus\_dr1}\\
ALHAMBRA Survey & 
\citet{molino2014} & 
J/MNRAS/441/2891/photom &   
4" &
\texttt{xmatch\_alhambra}\\
Gaia DR2 & 
\citet{gaia2018} &
I/345/gaia2 &
1.5" &
\texttt{xmatch\_gaia\_dr2} \\
PanSTARRS DR1 &
\citet{chambers2016} &
II/349/ps1 &
1.5" &
\texttt{xmatch\_panstarrs\_dr1} \\
SDSS DR12 &
 \citet{alam2015} &
V/147/sdss12 &
1.5"  & 
\texttt{xmatch\_sdss\_dr12}  \\
     All WISE & 
\citet{cutri2014} &
II/328/allwise  & 
4"  & 
 \texttt{xmatch\_allwise} \\
     GALEX AIS &
\citet{bianchi2011} & 
 II/312/ais  & 
4"  & 
 \texttt{xmatch\_galex\_ais} \\
     CFRS & 
\citet{lilly1995} &
VII/225B/catalog  & 
4"  & 
 \texttt{xmatch\_cfrs} \\
     DEEP2 & 
\citet{coil2004} &
 II/301/catalog  & 
1.5"  & 
 \texttt{xmatch\_deep2\_photo} \\
     DEEP2 All &
\citet{matthews2013}& 
III/268/deep2all  & 
1.5"  & 
 \texttt{xmatch\_deep2\_spec} \\
HSC-SSP PDR2 &
\citet{aihara2019}  &
-- &
1.5" &
\texttt{xmatch\_subaru\_pdr2} \\
\hline
  \end{tabular}
\end{table*}

\subsection{Data access} \label{sec:data_access}

Large  projects  like  \jp\ require  an  easy  and  agile  access to the data. For this purpose, CEFCA has developed a powerful \href{http://archive.cefca.es/catalogues}{Science Web Portal}\footnote{\url{http://archive.cefca.es/catalogues}} offering advanced tools for data search, visualization and download, each suited to a particular need~\citep{civera2020}. The data of this Public Data Release (PDR-201912) of \mjp\ can be accessed here: \url{http://archive.cefca.es/catalogues/minijpas-pdr201912}.

The portal includes a user-friendly sky navigator service including colour images of the survey. Sources are highlighted, and clicking on an object allows the user to visualize a summary of its  properties and  \js. Further object exploration is possible through an Object Visualization tool. This tool displays the detailed information about the selected source and the image thumbnail in each filter. It also provides tools to download reports and custom object fits cutouts and to perform cross matches with other catalogues.

An object list search tool is also offered, which lets the user upload a list of sky positions, object names or object identifiers and then returns a list of objects near those positions, displaying a customizable summary, \js\ and thumbnail images for all of them.
Users can also retrieve a list of objects within a certain angular distance to a given sky position, fulfilling additional brightness and photometric redshift criteria if needed, using the cone search service.

To download the full coadded images and derived products, an image search service has been implemented where users can select and  download the desired data using a variety of search criteria.
A coverage map service is also available which helps users to visualise the sky area covered and the distribution of the pointings of the survey. With the Multi-Order Coverage maps (MOC) download service \citep{moc}, users can download the MOC file which describes the area covered by the survey and can be used to compute very fast data set operations (e.g., unions, intersections) or query data (e.g., sources, images) of other data releases only inside this data release area, using external tools like VizieR\footnote{\url{http://vizier.u-strasbg.fr/}}, Aladin\footnote{\url{https://aladin.u-strasbg.fr/aladin.gml}} or Topcat\footnote{\url{http://www.star.bris.ac.uk/~mbt/topcat/}}.

An asynchronous queries interface based on Virtual Observatory (VO)\footnote{\url{http://www.ivoa.net/}} Table Access Protocol \citep[TAP,][]{tap} has  also been implemented in the portal where users can perform Astronomical Data Query Language (ADQL) queries \citep{adql} directly to the database.  All of these services support the Simple Application Messaging Protocol \citep[SAMP][]{samp} that enables the catalogues portal to interoperate and communicate with VO-compatible applications \footnote{\url{http://www.ivoa.net/astronomers/applications.html}}.

Catalogue data and images are also accessible through VO protocols \citep{hernandez2020}. These VO services allow the users to access data in a standardized way using VO compatible applications or their own scripts. Images are available via the Simple Image Access Protocol \citep[SIAP,][]{siap} that allows not only to search for all images covering a sky region, but also to retrieve the full fits images, cutouts or colour images. Catalogue data is accessible both via Simple Cone Search \citep[SCS,][]{scs} and TAP. 
The first allows to search all the objects within a given radius around a specified location, while TAP offers a  more flexible access to data tables allowing perform complex searches (using ADQL) on the tables storing information of images, filters, objects and \photoz.

Below we provide a summary of the different tools available to access the data:

\begin{description}

\item{ \bf \weburl{navigator.html}{Sky Navigator}.} Sky exploration by panning and zooming. By clicking on an object, one obtains a summary of its properties and the options to explore it or search it in other catalogues. \vspace{1.5mm}

\item{ \bf \weburl{object_list.html}{Object List Search}.} Search for a list of  objects via sky positions, object names or objects identifiers. It  returns a list of sources near those positions and displays a summary, photo-spectra and thumbnail images for the list of objects. \vspace{1.5mm}

\item{ \bf \weburl{image_search.html}{Image search}.}  Search and download images by position or name. \vspace{1.5mm}

\item{\bf \weburl{cone_search.html}{Cone search}.} Search the database for objects near a certain sky position. Restrictions on colors, magnitudes and \photoz\ can be added. \vspace{1.5mm}

\item{ \bf \weburl{coverage_map.html}{Coverage map}.} Shows the sky area covered by the data release. The map is linked to the Sky Navigator. \vspace{1.5mm}

\item{ \bf \weburl{moc.html}{Multi-Order Coverage Map (MOC)}.} To download the Multi-Order Coverage map (MOC), which describes the area covered by the data release (FITS file). \vspace{1.5mm}

\item{\bf \weburl{vo_services.html}{VO Services}.} Access to images and objects data through the Virtual Observatory (VO) protocols using VO compatible applications. VO services offered are Simple Cone Search (SCS), Table Access Protocol (TAP) and Simple Image Access Protocol (SIAP). \vspace{1.5mm}

\item{\bf \weburl{tap_async.html}{VO Asynchronous Queries (ADQL)}.} Search the database via Astronomical Data Query Language (ADQL) queries. A help manual with examples is provided. \vspace{1.5mm}

\item{\bf \weburl{download_services.html}{Direct Download Services}.} Allow easy automatic access to most of the data. It is possible to download directly some data through simple HTTP access\footnote{\url{http://archive.cefca.es/catalogues/minijpas-pdr201912/download_services.html}}. The data currently available through this service include:
  \begin{itemize}
  \item Full images and weight maps in FITS format and PNG.
  \item ``On-demand'' cutouts of images and weight maps in FITS and PNG formats.
  \item Masks of individual images in MANGLE format.
  \item PSF models of full frames in \psfex\ format or as actual FITS images of PSF models in a given position of any image.
  \item ``On-demand'' PSF models for a given position and image in FITS format.
  \item \photoz\ catalogues.
  \item Information on the original individual images.
  \end{itemize}
\end{description}

\section{Data quality} \label{sec:dava}

In this section we show the results of a variety of tests aimed  at characterizing  the quality of the data.

\subsection{Homogeneity across the footprint}  \label{sec:overlap}

A powerful test of the consistency of the data reduction and photometric calibration is the comparison
 of the photometry of objects in the overlap area of adjacent pointings. For these objects the data reduction and calibration have been  performed twice, in an independent way in each pointing (see Sect.~ \ref{sec:calibration}). Based on the \mjp\ observation strategy, the overlap area is approximately 10\% of the whole area of the total footprint (see Sect.~\ref{sec:footprint}).

\begin{figure}[t]
\centering
\includegraphics[trim={10 0 40 30}, clip, width=0.49\textwidth]{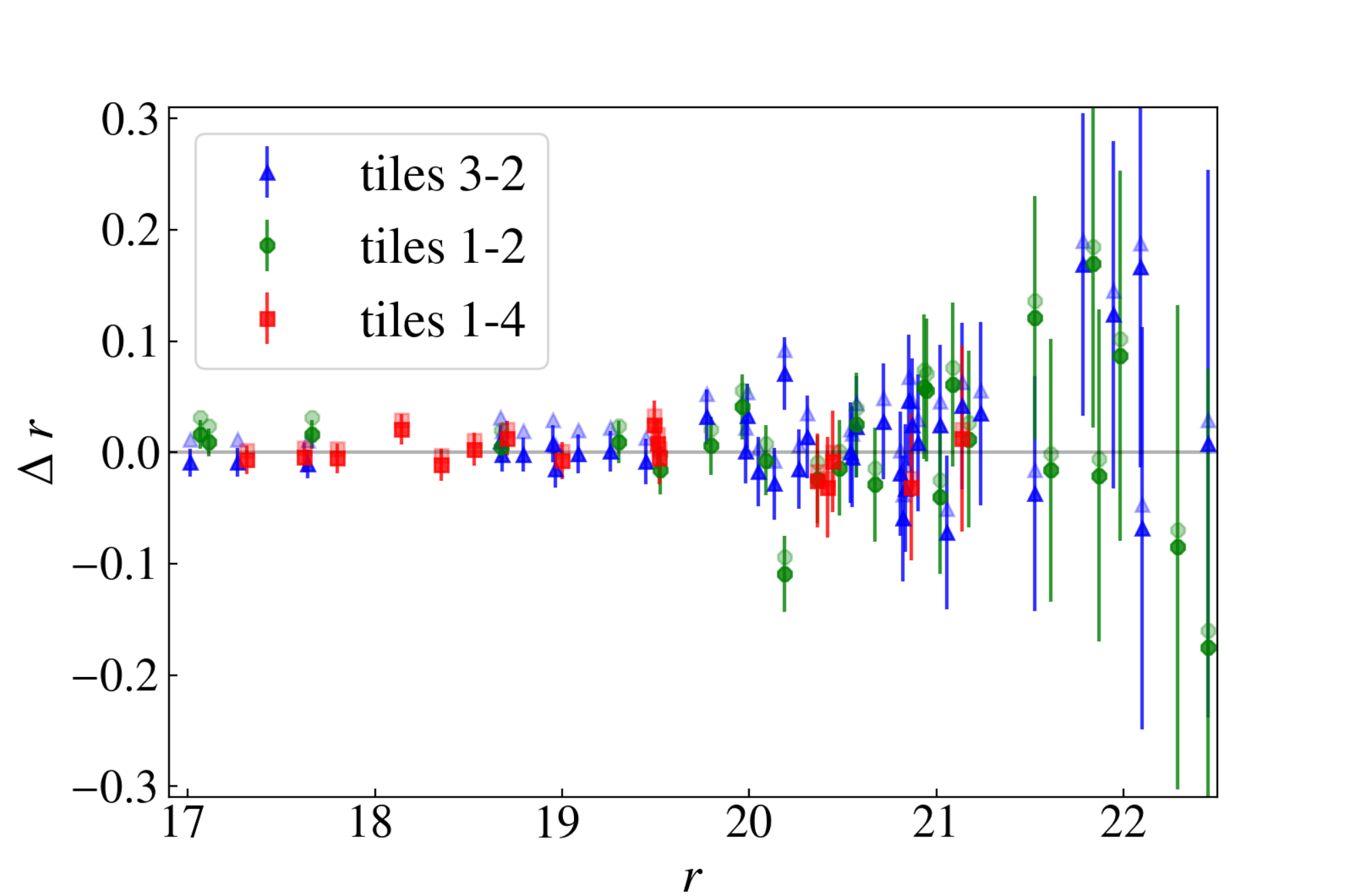}
\caption{Magnitude differences in objects in overlapping tiles for the $r$ band (rSDSS), using a fixed aperture of 6~arcsec. Each point corresponds to a point source in one of the overlapping areas (distinguished here by colors). The light symbols are the magnitude differences before correcting for the PSF of each tile, and the heavy symbols are the magnitude differences after that correction is applied. The error bars are those of the individual tiles, added in quadrature. }
\label{fig:overlapping_tiles}
\end{figure}

We have compared the magnitudes of point sources (CLASS\_STAR $> 0.5$) 
within a 6~arcsec aperture, taking into account the corrections to the flux driven
by the variations of the PSF between the different observations (i.e., the 
light profile of the point sources was used to extend the flux beyond the 
cut-off at 6~arcsec). The result of that comparison is shown in Fig.~\ref{fig:overlapping_tiles} for the $r$ 
band, where we plot the difference in the magnitudes for each object in the overalap area of two contiguous 
tiles, $\Delta r = r_{a} - r_{b}$, as a function of the mean magnitude,
$r = (r_a + r_b)/2$, with error bars given by 
$\sigma = (\sigma_a^2 + \sigma_b^2)^{1/2}$, where
$a, b =$ 1, 2, 3, 4 (the four \mjp\ pointings). In the figure
we show the differences in magnitudes before (light symbols) and after 
(heavy symbols) the PSF correction. From the scatter of the data points we can 
  estimate the calibration error for each filter: in the case of the 
$r$ band, this additional uncertainty due to the photometric calibration
should be of the order of $\sigma_{\rm zpt} \simeq 0.005$ in order to
obtain a reduced $\chi^2$ of 1. However, this calibration
error is different for each filter, with the upper limit being set by the filters in the blue-end
of the spectrum. These show ~a calibration error of approximately $0.04$ magnitudes
 (see Sect.~\ref{sec:calibration}). We stress, however, that the current statistics is too small to properly use this procedure to estimate calibration errors. This will not be a problem for \jp, where areas observed will be three order of magnitude larger, and the strategy will include a larger overlap area.

\subsection{Comparison with other surveys}

\begin{figure}
\centering
\includegraphics[trim={10 0 40 30}, clip, width=\columnwidth ]{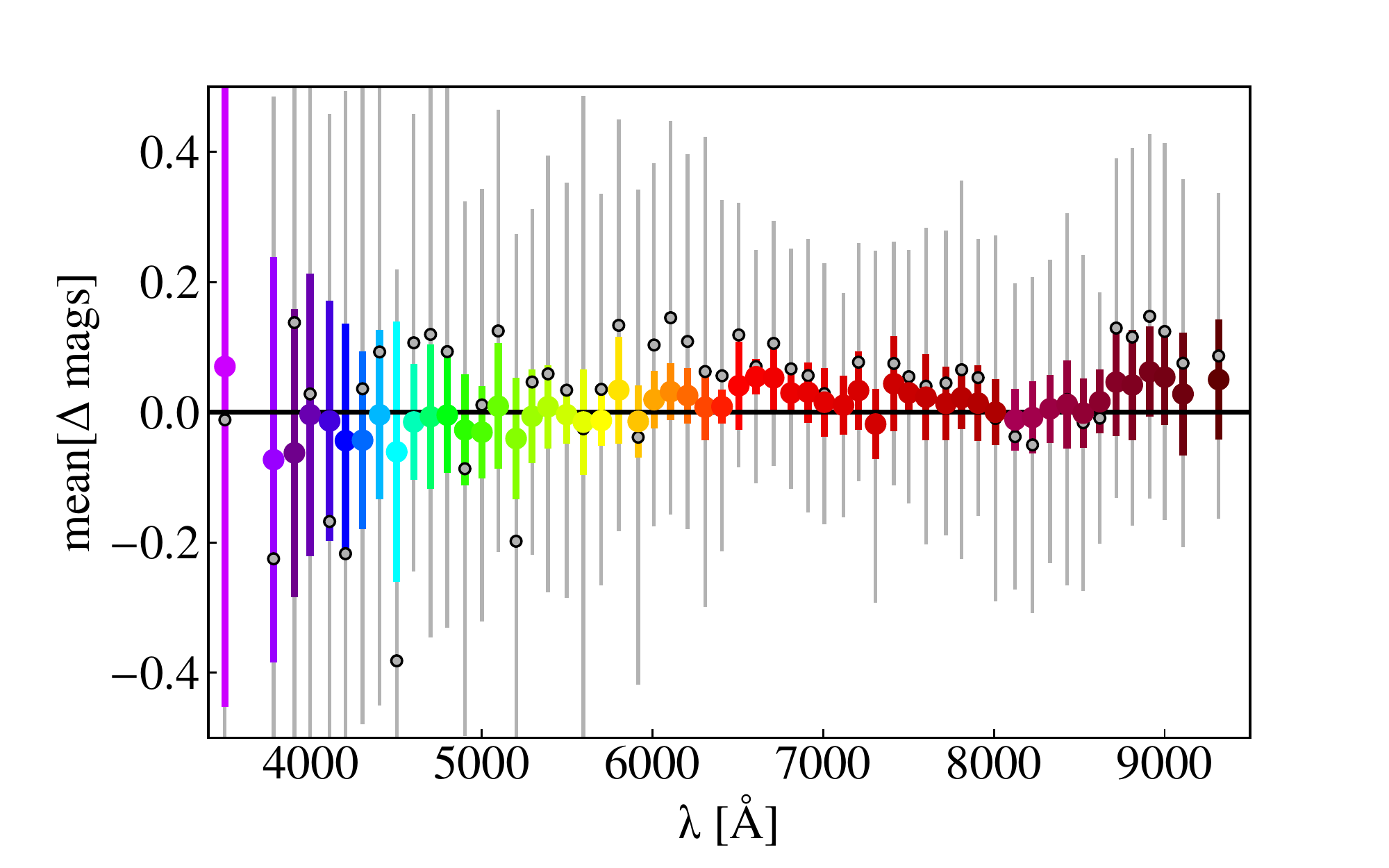}
\caption{Difference between \mjp\ photometry (\magpsfcor\ magnitudes) and synthetic photometry derived by convolving  SDSS spectra with the \jp\ filters. Coloured symbols, and thick error bars, correspond to galaxies with $r < 20$, while gray circles   with thin error bars to galaxies with $20 < r < 22.5$. The error bars refer to the scatter of the distribution within each filter. }
\label{fig:jpas_SDSSspectra}
\end{figure}

We further test the quality of the data by comparing our photometry with that from other surveys. The broad band  photometry   is compared with that from SDSS and 
\hsc, while the narrow band photometry is  compared with synthetic photometry derived by convolving SDSS spectra with the \jp\ filters, as detailed below:

\begin{description} 

\item[{\bf Broad band photometry}]{\ \\
We compare the broad band photometry of \mjp\ with the one of SDSS and \hsc.  We use  sources with no warnings or flags and the \magauto\ photometry, which is a proxy of total magnitude.  For this comparison we use colors, as they best reflect the SED of the sources. Colours are also very useful to check the quality of the relative calibration in \mjp. We show in Fig.~ \ref{fig:jpas_SDSS_HSC} the $(r-i)$ color difference of \mjp\ and SDSS/\hsc, as a function of the \mjp\ $r$-band for both point-like and extended sources, where point sources are defined to be the ones with  CLASS\_STAR > 0.9. We immediately see that colors in \mjp\ with respect to SDSS and \hsc\ are generally consistent. The lack of systematic shifts in the comparison with SDSS confirms the quality of the calibration.
As expected, the scatter is larger in the comparison  with  SDSS than in the one with \hsc, because of the shallower depths reached by SDSS.  
We get similar results when comparing the colors from other broad bands. The good agreement for both point and extended sources  confirms not only the quality of the calibration, but also that we can trust the SED of galaxies for extragalactic studies. 
 \vspace{1.5mm}}

\begin{figure}
\centering
\includegraphics[trim={22 52 40 30}, clip, width=\columnwidth ]{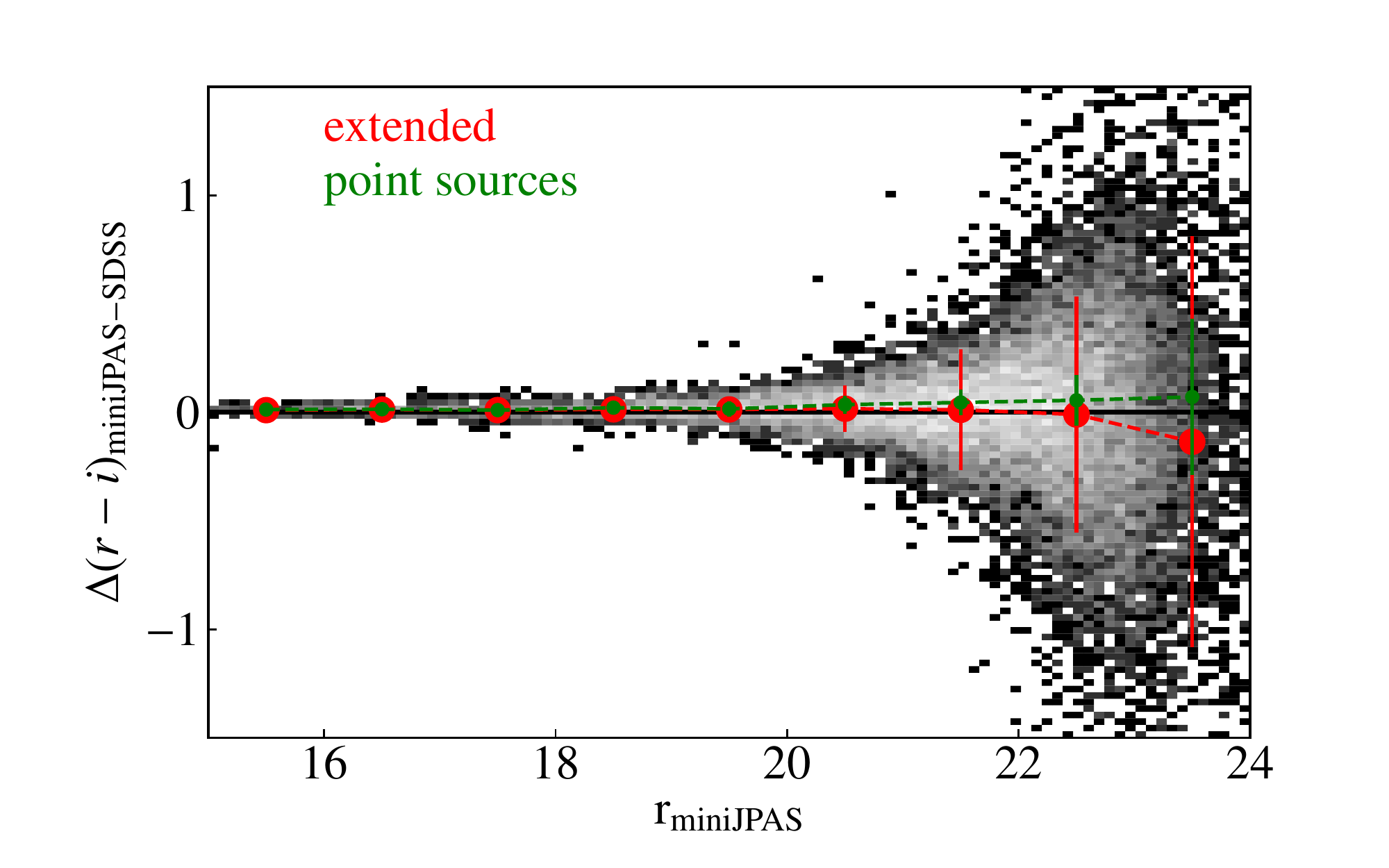}
\includegraphics[trim={22 0 40 32}, clip, width=\columnwidth ]{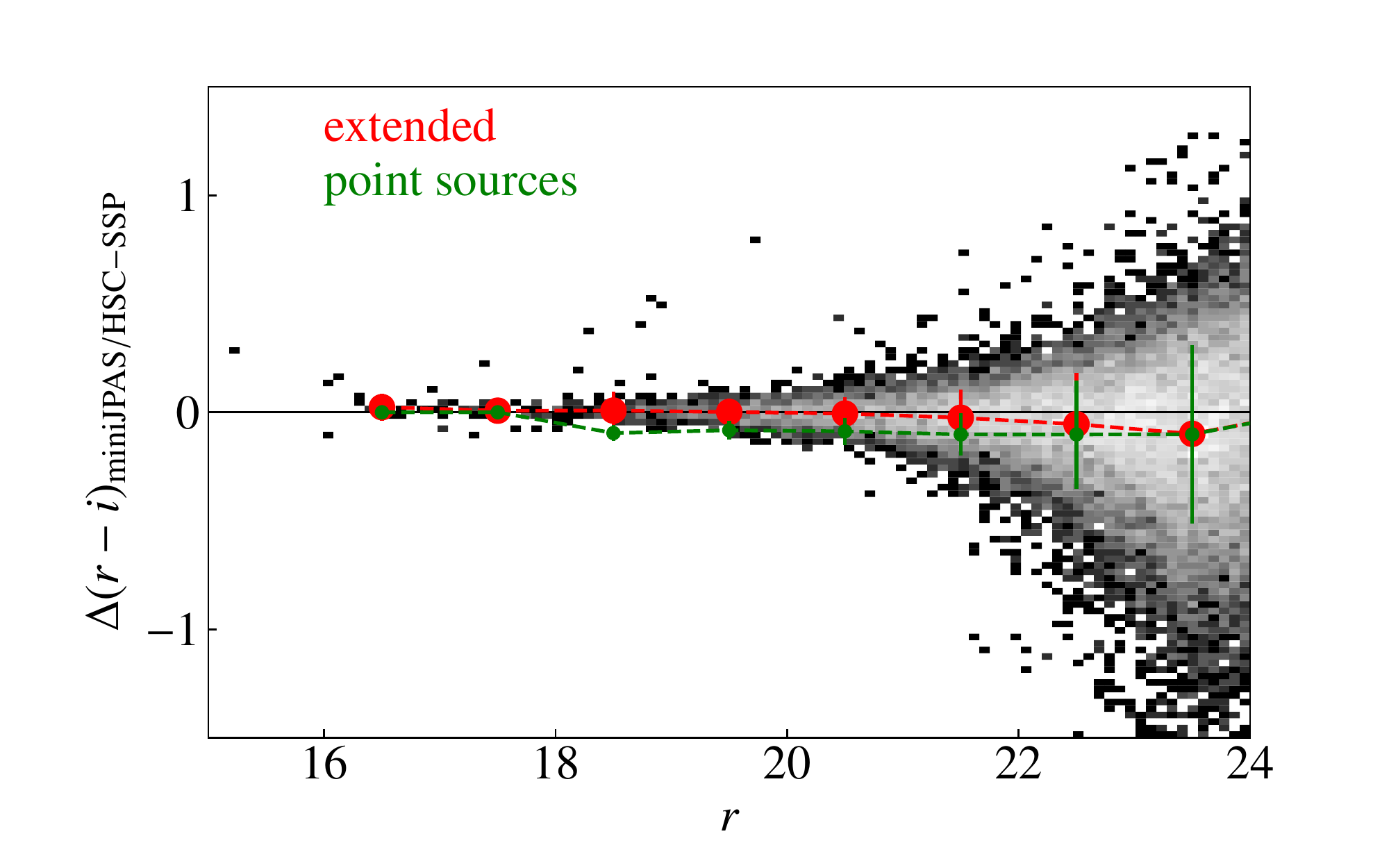}
\caption{Color $(r-i)$ difference of \mjp\ and SDSS (upper panel) and \hsc\ (lower panel) as a function of the \mjp\ $r$ band. The gray density map is for all objects. The coloured points indicate the median with 1-$\sigma$ error in different magnitude bins for extended (red) and point sources (green). }
\label{fig:jpas_SDSS_HSC}
\end{figure}

\item[{\bf Narrow band photometry}]{\ \\
We compare the narrow-band photometry of \mjp\ with the synthetic photometry obtained by convolving SDSS spectra with the transmission curves of the  \jp\ filters. We use the spectra available in the SDSS DR12 of galaxies with magnitude in the \mjp\ $r$-band $r \leq 20$ and  $20 \leq r \leq 22.5$ in the \magpsfcor\ photometry.  We used the \magpsfcor\ photometry to avoid aperture corrections. In any case, we scaled the spectra (in AB magnitudes) to the median of the $g, r, i$ \mjp\ \magpsfcor\ photometry to account for the differences in aperture between the SDSS fibre and the \mjp\ photometry.  In total, there are  405 spectra at $r \leq 22.5$, of which 122 at $r \leq 20$. Figure \ref{fig:jpas_SDSSspectra} shows the mean of the difference between the observed and the synthetic magnitudes for each narrow band filter. The mean of the shifts is $0.006 \pm 0.028$ and the median scatter is  $0.06$. These small values confirm the accuracy  of the \mjp\ narrow-band calibration. The scatter is larger for galaxies with 20 $\leq r \leq$ 22.5 than for brighter objects due to the shallower depth reached by SDSS, and it is larger in the blue than the red narrow-band filters.}

\end{description}

\subsection{\mjp\ completeness}

The definition of the limiting flux at which a survey is not biased in detection, a.k.a.~completeness, is one of the key points that must be addressed during the preliminary analysis and data validation phase. The \jp\ survey will include an automatic pipeline devoted to determining the completeness curves of the composite images or tiles. Here, we first briefly introduce the methodology and completeness curves obtained for \mjp, which will then be  confronted with the results obtained from the comparison with \hsc.

\begin{figure}
\centering
\includegraphics[trim={8 5 35 30}, clip, width=\columnwidth ]{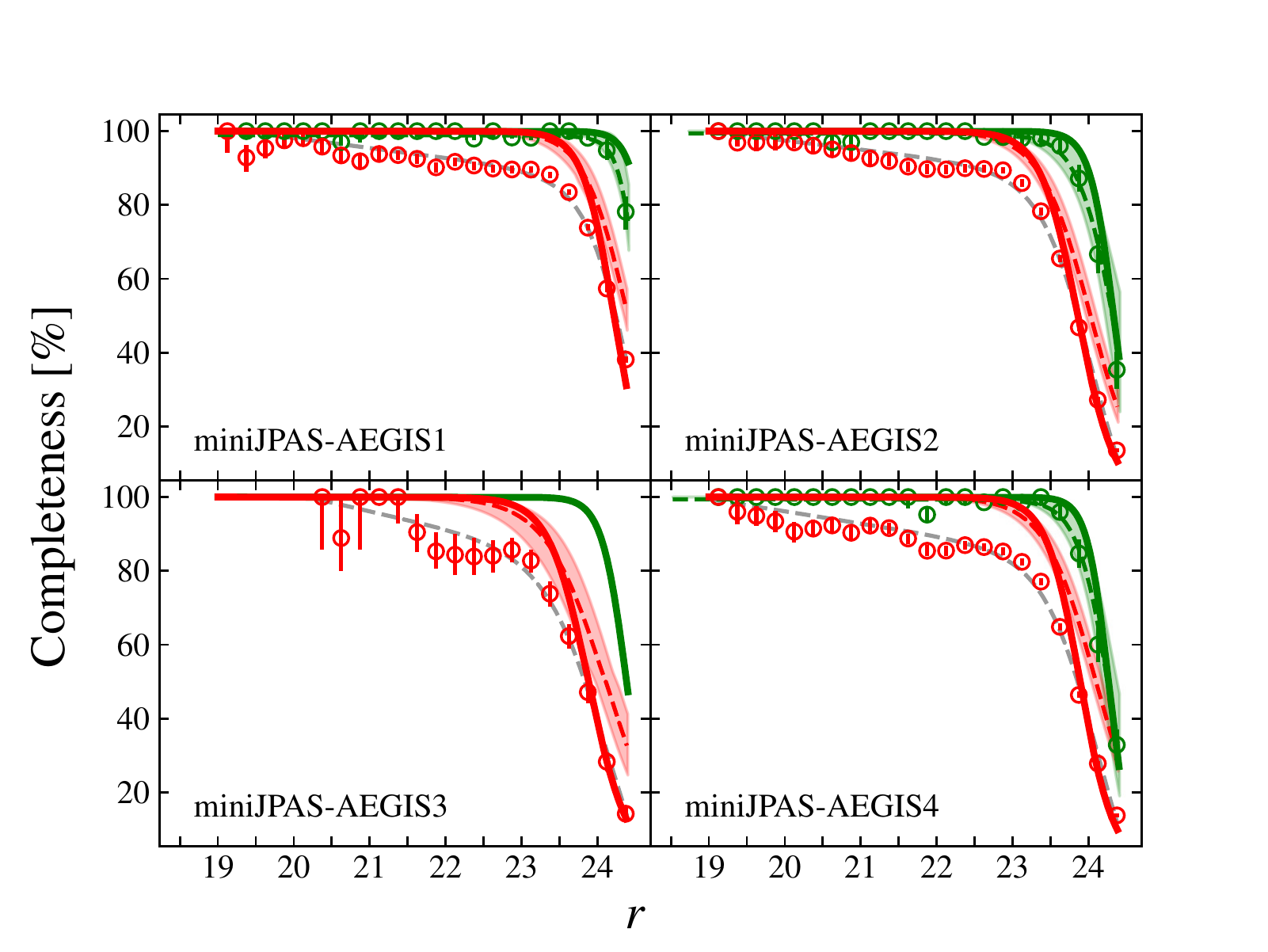}
\caption{Completeness  of extended and point-like  \mjp\ sources as a function of total magnitude in the \mjp\ $r$ band for the four \mjp\ pointings. The solid lines are the completeness curves of extended and point-like sources (red and green, respectively) obtained from a synthetic injection  of sources. The red (green) circles illustrate the observed fraction of common extended (point-like) sources in the \mjp\ and \hsc\ surveys (this comparison is not possible for miniJPAS-AEGIS3, given the  small overlap between  this pointing and \hsc). The dashed coloured lines are the completeness curves obtained from the comparison with the \hsc\ dataset, with uncertainties highlighted by the shaded area (see the text for the description of their derivation).  The dashed gray lines show the curve that best fits the fraction of common sources (Eq.~\ref{eqn:completeness} plus a linear term, details in text).}
\label{fig:completeness}
\end{figure}

Our pipeline is based on the use of synthetic images of galaxies and stars, which are treated separately. 
The completeness of each pointing is obtained after injecting in random
positions without previous detections a position-dependent PSFs for
point-like sources and light profiles of galaxies extracted from COSMOS
HST images in the $F814W$
for extended sources. In the latter case, an average colour term is
applied to transport the structure measured in $F814W$ to the appropriate
scale in the miniJPAS filters, and the original profile is convolved
with the corresponding position-dependent PSF.
The completeness curves are then calculated from the fraction of injected sources that the source detection process recovers (see Sect.~\ref{sec:photometry}). We note that this analysis must be made for each of the \mjp\ pointings to account for discrepancies related, for example, to different observational conditions. In Fig.~\ref{fig:completeness} we show the derived completeness curves, both for compact and extended sources. The curves are properly fitted by a Fermi-Dirac distribution function \citep[see e.g.][]{sandage1979, diaz2019} of the form:
\begin{equation}\label{eqn:completeness}
    \mathcal{C} = \frac{100}{\exp{\left\{(r_\mathrm{JPAS}-\mathcal{C}_{50})/\Delta_\mathcal{C}\right\}} + 1},\ 
\end{equation}
where $\mathcal{C}$ is the detection fraction (in per cent units), $\mathcal{C}_{50}$ is the magnitude at which the sample is $50$\% complete, and $\Delta_\mathcal{C}$ the decay rate on the fraction of detections. In Table~\ref{tab:completeness}, we show the parameters that result from fitting Eq.~(\ref{eqn:completeness}) to \mjp\ data. We estimate that the sample of point-like sources in the full \mjp\ catalogue is $99$\% complete up to $r \sim 23.6$ (\magauto). For extended sources this limit is constrained at $r \sim 22.7$. It is worth mentioning that the \mjp-AEGIS1 field is deeper than the other pointings with limiting fluxes at $r \sim 24.1$ and $r \sim 23.3$ for point-like and extended sources, respectively. 

With a typical seeing of $0.6-0.8$~arcsec in the optical range, along with a depth of $r\sim26$ ($5$~$\sigma$ limit within a $2$~arcsec diameter aperture), the \hsc\ survey can be used to test the \mjp\ completeness.  To carry out the comparison,  we used the second data release of the \hsc\ survey \citep[][]{aihara2019} as the reference catalogue for the \hsc\ photometry. The overlap area with \mjp\ is of $0.7$~deg$^{2}$. Taking as reference the $r$ band from \hsc\ (\texttt{cmodel} magnitudes, which are a proxy to total magnitudes), we computed the fraction of common detections between \hsc\ and \mjp\ at different magnitude ranges, following a similar procedure than the one described above for the injection of synthetic images. Owing to the lower seeing of the \hsc\ observations, we used as compactness discriminator the parameter \texttt{r\_extendedness\_value} provided by the \hsc\ catalogues. To perform a robust and fair determination of the completeness, we quadratically include the systematic uncertainties reported by \citet[][]{huang2018} for both \texttt{PSF} and \texttt{cmodel} magnitudes \citep[see Tables~1 and 3 in][]{huang2018}. The impact of these uncertainties on the fraction of common detections is determined via a Monte Carlo approach by assuming Gaussian errors for the $r_\mathrm{HSC}$ fluxes. Nevertheless, discrepancies between the \hsc\ and the \mjp\ observations (e.g.~seeing, deblending, etc.) may yield systematic discrepancies in the fraction of common sources, typically ranging from $5$ to $10$\%, even when both surveys are not biased by incompleteness. To account for this effect and obtain  alternative completeness curves based on \hsc\ data, we fitted the fraction of common detections to Eq.~(\ref{eqn:completeness}), adding a linear term of the form $\alpha \cdot r_\mathrm{HSC} + \beta$ (with $\alpha$ and $\beta$ as free parameters). 
Figure~\ref{fig:completeness} illustrates the fraction of common detected sources in the \hsc\ and the \mjp\ surveys for extended and point-like sources, as well as the fits following Eq.~(\ref{eqn:completeness}), with and without the added linear term. We find a good agreement between the  completeness curves obtained from synthetic sources and from the comparison with \hsc\ observations, within a 1$\sigma$ uncertainty level (see also Table~\ref{tab:completeness}). In addition, there is no discrepancy between the detection of common point-like sources, and therefore, $\alpha$ and $\beta$ are fully compatible with a null value. Regarding the \mjp-AEGIS3 pointing, we are not able to observationally determine the completeness curve of point-like sources owing to the very low overlapping area. For extended sources, there exist systematic discrepancies at $r_\mathrm{HSC}\gtrsim 20$ in all the \mjp\ pointings, which were properly taken into account by the linear term added to Eq.~(\ref{eqn:completeness}).

We note that the completeness curves were obtained from total
(synthetic) magnitudes in \mjp\ and from \texttt{cmodel} magnitudes in \hsc.
Regarding the practical application of the completeness curves, the
\texttt{AUTO} magnitude is our best proxy for a total magnitude and
should be used as reference to define a complete flux-limited sample.
Consequently, the limiting magnitudes in all the fields of \mjp\ for
the dual mode catalogues are $r_{\texttt{AUTO}} = 23.6$ and
$22.7$ for point-like and extended sources, respectively. The
completeness curve parameters are available in the \texttt{ADQL} table
\texttt{minijpas.TileImage}.

\begin{table*}
    \caption{Parameters describing the completeness function of the mini-JPAS survey obtained from injection of synthetic sources ($\mathcal{C}_{50}$ and $\Delta_\mathcal{C}$, see Eq.~(\ref{eqn:completeness}) ) and after comparing with the fraction of common sources with respect to \hsc\ photometry catalogues ($\mathcal{C}_{50}^\mathrm{HSC}$ and $\Delta_\mathcal{C}^\mathrm{HSC}$). In the last column, we present the limiting values for a completeness level of $99$\% for the \mjp\ detection band ($r$, \magauto).}
    \label{tab:completeness}
    \centering
    \begin{tabular}{lccccc}
    \hline
    \hline
    Field & $\mathcal{C}_{50}$ & $\Delta_\mathcal{C}$ & $\mathcal{C}_{50}^\mathrm{HSC}$ & $\Delta_\mathcal{C}^\mathrm{HSC}$ & $r_\mathrm{99}^\mathrm{AUTO}$ \\
    \hline
    Point-like &&&&& \\
    \hline
    \mjp-AEGIS1 & $24.74$ & $0.14$ & $24.57^{+0.09}_{-0.09}$ & $0.13^{+0.06}_{-0.06}$  & $24.10$ \\
    \mjp-AEGIS2 & $24.33$ & $0.16$ & $24.33^{+0.16}_{-0.12}$ & $0.23^{+0.11}_{-0.08}$  & $23.59$ \\
    \mjp-AEGIS3 & $24.39$ & $0.16$ & --                      & --                      & $23.65$ \\
    \mjp-AEGIS4 & $24.26$ & $0.14$ & $24.26^{+0.12}_{-0.10}$ & $0.20^{+0.08}_{-0.07}$  & $23.62$ \\
    \hline
    Extended &&&&& \\
    \hline
    \mjp-AEGIS1 & $24.22$ & $0.21$ & $24.39^{+0.07}_{-0.08}$ & $0.29^{+0.09}_{-0.06}$ & $23.26$ \\
    \mjp-AEGIS2 & $23.86$ & $0.24$ & $24.03^{+0.07}_{-0.07}$ & $0.33^{+0.04}_{-0.03}$ & $22.76$ \\
    \mjp-AEGIS3 & $23.88$ & $0.26$ & $24.11^{+0.14}_{-0.15}$ & $0.39^{+0.09}_{-0.07}$ & $22.69$ \\
    \mjp-AEGIS4 & $23.89$ & $0.22$ & $24.09^{+0.07}_{-0.07}$ & $0.34^{+0.04}_{-0.04}$ & $22.88$ \\
    \hline
    \end{tabular}
    \tablefoot{We find that total magnitudes, used for the injection of synthetic sources, are essentially equivalent to \texttt{AUTO} magnitudes at the $99$\% completeness limit.}
\end{table*}

\subsection{Number counts} \label{sec:number_counts}

\begin{figure}
\centering
\includegraphics[width=\columnwidth]{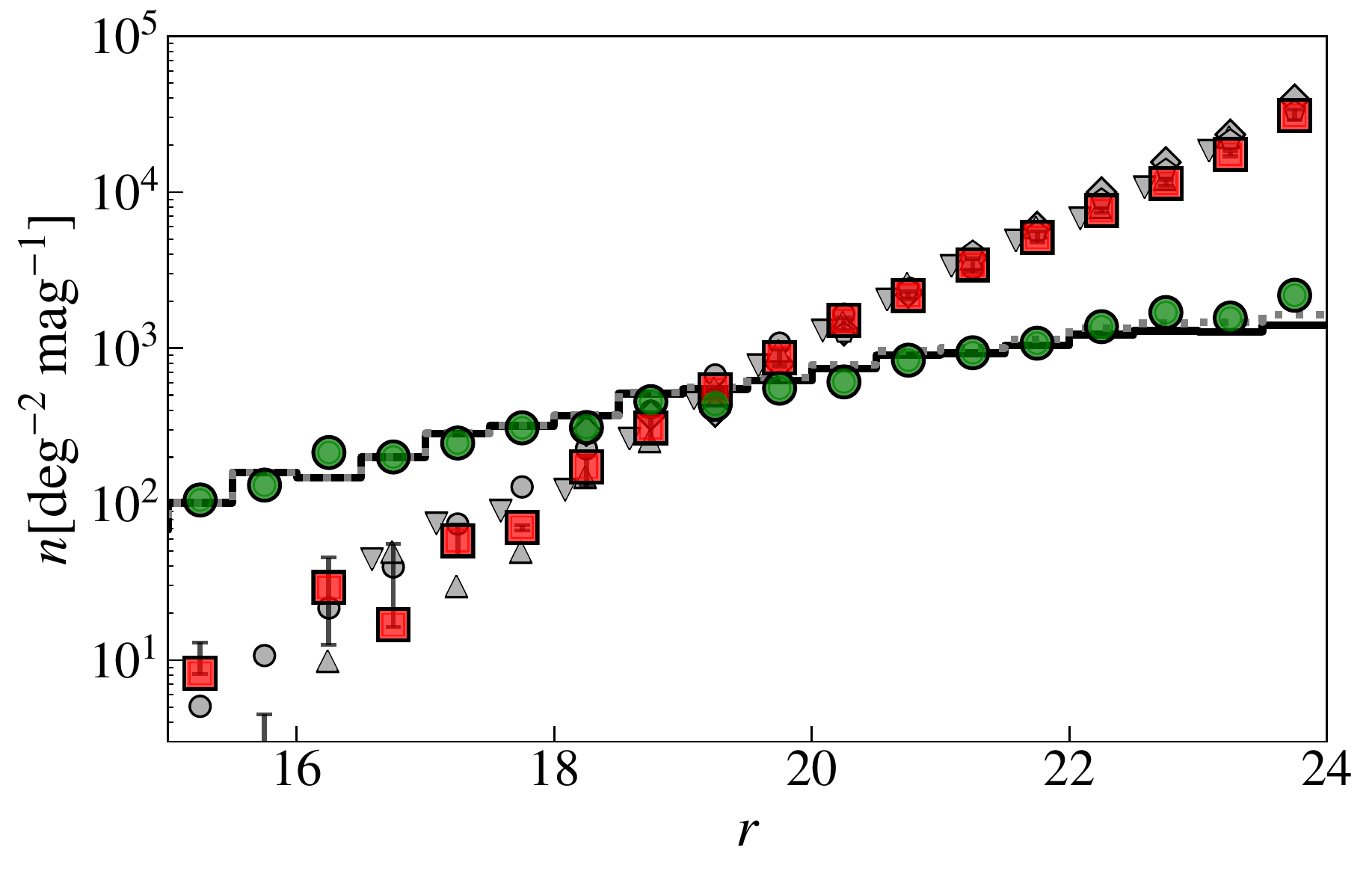}
\caption{Stellar (green circles) and galaxy (red squares) number counts as a function of the $r$ band magnitude. The black solid curve is the stellar number counts at the EGS position estimated with the default TRILEGAL model of the Milky Way \citep{trilegal}. The dotted gray curve 
includes the expected contribution for
point-like quasars \citep[from the BOSS survey,][]{pdlb2016} to the compact
population. Gray symbols are galaxy number counts from the literature: \citet[][circles]{yasuda01};  \citet[][triangles]{huang01}; \citet[][inverted triangles]{kummel01}; \citet[][diamonds]{kashikawa04}; and \citet[][pentagons]{capak04}. }
\label{fig:number_counts}
\end{figure}

We present the stellar and galaxy number counts in Fig.~\ref{fig:number_counts}. To derive the number counts separately for point-like and extended sources 
 we used the stellar-galaxy locus classification presented  in Sect.~\ref{sec:star_galaxy}, corrected for the completeness derived by synthetic images as described above.
The number counts derived from the Bayesian classification agree with
the expectations from the literature both for stars and galaxies up to 
$r= 23.5$, as shown in Fig.~ \ref{fig:number_counts}. The derived probabilities are publicly available in the ADQL table \texttt{minijpas.StarGalClass}.

\subsection{Caveats and known problems} \label{sec:problems}

We provide below a brief list of issues and caveats to keep in mind when using the data-release presented in this paper: 

\begin{description}
\item[{\bf Inhomogeneous exposure times.}] {
Narrow bands present heterogeneous total exposure times from band to band and from pointing to pointing as, in some cases, more than the planned number of images was  used to produce the co-adds.
For  JPCam observations the strategy will be defined to obtain more homogeneous depths. \vspace{1.5mm}}
\item[{\bf Extended faint sources.}]{
The superbackground subtraction described in Appendix~\ref{sec:app_bgk}, together with the small dithering pattern of \mjp, makes extended sources to be potentially confused with instrumental background. Therefore, for \mjp\ data we cannot discard that some faint extended real sources have been totally or partially removed from the final images. For \jp\ data, this issue is not expected because of its high dithering pattern of 1/2 CCD.
  \vspace{1.5mm}}
\item[{\bf Image quality.}] {
Since the time window to complete the \mjp\ project was limited, the initial requirements in terms of seeing, airmass and atmospheric stability were at times relaxed to guarantee the completion of the whole project on time. This has led to PSFs which are on average larger than what the system can typically provide (<1~arcsec), and to the presence of some thin clouds for the latest tray of reddest filters (mainly in J0790 filter and redder), which were detected only after the data reduction. \vspace{1.5mm}}

\item[{\bf Fringing correction.}]{
 The time limitation in the observations of \mjp\ resulted in a number of scientific images which turned out to be scarce to generate high-quality, master fringing images for all red filters. This means that the final fringing removal is not optimal in all cases.\vspace{1.5mm}}

\item[{\bf Astrometric accuracy in coadded images.}] {
Despite single individual images exhibit astrometric solution accuracies with respect to Gaia DR2 of the order of $\sim0.03$~arcsec \footnote{Measured as the RMS of the differences between
    Gaia DR2 coordinates and the final astrometric solution for the
    stars used in the astrometric calibration.}, these  degrade to up to $\sim0.3$~arcsec in the case of coadded images and dual-mode catalogues. Resampling and pixel rotations before coadding seems to be the main source of this effect, which will certainly be revisited in future releases.
    }

\end{description}

\section{A foretaste of \jp\ science }  \label{sec:science}

In this section we give a brief overview of the wide variety of scientific cases that can be addressed thanks to the \jp\ filter system. We use \mjp\ data to provide concrete examples and we discuss results in view of the upcoming  \jp\ data.  Representative \js\ of the wide variety of astrophysical objects in the 
surveyed area are presented in \ref{fig:photospectra}.

\subsection{Galactic Science} \label{sec:stars}

\begin{figure*}[ht]
\centering
\includegraphics[width=0.87\textwidth]{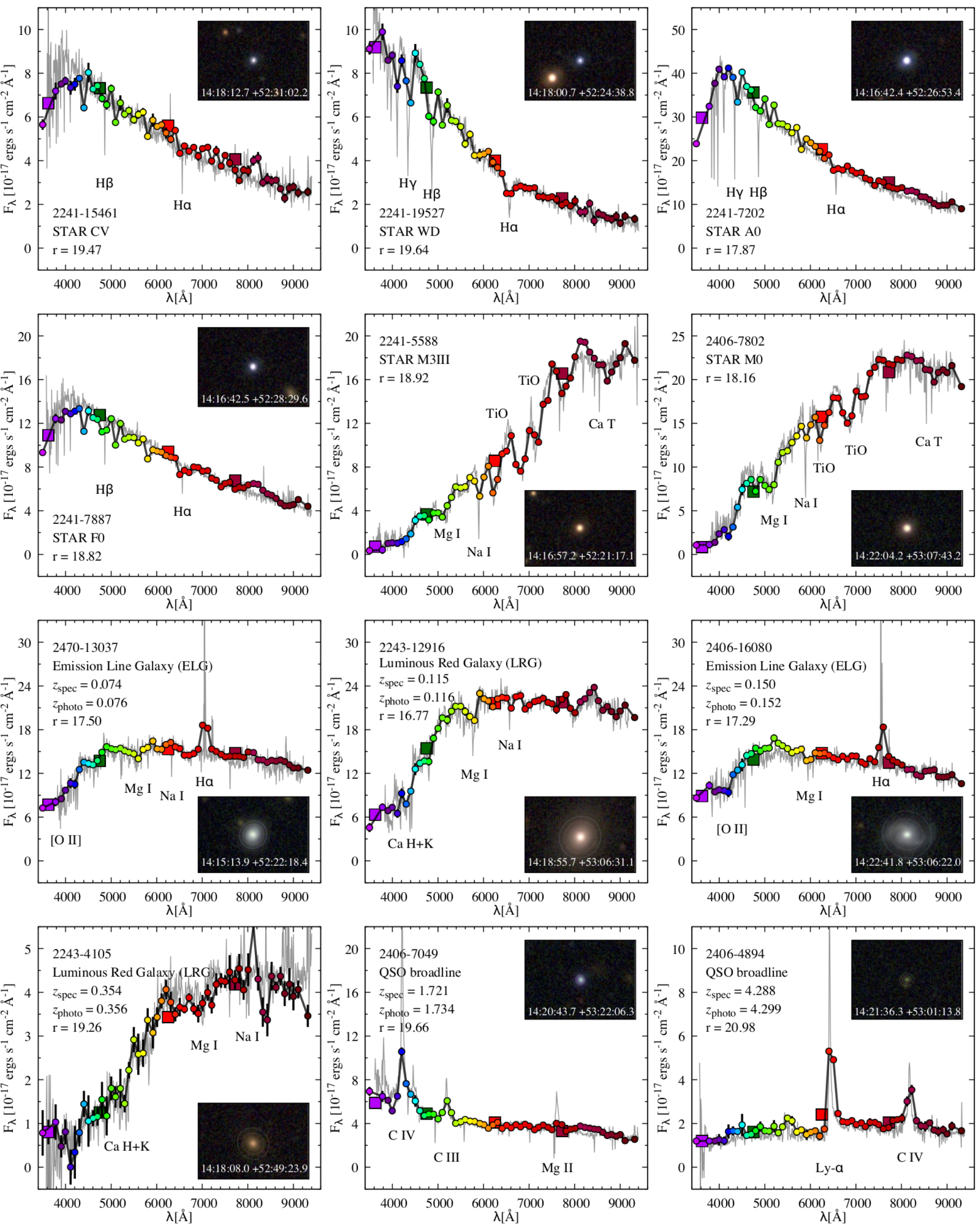}
\caption{The \js\ of different classes of stars, galaxies and quasars in the \mjp\ field (coloured symbols) compared with the SDSS spectra (gray lines).  The \mjp\ object ID, $r$ magnitude, and classification are listed in the legend.   Extragalactic sources are ordered by increasing redshift, and both spectroscopic (SDSS) and photometric (\mjp) redshift are provided in the legend. The multi-color inset images are centred on the object and are $30$~arcsec across. The ellipses visible for extended sources are the <auto ellipse> (inner ellipses) and the <petro ellipse> (outer ellipses). 
 }
\label{fig:photospectra}
\end{figure*}

The unique \jp\ photometric system is able to provide the equivalent of a low-resolution spectrum for each stellar source within its footprint. Key absorption features, as well as the shape of the SED, can be used to identify, classify and characterize targets of interest. 
In addition, the \js\ allow for the determination of stellar atmospheric parameters (effective temperature, surface gravity, and metallicity) and selected chemical abundances. Basic approaches used to derive stellar parameters can be based on photometry \citep[][]{ivezic2008, an2015, starkenburg2017, huang2019}, the fitting of theoretical SED’s \citep[][]{allendeprieto2016}, or on the application of machine learning algorithms \citep[][]{li2018,thomas2019, whitten2019b}. Previous experience with the Stellar Photometric Index Network Explorer \citep[SPHINX,][]{whitten2019b}, applied to the Early Data Release of \jplus\ \citep{cenarro2019} has shown the reliability of estimating effective temperature and metallicity based on narrow-band photometry, achieving $\sigma (T_{\rm eff})=91$~K and $\sigma([\rm{Fe}/\rm{H}])\sim 0.2$~dex. A key ingredient for a successful application of any machine learning technique is the definition of an adequate training dataset, which should ideally sample a parameter space as complete as possible. Examples of large medium-resolution spectroscopic databases in the northern hemisphere that can help address this issue are the SEGUE (Sloan Extension for Galactic Understanding and Evolution; Lee et al. 2008), LAMOST (Large Sky Area Multi-Object Fiber Spectroscopic Telescope; Li et al. 2018, Boeche et al. 2018), and APOGEE (Apache Point Observatory Galactic Evolution Experiment; Majewski et al. 2017). 

Different topics in stellar astrophysics can be addressed once the stellar parameters have been estimated. A few science cases include the search for metal-poor stars \citep[][]{youakim2017, yoon2018, placco2018,  placco2019}, blue horizontal branch stars \citep{santucci2015a, whitten2019a, starkenburg2019}, blue stragglers \citep[][]{santucci2015b}, ultracool dwarfs, and white dwarfs. Other topics include the identification and chemical characterization of Galactic stellar streams, classical dwarf galaxies and ultra-faint dwarf galaxies \citep[e.g.,][]{longeard2018, shipp2018, shipp2019, chiti2020, longeard2020}. Systems such as planetary nebulae and symbiotic stars, that involve circumstellar ionized gas, can be also identified and characterized based on the \js\, \citep[][]{gutierrezsoto2020}. \jp\ photometry can also contribute to the study of stellar populations of Globular Clusters as shown by \citet{bonatto2019} for the GC M15, based on \jplus\ Science Verification Data.
The PRISTINE survey \citep{starkenburg2017} has been successful in determining stellar atmospheric parameters and selecting metal-poor stars by combining SDSS photometry with a narrow-band filter centred on metallicity-sensitive Ca H\&K absorption features (3900-4000\AA). Since \jp\ has other filters in metallicity-sensitive regions besides the Ca H\&K lines, it is expected that the errors and identification thresholds will improve considerably. In addition, \jp\ can provide a powerful data-set to pre-select interesting stellar sources for large-scale wide-field spectroscopic efforts such as DESI \citep[][]{desi2016} and WEAVE \citep{dalton2012}.

Examples of stellar sources observed in \mjp\ are presented in the top 6 panels of Fig.~\ref{fig:photospectra}. The measured fluxes in the narrow filters are shown and compared with the SDSS spectra. The classes and spectral types listed in each panel were retrieved from the SDSS spectroscopic pipeline database. The  inset color images have $30$~arcsec across and are centred on the stars. It is possible to see that the \mjp\ photometry is capable of successfully capturing the continuum shape for these stellar sources and also the main spectral absorption features used to determine spectral types, atmospheric parameters, and selected chemical abundances (e.g., carbon, $\alpha$-elements). For the warmer stars (first four panels), the hydrogen Balmer lines are clearly traceable for $\lambdaup < 5000$~\AA, while for cooler stars (middle and right panels in the second row), several TiO molecular bands are present for $\lambdaup > 6000$~\AA.
The small footprint of \mjp\ does not allow the application of the aforementioned techniques that are currently being applied to the \jplus\ data 
\citep[][]{whitten2019b, bonatto2019}, as they require large training samples. This will be possible with the much larger areas scanned by \jp.

\subsection{Galaxy evolution studies} \label{sec:galaxies}

\begin{figure}
\includegraphics[width=0.5\textwidth]{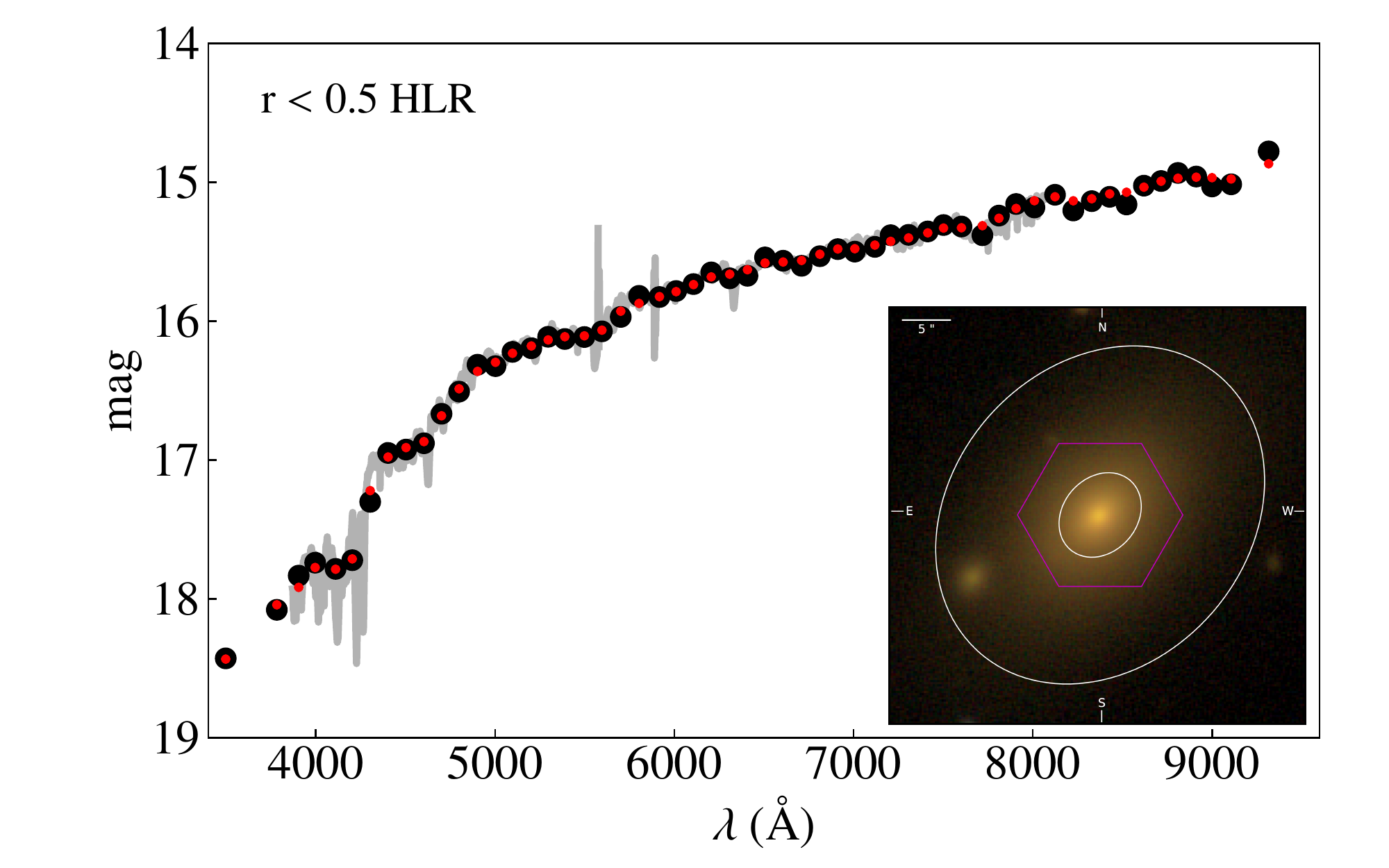}
\includegraphics[width=0.5\textwidth]{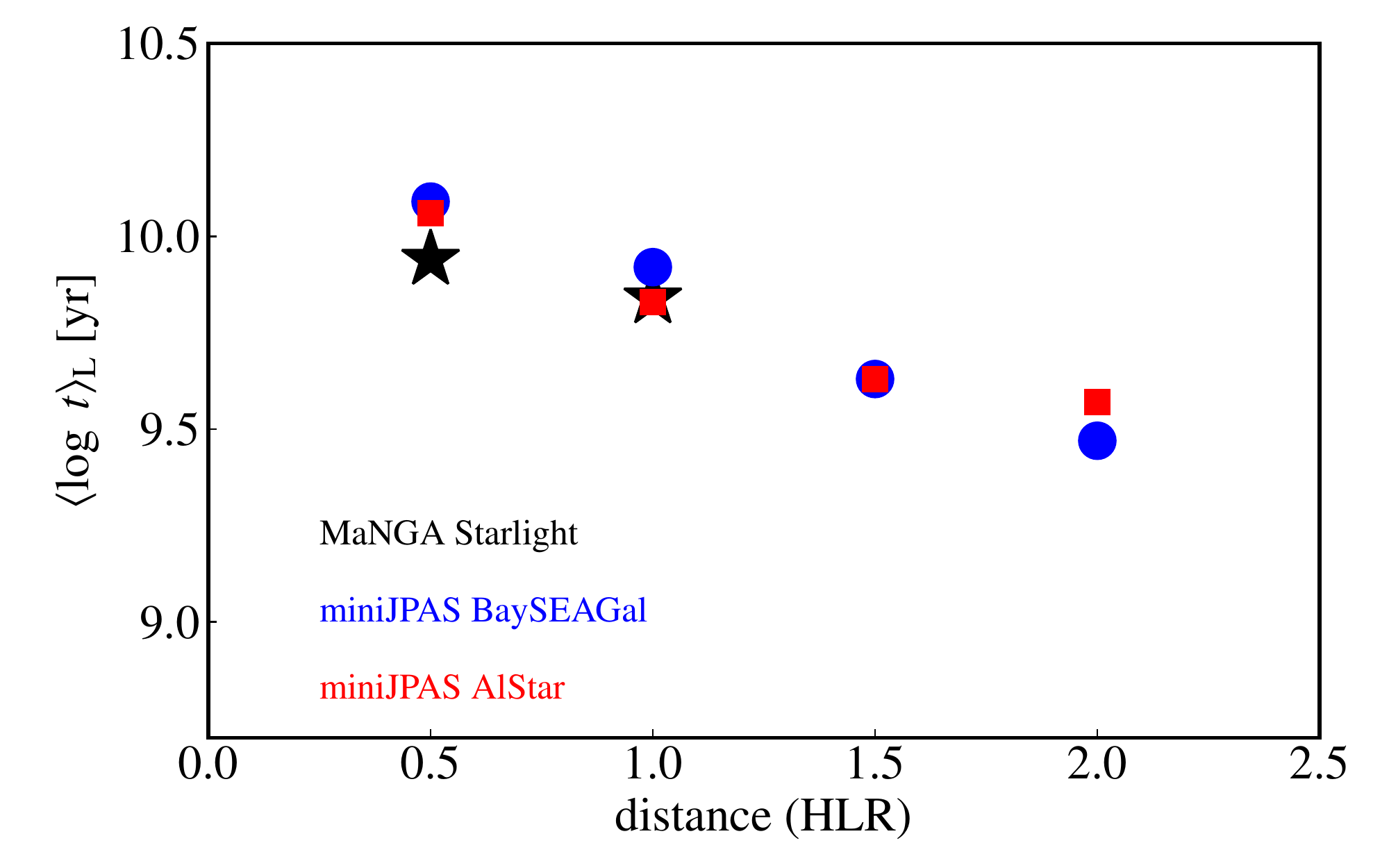}

\caption{Upper panel:  The MaNGA spectrum of the central ring of 0.5 HLR (grey line) of a galaxy at $z\sim 0.1$ is compared with
\mjp\ data (black dots). The result of the fitting of \mjp\ data is also plotted (red dots). The inset shows an image of the galaxy
in the $r$ band where two ellipses are overlaid at 0.5 HLR and 2 HLR. The FoV of the MaNGA
survey is over-plotted as a red hexagon. Bottom panel: Comparison of
the radial variation of the average age $\langle \log t \rangle_L$ derived from \mjp\ data, with the non-parametric code \texttt{AlStar} (red dots)
and the parametric code \texttt{BaySEAGal} (blue dots), and from the MaNGA data analysed with the \texttt{STARLIGHT} code (black stars). }
\label{fig:MaNGAgalx}
\end{figure}

The uniform and un-biased observations of \jp\ will offer a unique large dataset for galaxy evolution studies, from the IFU-like analysis of nearby sources, to stellar population studies across cosmic time. The narrow bands will allow  for the detection of strong line emitters, from star-bursting galaxies to quasars. The wide area  will also enable the detection of rare objects as well as the study of galaxies in all kinds of environments. 

Examples of \js\ of extragalactic sources observed by \mjp\ are provided in the bottom six panels of Fig.~\ref{fig:photospectra}. Ordered  by increasing redshift, the first four panels show examples of both  passive and star forming galaxies. Clearly visible are the $4000$~\AA\ break of red galaxies and the $\rm{H}\alpha$ emission of blue star forming objects.
The last two panels, instead, show examples of quasars, with the profiles of broad emission lines, such as Ly$\alpha$, CIV, CIII, clearly traced by multiple filters. 

In what follows  we provide few examples of galaxy studies based on the \mjp\ dataset.

\begin{figure*}
\includegraphics[width=\textwidth]{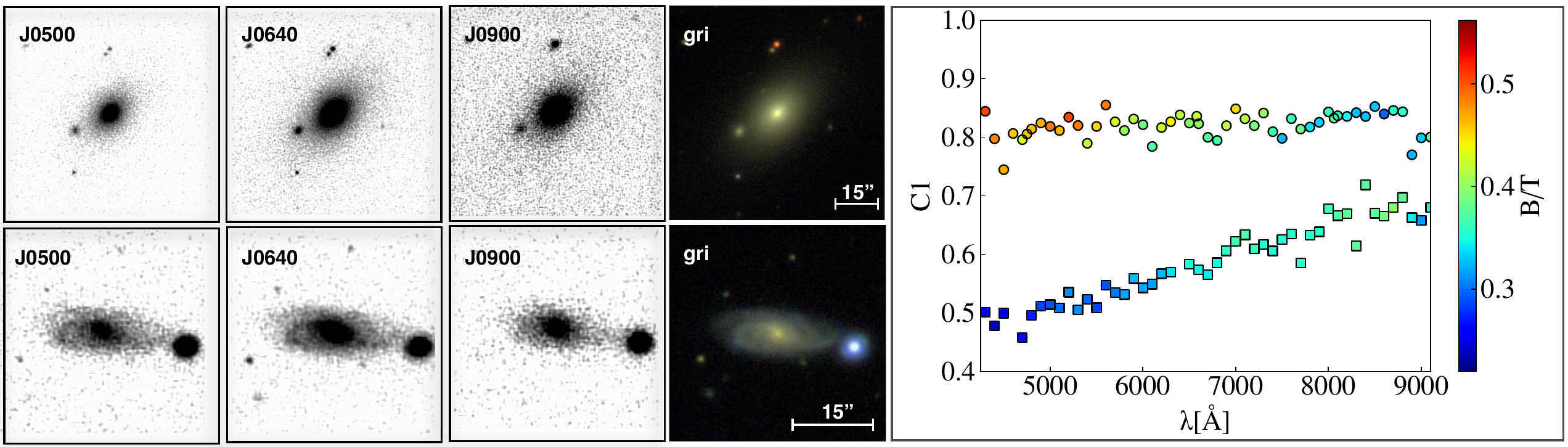}
\caption{{\it Left panel}: snapshots of an elliptical galaxy, ID 2470-10239 (upper row) and a spiral galaxy, ID 2470-10291 (bottom row), in three of the \jp\ narrow bands (J0430, J0640, J0900), as an example of the variation of galaxies' properties with wavelength. {\it Middle panel}:  $g,r,i$ broad band colour images of the two galaxies. The $15$~arcsec length, indicated on the images for reference, corresponds to $21.5$~kpc for the elliptical galaxy ($z \simeq 0.075$) and $20.8$~kpc  for the spiral galaxy ($z \simeq 0.073$).   {\it Right panel}:   variation of the concentration $C1$ parameter with wavelength, colour coded according to the bulge-to-total light ratio ($B/T$), for the early type galaxy  (circles) and the spiral galaxy (squares). The magnitudes are integrated up to 1 disk scale lengths.}
\label{fig:morphology}
\end{figure*}

\subsubsection{J-PAS: a low spectral resolution IFU survey} \label{sec:ifu}

The combination of the wide area and the CCD pixel size will make \jp\ a competitive IFU-like survey of extended galaxies at $z<0.15$, where the low spectral 
resolution will be compensated by the large statistics that the survey will provide. In this sense, \jp\ will be highly complementary to  CALIFA \citep{sanchez12, garciabenito15, sanchez16} and MaNGA \citep{bundy15, law15}, 
 by providing larger samples and having the ability of studying   galaxy properties to larger galactocentric distances, up to several half-light-radii (HLR)\footnote{We define it as the semi-major axis length of the elliptical aperture which contains half of the light of the galaxy at the rest frame wavelength 5600 \AA .},  with a significantly better spatial sampling ($0.23$~arsec~pixel$^{-1}$ for the broad bands and $0.46$~arsec~pixel$^{-1}$ for the narrow bands).
For example, even at  $z \sim 0.1$, where the spatial scale  is $\sim 1.8$~kpc~arcsec$^{-1}$, galaxies with a typical HLR of 
$4$~kpc can be studied 
every $\sim$0.5 HLR, assuming an average seeing of $1$~arcsec.

Moreover, \jp\ data will not suffer from FoV restrictions, so 
 the spatial properties of the largest galaxies of the Local Volume (distance $\leq 15$~Mpc) will be easily studied. 

In the small \mjp\ footprint there are  few extended galaxies with a HLR exceeding $5$~arcsec 
that can be useful to show the potential of \jp\ on this subject.
We have chosen, as example, a quiescent galaxy at  $z \sim 0.1$ with  $14.7$ mag in the $r$ band.
This galaxy ($\alpha$ = 14:15:20.37; $\delta$ = 52:20:45.19)   was also observed by  the MaNGA survey \citep{bundy15},
so that we can directly show how results based on the analysis of our \js\ compare with the results from spectroscopic  data. MaNGA data were retrieved from the SDSS DR14 \citep{abolfathi2018},
and the spectra were analysed using the \texttt{PyCASSO} pipeline  \citep{amorim2017}. From the \mjp\ data we define the nucleus as the maximum of the galaxy emission at the wavelength 5600~\AA, and we estimate the HLR to be 9.4~arcsec. The top panel of  Fig.~\ref{fig:MaNGAgalx} shows a comparison between the MANGA spectra  and the \mjp\ photometric  data within  $0.5$ HLR, demonstrating the very good agreement in terms of flux and SED.
We derive the spatially-resolved  stellar population
properties of the galaxy by analysing both \mjp\ and MaNGA data in rings centred on the nucleus. While the MaNGA FoV allows to study the galaxy properties in the first two rings (within $1$ HLR, see the inset in Fig.~\ref{fig:MaNGAgalx}), with \mjp\ data we can extend the analysis up to  $2$ HLR.   
The \mjp\ \js\ of the 4 rings 
 are fitted by  \texttt{BaySEAGal} (de Amorim et al. in prep), a Bayesian code that allows to derive a wide variety of intrinsic galaxy properties, such as the stellar mass, the luminosity-weighted age of the stellar population, the metallicity and stellar extinction. 
For comparison, we also fit the \js\ with the non-parametric code \texttt{AlStar} (Cid Fernandes et al. in prep.).
Both codes use the same set of single stellar populations (SSP) models by Bruzual \& Charlot (2019, in prep.) to build the star formation
histories. The magnitudes obtained from the fitting with \texttt{BaySEAGal} are plotted in  Fig.~\ref{fig:MaNGAgalx}. 
The MaNGA spectra of the two inner rings are instead fitted with the non-parametric code \texttt{STARLIGHT} \citep{cidfernandes05, lopezfernandez2016},
and following the same processes used for CALIFA data \citep{amorim2017}, using the same set of SSPs.
We find a very good agreement in the stellar population properties derived from the parametric and non-parametric codes (\texttt{AlStar} and \texttt{BaySEAGal}) with \mjp\ data, and these results also agree with the ones derived from the MaNGA spectroscopic data. In the bottom panel of  Fig.~\ref{fig:MaNGAgalx} we show, as an example, the results for the radial distribution of luminosity-weighted stellar ages  $\langle \log  t \rangle_L$. 
  Note that this negative age gradient was also found in the CALIFA sample \citep{gonzalezdelgado15}, and in a few extended galaxies in the ALHAMBRA survey \citep{sanroman2018}. The stellar mass in the inner $0.5$~HLR is estimated to be $10^{11.10} \rm{M}_{\odot}$ and  $10^{11.06} \rm{M}_{\odot}$  from the MaNGA and \mjp\ analysis, respectively.

\jp\ will also be competitive for IFU-like studies of star forming galaxies of the nearby Universe. The HII regions distributions and their ionized gas properties will be retrieved for galaxies of the Local Volume ($z< 0.015$), in similar way as it has been done for \jplus\ \citep[][ following the methodology of \citet{vilella2015}]{logrono2019}. Several methodologies can be developed to retrieve the emission line flux  for galaxies out of the Local Volume. 
One of these is based on a  Neural Network method that estimates the equivalent width of the emission lines of H$\alpha$, H$\beta$, [NII]$\lambda$6584, [OIII]$\lambda$5007 using a collection of spectra from the CALIFA and MaNGA surveys used  for training (Mart\'inez-Solaeche et al., in prep.).

\begin{figure*}
\includegraphics[width=\textwidth]{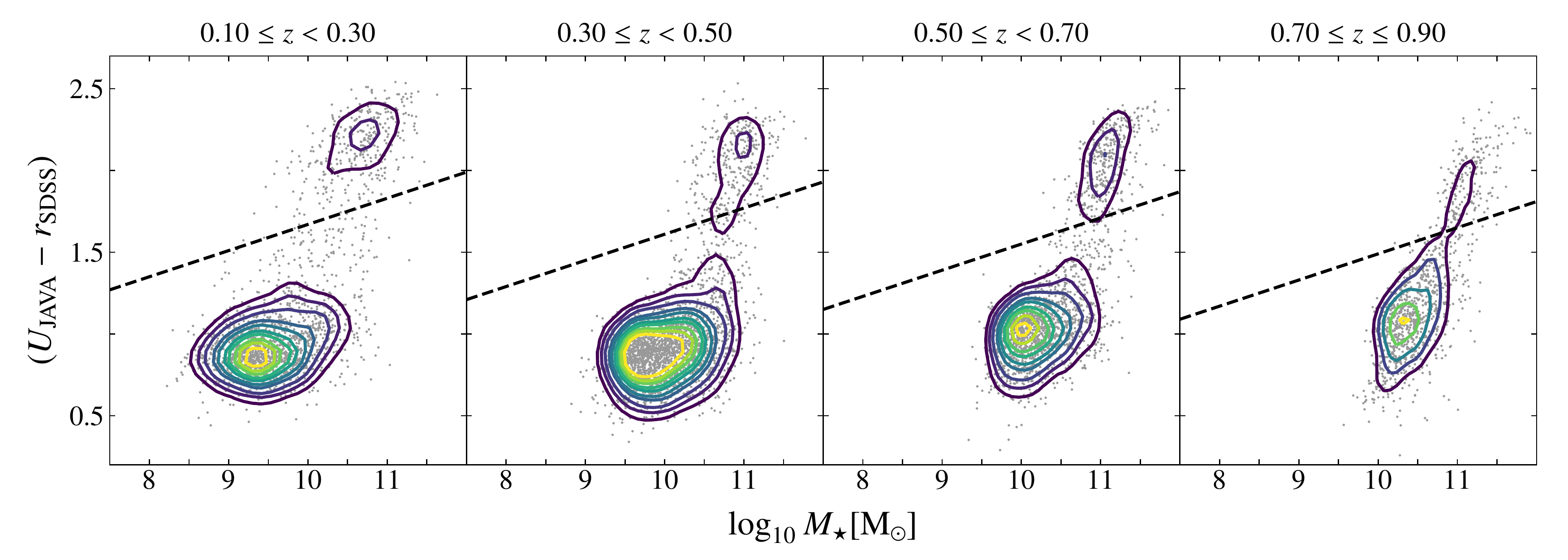}\\

\caption{Rest-frame stellar mass-colour diagram corrected for extinction, showing the distribution of star-forming (lower cloud) and quiescent (upper cloud) galaxies at four redshift intervals. The dashed black line illustrates the lower colour limit of quiescent galaxies as a function of stellar mass and redshift. Yellow (purple) contours illustrate higher (lower) densities.}
\label{fig:MCDE}
\end{figure*}

\subsubsection{Morphological studies in multi-color} \label{sec:morphology}

The morphology of galaxies, derived from the analysis of photometric data, can vary with wavelength, as different stellar populations present different colors. For example, if the disk of a galaxy is characterized by a young, star-forming population, while the spheroid is old and metal rich, the S\'ersic index  will increase for redder wavelengths \citep{vulcani2014,vika2015}. Several works have been already exploring the variation of  morphological parameters with wavelengths and recovered the star formation history and stellar population of the different galaxy components. The number of fitted  wavelengths has been increasing during the years, from, e.g., the 5 broad bands of the SDSS survey \citep[$ugriz$, ][]{botterell2019}, to the  9  of the GAMA survey \citep[SDSS + UKIDSS,][]{kennedy2016} and the $\simeq 15$ bands in the Dustpedia project \citep[GALEX, SDSS, 2MASS/UKIDSS, WISE,][]{davies2017}. 
  The \jp\ data set will be unique for this type of studies, as it will  provide a low-resolution spectrum for each galaxy component  and multicolor CAS (concentration, asymmetry, entropy) parameters. \citet{vika2013} showed, using SDSS data, that a multi-band bulge-to-disk light decomposition is reliable up to  $z\sim 0.3$, while the CAS parameters have been used to define  morphological properties up to $z\sim 0.25$ using SDSS data \citep{ferrari2015} and even up to $z\sim 3$  using the Hubble Space Telescope GOODS ACS and Hubble Deep Field images \citep{concelice2003}. 
  
Figure  \ref{fig:morphology}, generated with \mjp\ data, gives a taste of the \jp\ potential. From a purely qualitative perspective, the first three panels  show how an early- and a late-type galaxy appear at different wavelengths.   While the  isophotal intensity of the early type galaxy increases with increasing wavelength, since the majority of its mass comprises old and metal rich stars, for the late type  galaxy the spiral arms, where most star formation occurs, fade away at red wavelengths. The colors of the different components are also clearly visible in the middle $g,r,i$ color image. More quantitatively, we show in the right panel, for both galaxies, the variation with wavelength of the concentration $C1$\footnote{The concentration $C1$, first of the three CAS parameters, is defined as follows: $C1= \log (R_{80}/ R_{20})$, where $R_{80}$ and $R_{20}$ are the radii containing, respectively, $80\%$ and $20\%$ of the total light inside the Petrosian Region.}, which ranges from 0 to 1. This was  derived using the non-parametric code \texttt{MFMTK} \citep[][]{ferrari2015}. The symbols are color-coded according  to  the bulge-to-total light ratio, obtained  performing a bulge-to-disk light decomposition of the galaxy images using \texttt{GALFITM},  a multi-wavelength extension of \texttt{GALFIT} \citep{peng2011}.  
  For the early type galaxy the concentration  is nearly constant with wavelength, while for the late type galaxy $C1$  increases with increasing wavelength. The spiral galaxy is disk-dominated ($B/T < 0.25$), while in the early type galaxy the bulge and the disk contribute almost equally to the galaxy light. Interestingly, while for the spiral galaxy the B/T ratio increases with increasing wavelength, i.e. the percentage of light associated with the bulge is higher in redder bands, in the early type galaxy the bulge is slightly more prominent in bluer bands. Novel techniques, e.g., the Kurvature method of \citet{lucatelli2019}, allow to disentangle the structural multi-components of a galaxy and give indication of the transition regions and  allow to infer the presence of other features, such as bars and nuclear disks. All this information, together with the SED fitting of the \js\, allow to obtain a full picture of the build-up of  galaxy structures (Cortesi et. al. in prep).
  
In conclusion, the \jp\ filter system will allow to accurately recover galaxy morphologies as well as to study structural and stellar population properties.  The wide area of \jp\ will  also allow to observe and characterize peculiar and rare objects in all kinds of environment, such as jellyfish galaxies and green peas.

\subsubsection{Galaxy properties across cosmic time} \label{sec:gal_evol}

The accuracy of the photometric redshifts and the full \js\ allow  to characterize the galaxy population and their properties 
from the nearby universe up to $z\sim1$. 
By  fitting the continuum with parametric or non-parametric star formation histories, one can constrain  stellar population properties of galaxies, as well as  segregate  galaxy populations according to their star-formation activity \citep[see, e.g.,][]{cidfernandes05, mathis2006, walcher2011, moustakas2013}. The ability of the \jp\ narrow band-filters in extracting physical parameters of galaxies have been studied in \citet{mejia2017} for a variety of star formation histories. The authors used SED fitting codes \citep[TGASPEX and DynBaS][]{magris2015} to investigate biases, correlations and degeneracies affecting the retrieved parameters in mock galaxies. They also compared results obtained from SDSS galaxy spectra to narrow-band SEDs (synthesised from the same spectra), concluding that the \jp\ filter system yields the same trend in the age-metallicity relation as spectroscopy, for typical SNR values. With the data introduced in Sec.~\ref{sec:ifu}, González Delgado et al. (2020, in prep.) study a sample of galaxies with several SED fitting codes (MUFFIT by \citealt{diaz2015}; TGASPEX by \citealt{magris2015}; AlStar by Cid Fernandes et al. 2020, in prep.; BaySEAGal by de Amorim et al. 2020, in prep.), additionally showing that \jp\ photometry yields the same trend in the age-mass, color (rest and intrinsic)-mass as the analysis of spectroscopic data.  

In the example of Fig.~\ref{fig:MCDE}, we used the MUlti-Filter FITting code \citep[MUFFIT, ][]{diaz2015} to separate star-forming and quiescent galaxies via a rest-frame stellar mass-colour diagram corrected for extinction \citep[][]{diaz2019}. In this specific analysis we used the single stellar population models of \citet[][]{bruzualcharlot2003} to build two-burst composite stellar population models, assuming the extinction law of \citet[][]{fitzpatrick1999}. The accurate determination of the intrinsic extinction of galaxies, which needs precise and well-calibrated photospectra, is essential to differentiate quiescent galaxies from dusty star-forming galaxies and it is also important for the selection of the well-known green valley galaxies to avoid contamination from obscured star-forming galaxies \citep[][]{brammer2009, cardamone2010, moresco2013, diaz2019}. At the same time, other interesting properties related to the stellar content of galaxies and its evolution (such as formation epoch and age, environment effect, quenching mechanisms, metallicity, initial stellar mass function, etc.) can be studied in a self-consistent way (Gonz\'alez Delgado et al. 2020, in prep.), making possible a proper analysis and interpretation of the correlations between different physical parameters \citep[see][and references therein]{belli2015, martin2015, diaz2019b,diaz2019c}.

\begin{figure}
\centering
\includegraphics[width=\columnwidth ]{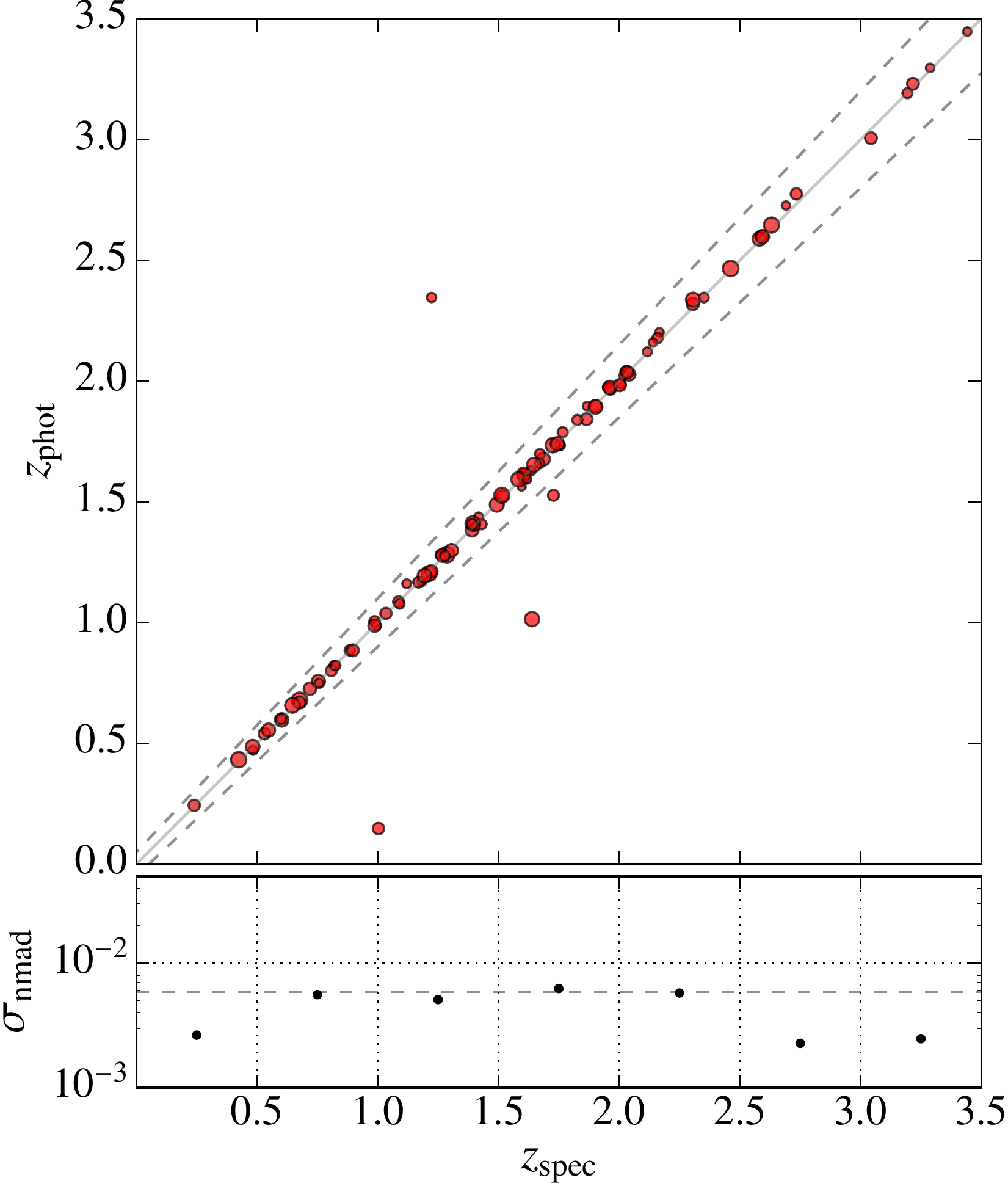}
\caption{
Upper panel: photometric versus spectroscopic redshifts for a sample of $97$ DR14 quasars with $z_{\mathrm{spec}}<3.5$ detected in the \mjp\ field. Larger symbols represent higher $\mathrm{med}(\mathrm{SNR})$. The solid diagonal line indicates $z_{\mathrm{phot}}=z_{\mathrm{spec}}$, and the dashed lines correspond to $z_{\mathrm{phot}}=z_{\mathrm{spec}}\pm 0.05(1+z_{\mathrm{spec}})$. The average \photoz\ uncertainty, $\sigma_{\mathrm{nmad}}$, is $0.0059$  and the fraction of outliers is $4.1\%$. Bottom panel: \photoz\ precision  as a function of redshift, with the horizontal dashed gray line indicating the average \photoz\ uncertainty.
}
\label{cqueiroz_qso_zphot}
\end{figure}

\subsubsection{Emission line objects} \label{sec:elg}

As visually evident from the examples of Fig.~\ref{fig:photospectra},  the \jp\ filter system allows the detection and characterization of emission lines with equivalent widths larger than  a few \AA\ (Mart\'inez-Solaeche et al. 2020, in prep.), from star forming galaxies to AGN and quasars.  

H$\alpha$ and [OII]$\lambda$3727 can be used to trace the star forming population from the local universe  up to $z \sim $ 1.  Weaker lines, such as  H$\beta$ and [OIII]$\lambda$5007,  can also be easily detected for galaxies with $r < 20$ and $z<0.35$, the same magnitude and redshift range of the GAMA survey \citep{driver11}. 
This capability is very important to identify star forming galaxies without, on one hand, the limitations on redshift and/or area coverage typical of  narrow-band emission lines surveys, and, on the other hand, the limitations on magnitudes or pre-selection biases  of spectroscopic surveys.

AGN and quasars can also  be easily identified and their properties characterized with the \js\ \citep{abramo2012, chavesmontero2017}. 
 We used \mjp\ data to test a new method that estimates the \photoz\ of quasars by means of a PCA modeling of the spectral variations (Queiroz et al., in prep.). The eigenspectra of the PCA are computed using as basis a selection of SDSS spectra of broad-line quasars  \citep[][but see also \citet{abramo2012}]{yip2004}, and they correspond to the most relevant modes of variation of broad-line quasars.
In addition to these eigenmodes we also include a reddening law to allow the eigenspectra's slope to adjust and better fit reddened quasar spectra. Therefore, our method employs an optimization routine that is tuned to extract the full redshift probability distribution.
The quality of the derived photometric redshifts for quasars at $z_{\mathrm{spec}}<3.5$ with $r<22$ is estimated using a subsample of $97$ quasars in the \mjp\ footprint with $\mathrm{med}(\mathrm{SNR})$\footnote{where the $\mathrm{med}(\mathrm{SNR})$ is defined as the median value of the ratio flux/flux$\_$err for all filters with detection}$\geq 5$, reliable spectral identification in SDSS from \citet{DR14Q} (\texttt{mask} flag and \texttt{zWarning} flag equal to zero), and for which a visual inspection did not show relevant calibration issues within the \mjp\ observations.

The result is shown in Fig.~\ref{cqueiroz_qso_zphot}. The estimated average \photoz\ uncertainty is $0.0059$ and the fraction of outliers is $4.1\%$. We notice that \photoz\ uncertainty decreases for $z>2$, thanks to the fact that  multiple strong emission lines are present within the \jp\ spectral coverage at those high-$z$.  With the larger samples provided by \jp\, we expect to be able to further fine-tune our methods and reach even higher redshift precision. 

At $z>2$ \jp\ will also be able to detect and characterize large samples of bright Ly-$\alpha$ emitters (LAEs). The wide spectral coverage of the \js\ will allow to distinguish those from quasars, making \jp\ highly complementary to surveys studying LAEs with a few narrow bands \citep[e.g.,][]{ouchi2018} and/or with spectroscopic instruments with limited spectral coverage, such as HETDEX \citep[][]{hetdex}.
In the next section we will discuss the power of \jp\ in performing clustering studies using the expected large samples of emission line galaxies at $z<1$ and quasars and LAEs at $z>2$.

\begin{figure*}[th]
\centering
\includegraphics[width=0.9\textwidth]{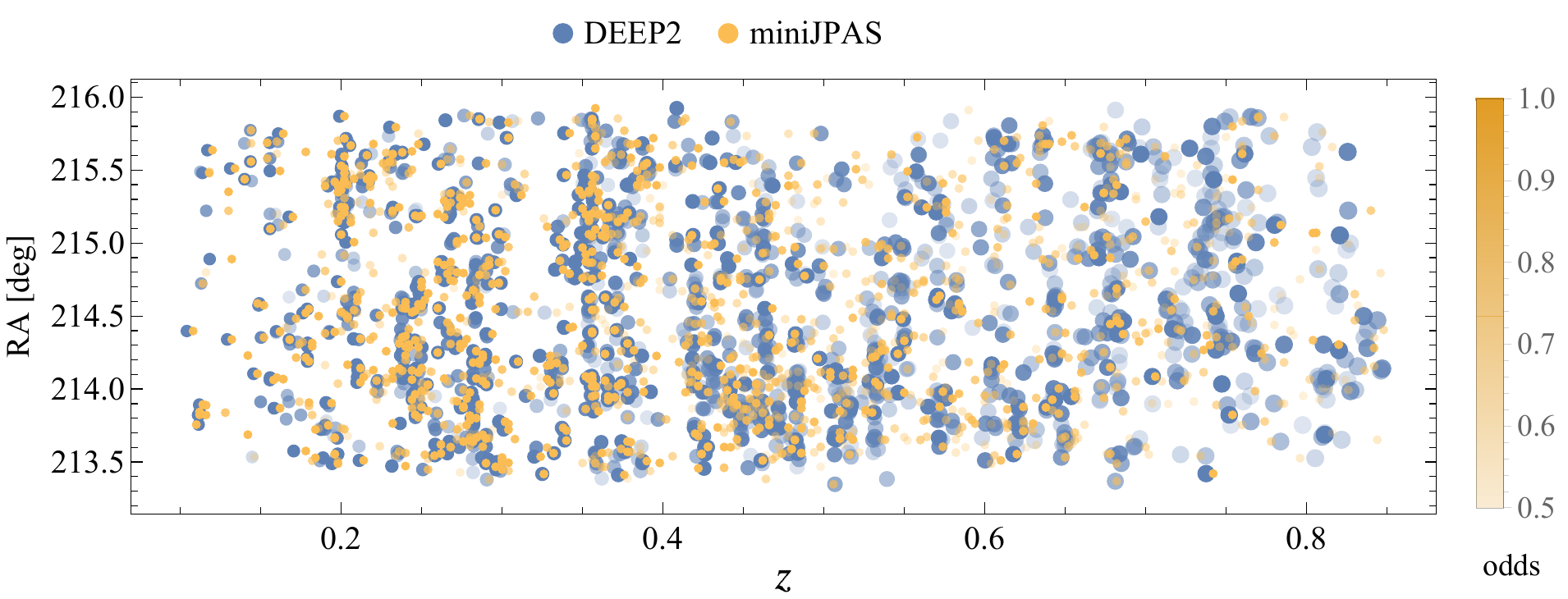}
\caption{
A slice of \mjp\ as mapped by the spectroscopic sample of DEEP2 (blue symbols) and the \mjp\ counterparts (orange symbols), for $r<22.5$.  \mjp\ successfully reproduces the pattern of spatial clustering of galaxies unveiled by the spectroscopic DEEP2 survey.
The redshifts of \mjp\ sources are given by the best \photoz\ solution, as described in Sect.~\ref{sec:photoz}.
The color shade of the \mjp /DEEP2 galaxy pairs is related to the ODDS value of the \mjp\ \photoz\ solution.
The size of the symbols of DEEP2 sources is given by the target \photoz\ precision, $\sigma_{\rm{NMAD}} = 0.003 \times (1+z)$, which is reached when the constant-size \mjp\ symbol lies within the DEEP2 one.
}
\label{fig:fig_LSS}
\end{figure*}

\subsection{Large scale structure} \label{sec:lss}

As already discussed and demonstrated in previous sections, \jp\ was designed as a high-completeness astrophysical survey \citep{benitez2009,benitez2014}, with a set of narrow-band filters able to provide photospectra with enough resolution to result in excellent \photoz\ (see Sect.~\ref{sec:photoz}).

An example of the ability of \jp\ to map the cosmic web is presented in Fig.~\ref{fig:fig_LSS}, where we show a slice of the  common volume mapped by the spectroscopic sample of DEEP2 and by the photometrically  characterized \mjp\ counterparts.
Across a wide redshift range,  \mjp\ is able to reproduce the pattern of spatial clustering of galaxies unveiled by the spectroscopic DEEP2 survey. The impact of the line-of-sight blurring due to the photo-$z$ errors becomes visible in particular for objects with low values of \photoz\ accuracy (ODDS). This typically affects scales along the line-of-sight below $\lesssim 50\,h^{-1}$~Mpc.  
Therefore, \jp\ will be able to accurately map the cosmic structures, from the largest scales to mildly non-linear scales, where most of the cosmological information is encoded.
 In particular, it will be possible to constrain cosmological models using the filamentary structure of the universe~\citep{tempel2014,kitaura2019}. 

Due to its large area, high density of objects, as well as accurate \photoz, \jp\ will impact cosmology in several different ways.
First, since the entire sample of galaxies has \photoz\ errors lower than $< 0.8 \%$  (see Sect.~\ref{sec:photoz}), we will be able to extract the scale of baryonic acoustic oscillations (BAO) from galaxy clustering up to $z \sim 1$, both in the transverse and in the radial directions \citep[see, e.g.,][for the impact of photometric redshifts on clustering measurements]{chaves-montero18}. This analysis will produce accurate measurements of  the angular-diameter distance $D_a(z)$ and the Hubble parameter $H(z)$ \citep{blake2003,seo2003}, allowing us to tightly constrain the dark energy equation of state and the other cosmological parameters.
Second, being an imaging survey which does not need to rely on the pre-selection of targets,
we will be able to measure the shape of the power spectrum (or, equivalently, the correlation function) using many different tracers of the large-scale structure \citep{seljak2009,mcdonald2009,abramo2016}: galaxies of different spectral types, colors, luminosities and stellar masses \citep{arnalte2014, hurtado2016, montero2020}, quasars \citep{abramo2012}, Ly-$\alpha$ emitters \citep{gurung-lopez2019a, gurung-lopez2019b} or even groups and clusters. The multi-tracer character of \jp\ will lead to improved constraints not only on dark energy but also on modified gravity models, primordial non-Gaussianities, and neutrino masses \citep{abramo2017}. In particular, \jp\ has the potential to provide the most precise determination of the Hubble parameter $H(z)$ and the growth rate  $f\sigma_8(z)$ for 
$0.2\leq z \leq 0.6$, outperforming past and upcoming surveys in this redshift range.  Dedicated forecast analysis can be found in  \citet{costa2019}, \citet{aparicioresco2019}
 and  Aparicio Resco \& Maroto (2020, in prep.).

\begin{figure}
\centering
\includegraphics[width=\columnwidth ]{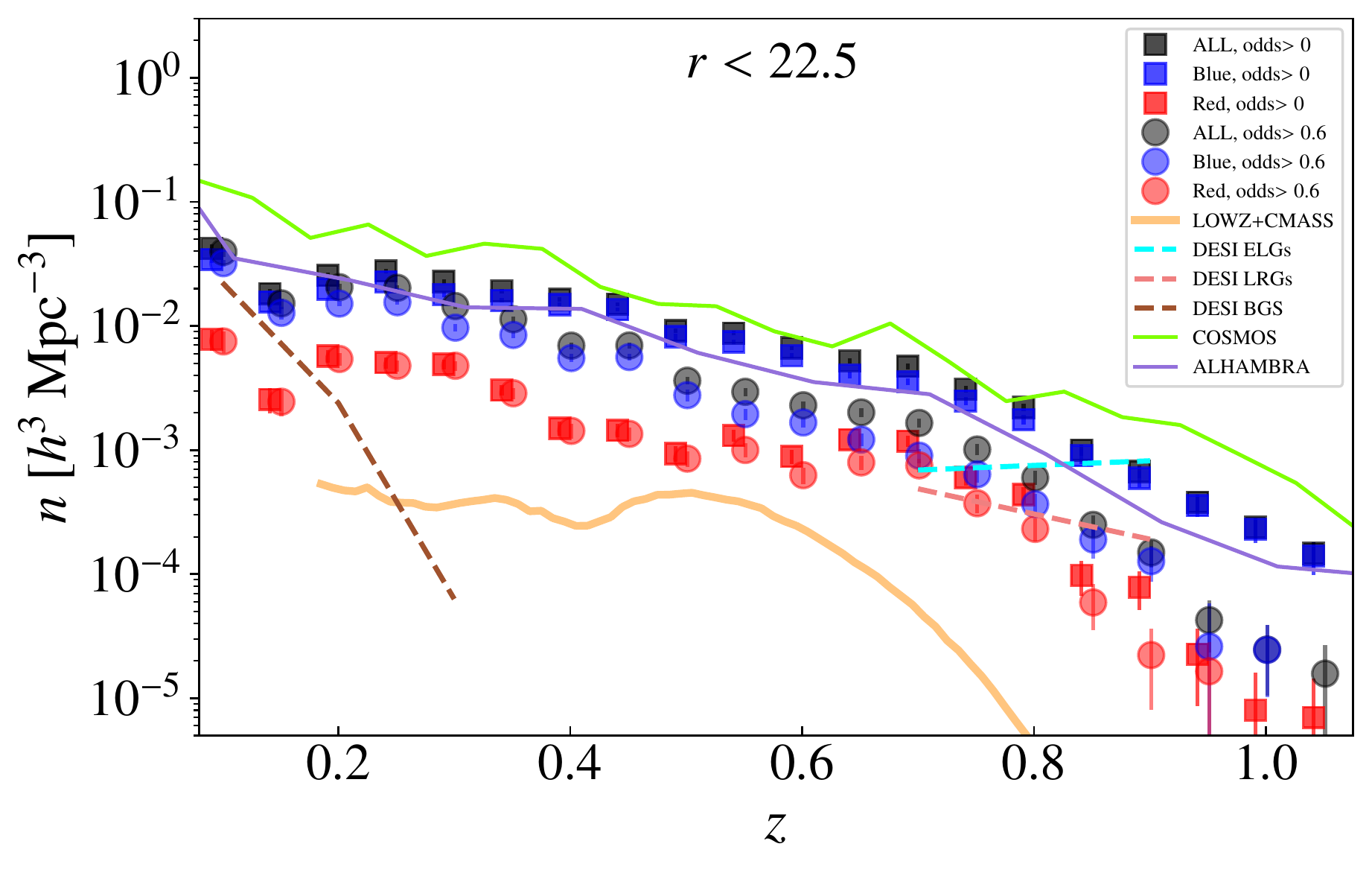}
\caption{Comoving number density of galaxies in \mjp, as a function of redshift. The total galaxy population (dark gray  squares) is broken into star-forming (blue squares) and quiescent (red squares). We also apply a cut in \photoz\ precision (ODDS $>0.6$) (circles, with the same color coding as the squares). We show the number density of galaxies for other surveys like LOWZ+CMASS (orange line), DESI \citep[][brown, cyan, and magenta solid lines for, respectively, BCGs, ELGs, and LRGs]{desi2016}, ALHAMBRA \citep[][purple line, with an applied cut of $F644W<22.5$, where  $F644W$ is a medium-band centred at $6440$\ang]{molino2014}   and COSMOS \citep[][green line, with an applied cut of $r<22.5$ ]{laigle2016}. }
\label{fig:n3d_vs_z}
\end{figure}

\begin{figure}
\centering
\includegraphics[trim={10 0 40 40}, clip, width=\columnwidth ]{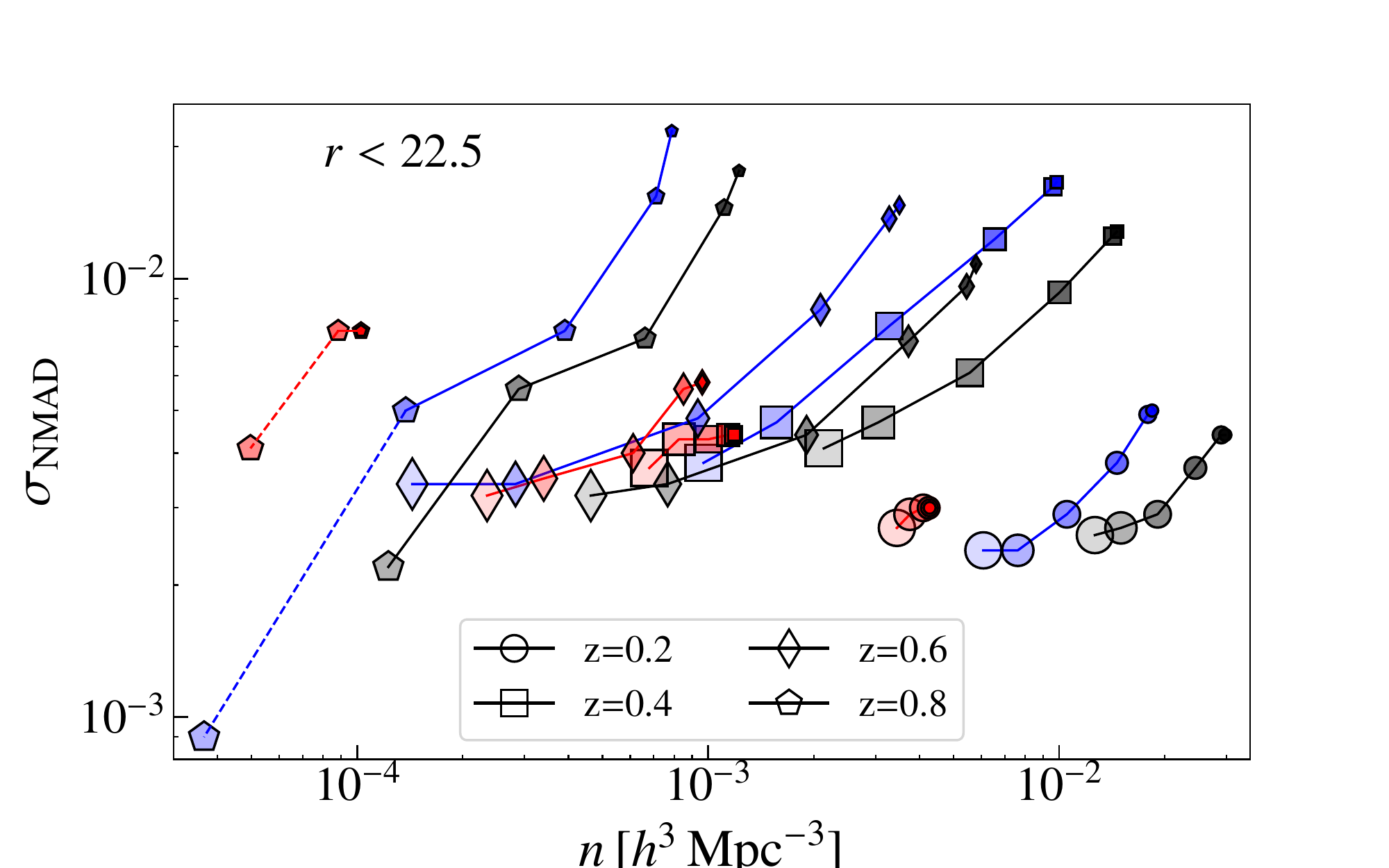}
\caption{Relation between galaxy number density and corresponding redshift error, as a function of redshift, ODDS cut and galaxy color.  Red galaxies, blue galaxies and the full sample are shown, respectively, by red, blue and black lines and symbols. Results are presented in redshift bins of $\Delta z=0.2$, centred at $z=0.2$ (circles), $z=0.4$ (squares), $z=0.6$ (diamonds) and $z=0.8$ (pentagons). For each redshift, the symbols  correspond to the cumulative number density for progressively increasing ODDS cuts, with cuts at $ODDS>$0.0, 0.2, 0.4, 0.6, 0.8, 0.9 (for some cuts the symbols are overlapping). The larger the size of the symbols, the larger the ODDS cut (thus the redshift precision). Dashed lines correspond to the regime where we have, in the \mjp\ field, less than 15 galaxies.}
\label{fig:sgnmad_vs_n}
\end{figure}

Another indicator of the power of \jp\ to constrain cosmology is the signal-to-noise with which we can measure galaxy clustering. For a galaxy with bias $b_g$, the amplitude of its clustering in Fourier space is given by $P_g = b_{g}^2 P_m(k)$, where $P_m(k)$ is the matter power spectrum, whereas the main source of noise is shot (count) noise, $1/\bar{n}_g$, where $\bar{n}_g$ is the comoving galaxy number density.
Since the matter power spectrum at the typical BAO scales ($k_{BAO} = 0.1 \, h$ Mpc$ ^{-1}$) is of order $P_m(k_{BAO}) \sim (1+z)^{-2} \times 10^4 \; h^{-3}$ Mpc$^3$, and galaxy bias is of order $b_g~\sim~1~-~2$, in order to rise above the level of shot noise a galaxy survey should achieve number densities of $\bar{n}_g \gtrsim (1+z)^2 \times 10^{-4} \, h^{3}$ Mpc$^{-3}$.
In Fig.~\ref{fig:n3d_vs_z} we show the estimated (comoving) spatial number density of galaxies at different redshifts for different color and ODDS cuts, based on the \photoz\ results of \mjp\ (see Sect.~\ref{sec:photoz}) for extended sources with $r<22.5$.
These estimates account for the masked regions in the \mjp\ footprint, and incorporate the statistical weights correcting for the stellar contamination \citep{clsj2019mor}.
As expected, we find that blue/late-type galaxies are significantly more abundant than red/early-type ones. Moreover, the photo-$z$ precision of blue galaxies depends significantly on redshift, while the one of red galaxies is approximately constant throughout cosmic ages. 
Typically, we find number densities in the range $[10^{-3},5 \times 10^{-2}]$~$h^{3}$~Mpc$^{-3}$, significantly higher than past spectroscopic surveys like LOWZ+CMASS and comparable to the photometric pencil-beam surveys like COSMOS or ALHAMBRA and the upcoming DESI spectroscopic survey.  We can select galaxy samples of progressively higher \photoz\ quality by applying cuts in the ODDS parameter (see details in Section \ref{sec:photoz}), at the cost of reducing the sample completeness.   In  Fig.~\ref{fig:n3d_vs_z} we show that for a cut in ODDS of 0.6, we still obtain fairly large number densities, in the range  $[10^{-4},10^{-2}]$~$h^{3}$~Mpc$^{-3}$. The redshift precision associated to each ODDS cut depends on redshift, as shown in  
 Fig.~\ref{fig:sgnmad_vs_n}, where the complicated relation between redshift precision and expected number density of sources is shown for different ODDS cuts and  redshift slices. An ODDS cut of 0.6, as used in Fig.~\ref{fig:n3d_vs_z}, generally ensures a relative redshift error of $0.004$ or lower. 
The previous results and the completeness shown in Fig.~\ref{fig:completeness} suggest that \jp\ will be able to perform very well in the redshift range $0.2<z<0.6$ \citep[][]{aparicioresco2019,costa2019}. Indeed, thanks to its unique combination of speed and photo-$z$ accuracy, \jp\ will be able to observe a large number of ELG in this very important redshift range.

In addition to measuring clustering on large scales, \jp\ will also study galaxy correlations at very short scales (a few tens of kiloparsecs, sampling the so-called 1-halo term) due to the high number densities of detected sources. In particular, since \jp\ is not affected by fiber-collision problems in the way spectroscopic surveys are, the small scale limit is set only by the combination of seeing and photometric redshift accuracy.
Beyond the standard clustering analyses, the quality of \jp\ data will enable a family of alternative cosmological tests. For instance, the redshift precision of \jp\ will allow us to analyse angular redshift fluctuations \citep[ARF,][]{chm2019} under narrow redshift shells ($\Delta z\sim 0.01$). This new cosmological observable contains information about the bias and peculiar velocities of the probes, and thus is sensitive to gravity and non-Gaussianity parameters from inflation, among others. A tomographic analysis \citep[like that of][]{chm2020} would hence provide a detailed view of the redshift evolution of source bias, peculiar velocity amplitudes, and a number of other cosmological quantities that can be derived thereof. At the same time, the combination of ARF with standard 2D angular clustering maps can be naturally cross-correlated with current and future CMB/sub-millimeter maps, and other 2D maps derived from them, like the CMB lensing convergence map, maps of the Cosmic Infrared Background, or maps of the thermal Sunyaev-Zel’dovich effect \citep[][]{sunyaev1972}. 
Given its high number density of galaxies (reaching $>10^{-2} \, h^3$ Mpc$^{-3}$ at $z<0.5$), \jp\ will also be able to map the cosmic web in rich detail, revealing the pattern of sheets, filaments and voids over an unprecedented volume, and providing a powerful test of the cosmological model as well as of our theory of structure formation \citep{kitaura2019}.
Another potential field where J-PAS may contribute significantly is the study of emission from unresolved sources in different (and narrow) redshift shells: the narrow bands should allow for the isolation of line emitters placed at particular redshifts, even when these lack the required S/N to be identified as sources in the catalogue. Crucial for this {\it intensity mapping} (IM) approach \citep[see, e.g.,][]{IM} are the adopted techniques for background removal, which may impact IM signal on the largest scales.
Finally, \jp\ will have the unique ability to  measure the cosmic shear (see Sect.~\ref{sec:wgl})
in the same volume in which the galaxy power spectrum is measured, thus probing  different combinations of the gravitational potentials with two independent observables.

At the highest redshifts ($z>2$), cosmological studies will be possible via quasars and Ly-$\alpha$ emitters (LAEs). Because of their strong emission features, these objects can be easily detected and characterized with the \js. Using \mjp, we tested the quasar  \photoz\ uncertainty for objects at $r<22$, being on average $\sim 0.5\%$, but decreasing to $0.3\%$ for objects at $z>2$ (see Sect.~\ref{sec:elg}). Even at fainter magnitudes (up to $r\sim23$) we expect to recover complete quasar samples (with a density of $\sim 200$~deg$^{-2}$) with $\sim 1\%$ redshift precision, allowing efficient spectroscopic follow-up campaigns. The WEAVE-QSO survey \citep{pieri2016} will target $\sim400k$ \jp\ quasars at $z>2$ which will allow precise Ly-$\alpha$ forest and IGM studies. Finally, at $z>2$, \jp\ will select large samples of LAEs to perform cosmological studies. This will be highly complementary to the HETDEX spectroscopic survey \citep{hetdex}, as the lower redshift precision of \jp\ will be compensated by the larger area and wider spectral coverage, which allows to easily distinguish LAEs from quasars.

\subsection{Galaxy clusters} \label{sec:cl}

Being the largest gravitationally bound structures in the Universe, galaxy clusters are key to discriminate among the various cosmological models. In particular, they can be used to place constraints on several fundamental parameters such as $\Omega_m$ and $\Omega_{\Lambda}$, $w$ and $\sigma_8$ \citep[e.g., see][ for a review]{allen2011}.
\citet{ascaso2016} applied the Bayesian Cluster Finder \citep[][]{ascaso2012}  and calculated the selection function for \jp\ from N-body+analytical mock catalogues similar to the ones of \citet{merson2013} adapted to \jp. Thanks to the exquisite \photoz\ that \jp\ can deliver, they found the minimum halo mass threshold for detection of galaxy systems with both $>$80\% completeness \textit{and} purity to be  M$_{halo} \sim 5 \times 10^{13} M_\odot$ up to z $\sim$ 0.7, outperforming DES \citep{DESY1}, ACTpol \citep{hilton2018} \& SPTpol \citep{bleem2020}, and comparable to eROSITA \citep{merloni2012} and LSST \citep{lsst2009} at lower-intermediate redshifts. This is a unique feature of \jp, that will allow to expand the study of the evolution of galaxy systems down to galaxy groups. 
These low-mass systems are particularly interesting and poorly observed at higher redshifts and \jp\ will allow us to clarify many of their unusual properties, among them, to analyse the relative importance of the different mechanisms responsible for the reduction of their measured low baryon fraction \citep[e.g.,][]{dai2010}, for example by comparing stellar loss to intracluster light \citep[][]{jimenez-teja2018, jimenez-teja2019}  to IGM ejection \citep[e.g.,][]{bower2008}  due to the energetic excess by central AGN. 
Moreover, galaxy star formation activity has been suggested to vary over an extended range of environments, measured as a function of galaxy density, and redshift \citep[e.g.,][]{elbaz2007, paulinoafonso2020}. Therefore, the statistical study of galaxy properties in groups will allow us to probe the evolution of galaxies in intermediate environments.

\begin{figure}
\centering
\includegraphics[trim={0 0 0 20}, clip, width=\columnwidth ]{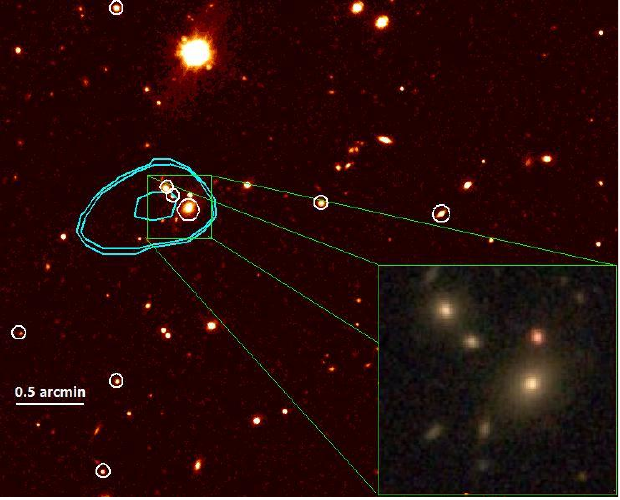}
\caption{Example of a low-mass cluster (group), mJPC 2241-16656, detected by all three cluster finder algorithms in the \mjp/Chandra AEGIS field. The system, dominated by the galaxy SDSS J141719.86+523537.6 is at RA=14:17:20 DEC=+52:35:38 and has \photoz=0.238.  Overlaid on the i-band image, white circles mark the position of galaxy members (within a window of $\pm$ 4$\sigma_{los}$ - or ~840 km/s) within R$_{200}$ (507 h$_{70}^{-1}$ kpc) \citep{evrard1996}. These galaxies have  membership probability larger than $88\%$ as provided by the AMICO code. The contours in cyan show the extended X-ray emission in the 0.5-2~keV range obtained by the analysis of Chandra data in 0.5dex surface brightness levels levels \citep[for details refer to][]{erfanianfar2013}. In the bottom right we show a zoom-in of the BCG surroundings for illustration. 1~arcmin corresponds to $\approx$ 230~kpc at the clusters redshift.}
\label{fig:cluster2}
\end{figure}

To demonstrate the strength of our multi-band data for the selection of groups and clusters (which hereafter we refer to only as ``clusters''), 
we searched for clusters in \mjp\ within the redshift range $0.05<z<0.8$ with the Voronoi Tessellation (VT) technique \citep[][]{ramella2001, lopes2004}, the Adaptive Matched identifier of Clustered Objects \citep[AMICO,][]{bellagamba2018, bellagamba2019, maturi2019} and the Photo-$z$ Wavelet Cluster Detection Code (PZWav), as described in \citet{euclid2019}. 
AMICO and PZWav are the detection algorithms adopted for the upcoming Euclid survey \citep[][]{euclid2019}. Their use in \jp\ will allow to compare and cross validate the results of the two surveys. All cluster finder algorithms mentioned above found a minimum of 30\%-43\% of the systems detected in X-rays with a $\leq$ 0.5 Mpc matching scale for $z \leq 0.5$. From these first tests we can detect clusters down to a mass of $\sim 10^{13}M_\odot$ and clusters up to redshift $z=0.76$ \citep[mass based on the Chandra X-ray Luminosity scaled mass of][]{erfanianfar2013}.  The cluster catalogue based on \mjp\ data will be presented in an upcoming paper (Maturi \& the \jp\ Collaboration 2020, in Prep.). We show here only two illustrative examples. 
An example of a low mass system is shown in Fig.~\ref{fig:cluster2}. It is dominated by a bright group galaxy at $z=0.24$ and it has a mass (scaled from X-ray Luminosity) of 2.9$\times 10^{13} M_{\odot}$, typical of groups or poor clusters.

In Fig.~\ref{fig:cluster1}, instead, we show the most massive cluster detected in the \mjp\ footprint. 
This is the first cluster that stands out in all cluster finding algorithms used (VT, AMICO, PZWav). It is roughly centred around the Bright Cluster Galaxy at RA:$213.62543$ DEC:$+51.93786$ (one of the two galaxies with spectroscopic redshift in the region, with $z_s=0.289$), and we will refer to this cluster as \mjp\ Pathfinder Cluster $2470-1771$, or mJPC $2470-1771$ \footnote{The identifier number corresponds to that of the BCG in the \mjp\ data public data release. Its identifier in SDSS is 
SDSS J141430.10+515616.3}. It is in the Southwest edge of the \mjp\ field and it has not been analysed previously. The cluster has an unusual central alignment that spans about 500 h$_{70}^{-1}$ kpc in the plane of the sky, as shown in Fig.~\ref{fig:cluster1}.
Even though this cluster was barely missed by the Chandra observations of the AEGIS field,  the XMM-Newton program covering a large area around the Chandra field (PI Merloni) allowed us to derive some X-ray properties relatively well through a a very off-centered 18 ksec XMM pointed observation.
We used standard data reduction protocol using the XMM-Newton Science Analysis System (SAS) version \texttt{xmmsas-v18.0.0}.
 The X-ray spectrum of the hot gas was modeled with an absorbed thermal Bremsstrahlung with emission lines (\textit{phabs apec}), with redshift fixed on 0.289, corresponding to the Bright Cluster Galaxy (BCG) and hydrogen column density of 1.9 $\times$ 10$^{20}$  cm$^{-2}$.
Simultaneous spectral fitting of all imaging spectrometers showed an intracluster gas temperature of $(2.8 \pm 0.8)$~keV for a core excised (30~arcsec) region of 80~arcsec radius centred in the X-ray peak. 
The R$_{200}$, scaled from that X-ray temperature is $\sim 1200 \rm{h}_{70}^{-1}$~kpc and a mass M$_{200c}$ is $\sim$ 3.26$\pm$1.4$\times 10^{14}$~M$_{\odot}$. Within that radius \mjp\ finds 29 galaxies within 2000~km~s$^{-1}$ of the BCG redshift (scale is 4.378 kpc~arcsec$^{-1}$).
Using \mjp\ data alone the mass estimated using richness scaling relation \citep{lopes2009}, photo-z derived velocity dispersion \citep{heisler1985} and the cluster model amplitude returned by the AMICO code is found to be (2.6,4.8,3.7)$\times 10^{14} M_{\odot}$ with (50\%,16\%,17\%) confidence errors respectively. 
A more detailed analysis of this cluster, including the physical properties of its member galaxies, will be presented in a separate work.

\subsection{Lensing} \label{sec:lensing}

The excellent conditions at the OAJ open the window to carry out many studies using gravitational lensing, since not only the shapes of several hundred millions of galaxies will be measured, but their redshifts will also be known with high accuracy.

\subsubsection{Strong Lensing}
\label{sgl}

\begin{figure}[t]
\centering
\includegraphics[width=0.48\textwidth ]{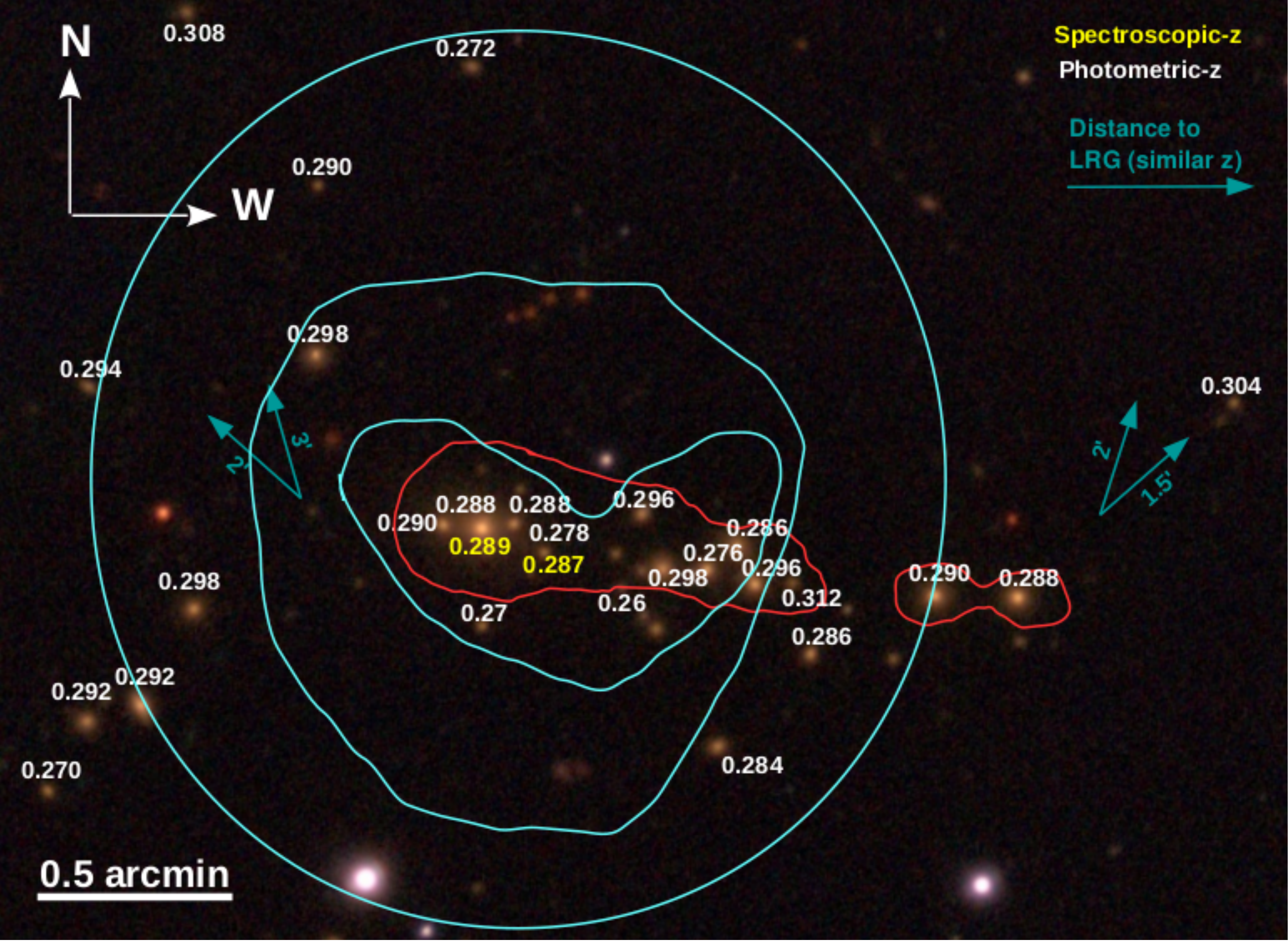}
\caption{The most massive cluster found in the \mjp\ footprint, centred at  RA=213.6254, DEC=51.9379. This cluster is also part of the \texttt{redMaPPer} catalogue where it is listed as a cluster with richness $\lambda = 33$. The brightest galaxy has a spectroscopic redshift (from SDSS) of 0.289. Numbers in white next to galaxies indicate the photometric redshift derived from \mjp\ for galaxies with photometric redshifts between $\sim 0.26$ and 0.31. More galaxies in this redshift interval (including some luminous ellipticals) are found at larger distances. The arrows indicate the direction and approximate distance of luminous galaxies found nearby and in this redshift interval. In red are overdrawn the strong lensing critical curves, estimated for fiducial sources at redshift $z_s=2$ with the \texttt{EasyCritics} code applied to our data. We also show the X-ray iso-contour emission in the 0.5-2 keV range obtained by the analysis of XMM-Newton (cyan) in steps of 0.5~dex in surface brightness. 1~arcmin  corresponds to $\approx$ 200 kpc at the clusters redshift. }
\label{fig:cluster1}
\end{figure}

Strong lensing by galaxy clusters offers the opportunity to map the central distribution of dark matter in the cluster, invisible otherwise. Our team has two unique tools to automatically analyse potential cluster strong-lenses found within the \jp\ footprint. The first method was formulated by \citet{zitrin2012}, who showed that by calibrating the effective mass-to-light ratio of cluster members (e.g., using other well-known clusters analysed in HST data), their light-traces-mass method could be operated automatically on large sky surveys, without requiring multiple images as input and relying only on the photometry of identified cluster members  \citep[see also][]{carrasco2020}. This is particularly important because multiple-image constraints for building strong-lens models are typically very hard to identify from the ground.
Built on the foundations of this method, \citet{stapelberg2019}  formulated \texttt{EasyCritics}, the second method we use here, which not only estimates the lensing strength of clusters, but also finds them in the data. In Fig.~ \ref{fig:cluster1} we show the critical curves from the \texttt{EasyCritics} method for a fiducial source at redshift $z_s=2$. Note that the calibration used to derive these curves relied on CFHTlens data, not on the \mjp\ data, and thus these curves should only be referred to as an illustration. 

In addition to the possibility of \jp\ to find relatively small galaxy-cluster lenses, \jp\ offers unique opportunities to find galaxy-galaxy lenses. Most of these systems are unresolved from the ground with $2.5$~m class telescopes, making the task to find them very difficult. However, using the \js\ it is in principle possible to find some of these rare systems by looking for outliers corresponding to the overlapping spectra of a foreground lens (the most likely lenses are ellipticals with no emission lines) and a background high-redshift galaxy (common background sources are star forming galaxies at $z>1$, with emission lines). A similar, and successful, effort was conducted using real spectra from SDSS in the SLACS project \citep{auger2009}

\subsubsection{Weak Gravitational Lensing} \label{sec:wgl}

Clusters as massive as mJPC 2470-1771 (Fig.~\ref{fig:cluster1}) are predicted to show strong lensing effects. Such clusters also produce weak gravitational lensing signals. The photometric depth of \jp, in general, makes weak lensing measurements of individual clusters challenging. However, this can be alleviated by performing joint analyses of large samples such as stacked analysis. This has the potential to extract valuable astrophysical and cosmological information from not only massive clusters, but also from low mass ones (see e.g., Fig.~\ref{fig:cluster2}) and even from individual galaxies. 

The OAJ site has shown to have seeing better than  0.8$^{\prime\prime}$ during 68\% of the night time \citep{moles2010} and thus can produce images with quality for shear estimation from faint galaxy shape measurements. Given its area and depth, \jp\ has the potential to be one of the main contributors on the fields of cosmic shear, cluster and galaxy-galaxy lensing, prior to the LSST/Euclid era. By using the same procedures and selection criteria as the first year of the \hsc\ survey \citep{mandelbaum2018}, we estimated with \mjp\  that we have an effective number density of source galaxies of  $\sim$2.7 arcmin$^{-2}$, using the co-added $r$
band images alone, which is a lower limit to what we expect to obtain with \jp. This is because the broad band images of the future \jp\ observations are planned to be taken under better conditions (airmass and seeing) than those of \mjp. Moreover, with the use of multiple filters for shape measuring we expect to significantly increase the effective number density of source galaxies.

Lensing magnification provides complementary, independent observational
alternatives to gravitational shear
\citep[][]{broadhurst1995,umetsu2011,umetsu2013}.
In contrast to the shear-based analysis, measuring the effects of
magnification does not require source galaxies to be spatially resolved,
whereas it does require accurate photometry across the sky and a
stringent flux limit against incompleteness effects. This is  where
J-PAS has unique advantages against other wide-field surveys.

In the presence of lensing magnification $\mu$ ($\approx1+2\kappa$ in
the weak-lensing regime), the source counts $n(<m)$  for a given
magnitude cut $m$ are  modified as $n(<m)/n_0(<m) = \mu^{2.5s(m)-1}$,
with $n_0(<m)$ the intrinsic (unlensed) source counts and $s(m)$ the
intrinsic count slope, $s(m)=d\log_{10}n_0(<m)/dm$.
By analysing \mjp\ data, we find that the source counts $n_0(<m)$ for a
$r$-magnitude-limited sample of background galaxies  increase
progressively with magnitude down to the completeness limit ($r\sim
22.7$).
Since the net effect of magnification bias depends on the count slope $s(m)$,
while its signal-to-noise ratio is limited by $n_0(<m)$,  we will make use of the precise \photoz\ delivered by \jp\ to 
optimize the selection of background galaxies in magnitude and redshift
space so as to maximize the sensitivity.
Therefore, with \jp\ data, we will measure the magnification signal around a large
statistical sample of galaxy clusters to obtain a
robust mass calibration for cluster cosmology.

\section{Summary and Conclusions} \label{sec:summary}

In this paper we have presented \mjp, a unique 3D survey using the full \jp\ filter system that we have carried out to evaluate the scientific potential of the large \jp\ survey to follow. This first data has allowed us to establish the photometric depth and redshift accuracy that the system can achieve and to test  our reduction pipelines for \jp.
The footprint of \mjp\ was chosen to cover $\sim 1 \rm{deg}^2$ of the well studied AEGIS field, in all the 56 filters of the \jp\ filter system, plus the four broad bands $u,g,r,i$. The observational strategy adopted allowed us to reach, and often  surpass, the desired minimum depths  as described in \citet{benitez2014}, obtaining $\rm{mag}_{\rm {AB}}$ between $\sim 22$ and $23.5$ for the NB filters and to $24$ for the BB filters to $5\sigma$ in a $3$~arcsec aperture.

  The \mjp\ primary catalogue contains over 64k sources, extracted in the primary $r$ detection band with forced photometry in all other bands. In addition, over 600k sources have been extracted separately in the individual bands, generating catalogues that are particularly interesting for the detection of extreme line emitters over the wide spectral range covered by the narrow band filters. We estimate the full \mjp\ catalogue to be complete up to $r=23.6$~AB for point-like sources and $r=22.7$~AB for extended sources.
Photometric redshifts have been derived using a customised version of \textsc{LePhare}, modified to work with a larger number of filters and higher resolution in redshift than typically required for broad-band photometry. We have demonstrated here that the \jp\ filter system is able to reach sub-percent precision in \photoz\  for all sources brighter than mag$\sim 22.5$. The target precision of $0.3\%$  is reached for about half the sources, and  subsamples with redshift precision as high as $0.2\%$ can also be extracted by performing appropriate cuts from the \photoz\ outputs.

All data, from images to catalogues and value-added data products from this \mjp\ survey have been  publicly released at the end of 2019 and are available in the \jp\ data portal\footnote{\url{http://archive.cefca.es/catalogues/minijpas-pdr201912}}. Data can be accessed via a variety of tools: sky navigator , object list searches and ADQL queries.

The analysis of  \mjp\  has highlighted several examples of the many scientific applications possible with the upcoming large \jp.  We have shown how the \js\ can be used to characterize a very wide variety of sources, from galactic objects to high redshift quasars, thanks to the photometric detection of absorption and emission features readily visible in our spectral energy distributions. We have demonstrated that stellar population studies can be carried out up to $z \sim 1$ and the that detailed morphology and structural properties of galaxies can be analysed combining the multi-color information.  We have shown the capability of \jp\ as effectively a massive IFU survey, of cosmologically useful spectral resolution over an unprecedentedly large FoV.  The accurate \photoz\ achieved are key to the study of the large scale structure, including baryon acoustic oscillations as well as the shape and evolution of the power spectrum of galaxies and clusters. It is now clear from our survey that galaxy samples with sub-percent redshift precision and a number density as high as $>10^{-2} \, h^3$ Mpc$^{-3}$ can be selected at $z<0.4$, and with a number density exceeding $10^{-4} \, h^3$ Mpc$^{-3}$ at $0.4<z<0.9$, as well as a near-complete sample of $r<23$ quasars at redshifts $z>2$. We have shown that secure identification of clusters and groups of galaxies can be made within this redshift range because  our redshift precision means there is minimal contamination by non members. Furthermore, we have shown with \mjp\ that weak lensing is detectable for individual massive clusters and can be stacked for groups, allowing us to track the mass evolution of clusters to unprecedented depth, providing a self-contained evolution of the mass function with the \jp\ survey.

  We conclude that \mjp\ has demonstrated the capability of the \jp\ filter system to deliver accurate classification of stars and extragalactic sources, and the unambiguous redshift determination of tens of thousands of galaxies per squared degree. \mjp\ is only a tiny wedge of the sky that \jp\ will probe, as \jp\ will surpass \mjp\ by more than three orders of magnitude in terms of area and hence number of sources and volume surveyed.  The full survey will provide a contiguous 3D map of the Northern sky to $z\sim 0.9$, offering the opportunity to study galaxy evolution and the large scale structure of the Universe over a large fraction of the cosmological volume and opening the window to  serendipitous discoveries.
  
   \vspace{1cm}

\begin{acknowledgements}

{ \tiny 
SB acknowledges PGC2018-097585-B-C22, MINECO/FEDER, UE of the Spanish Ministerio de Economia, Industria y Competitividad. CEFCA researchers acknowledge support from the project PGC2018-097585-B-C21.  R.A.D. acknowledges support from the Conselho Nacional de Desenvolvimento Cient\'ifico e Tecnol\'ogico - CNPq through BP grant 308105/2018-4, and the Financiadora de Estudos e Projetos - FINEP grants REF. 1217/13 - 01.13.0279.00
and REF 0859/10 - 01.10.0663.00 and also FAPERJ PRONEX grant E-26/110.566/2010 for hardware funding support for the J-PAS project through the National Observatory of Brazil and Centro Brasileiro de Pesquisas Físicas. LRA acknowledges financial support from CNPq (306696/2018-5) and FAPESP (2015/17199-0). VM thanks CNPq (Brazil) and FAPES (Brazil) for partial financial support. L.A.D.G. and K.U. acknowledge support from the Ministry of Science and Technology of Taiwan (grant MOST 106-2628-M-001-003-MY3) and from the Academia Sinica (grant AS-IA-107-M01).
J.M.D. and D.H acknowledge the support of project PGC2018-101814-B-100 (MCIU/AEI/MINECO/FEDER, UE). V.M.P. acknowledges partial support from grant PHY~14-30152 (Physics Frontier Center/JINA-CEE), awarded by the U.S.\ National Science Foundation (NSF). MQ thanks CNPq (Brazil) and FAPERJ (Brazil) for financial support. PC acknowledges financial support from Funda\c{c}\~{a}o de Amparo \'{a} Pesquisa do Estado de S\~{a}o Paulo (FAPESP) process number 2018/05392-8 and Conselho Nacional de Desenvolvimento Cient\'ifico e Tecnol\'ogico (CNPq) process number  310041/2018-0. AAC acknowledges support from FAPERJ (grant E26/203.186/2016), CNPq (grants  304971/2016-2 and 401669/2016-5), and the Universidad de Alicante (contract UATALENTO18-02). C.Q. acknowledges support from FAPESP (grants 2015/11442-0 and 2019/06766-1). P.B acknowledges support from Coordena\c{c}\~ao de Aperfei\c{c}oamento de Pessoal de N\'ivel Superior - Brasil (CAPES) - Finance Code 001. 
IAA researchers acknowledge financial support from the State Agency for Research of the Spanish MCIU through the ``Center of Excellence Severo Ochoa'' award to the Instituto de Astrof\'isica de Andaluc\'ia (SEV-2017-0709). RGD, GMS, JRM, RGB, EP   acknowledge financial support from  the project AyA2016-77846-P. TC is supported by the INFN INDARK PD51 and PRIN-MIUR 2015W7KAWC. MAR and ALM acknowledge support from the MINECO project FIS2016-78859-P(AEI/FEDER, UE). ET, AT and JL acknowledge the support by ETAg grants IUT40-2 and by EU through the ERDF CoE grant TK133 and MOBTP86. CK, JMV, JIP acknowledge financial support from project AYA2016-79724-C4-4P. PAAL thanks the support of CNPq, grant 309398/2018-5. LC thanks CNPq for partial support. Y.J-T acknowledges financial support from the Fundação Carlos Chagas Filho de Amparo à Pesquisa do Estado do Rio de Janeiro – FAPERJ (fellowship Nota 10, PDR-10) through grant E-26/202.835/2016, and from the European Union’s Horizon 2020 research and innovation programme under the Marie Sklodowska-Curie grant agreement No 898633 - CICLE. DMD acknowledges financial support from the Sonderforschungsbereich (SFB) 881 ``The Milky Way System'' of the German Research Foundation (DFG) and from the MINECO grant AYA2016-81065-C2-2. FP thanks the support of the Spanish Ministry of Science funding grant PGC2018-101931-B-I00. JC acknowledges support of the project E AYA2017-88007-C3-1-P, and co-financed by the FEDER. JIGs acknowledges support of projects  of reference AYA2017-88007-C3-3-P, and PGC2018-099705-B-I00 and co-financed by the FEDER.
EMG and PV would like to acknowledge finantial support from the project ESP2017-83921-C2-1-R (AEI, FEDER UE). GMS acknowleges finantial support from a predoctoral contract, ref. PRE2018-085523 (MCIU/AEI/FSE, UE).
 S.C. is partially supported by CNPq (Brazil). R.G.L. acknowledges CAPES (process 88881.162206/2017-01) and Alexander von Humboldt Foundation for the financial support. JSA acknowledges support from FAPERJ grant no. E26/203.024/2017, CNPq grant no. 310790/2014-0 and 400471/2014-0 and the Financiadora de Estudos e Projetos - FINEP grants Ref. 1217/13 - 01.13.0279.00 and Ref. 0859/10 - 01.10.0663.00. RvM acknowledges support from CNPq (Brazil). AFS, PAM, VJM and FJB  acknowledge support from project AYA2016-81065-C2-2. PAM acknowledges support from the ``Subprograma Atracci\'o de Talent - Contractes Postdoctorals de la Universitat de Val\'encia''. ESC  acknowledges support from Brazilian agencies CNPq and FAPESP through grants  308539/2018-4 and 2019/19687-2, respectively. CMdO acknowledges support from Brazilian agencies CNPq
(grant 312333/2014-5) and FAPESP (grant 2009/54202-8). LSJ acknowledges support from Brazilian agencies CNPq (grant 304819/2017-4) and FAPESP (grant 2012/00800-4). JMC acknowledges support from CNPq (grant 310727/2016-2).
JJBP and AMC would like to acknowledge the support from the Spanish Ministry MCIU under the MCIU/AEI/FEDER grant (PGC2018-094626-B-C21) and the Basque Government grant (IT-979-16). AMC acknowledges the postdoctoral contract from the University of the Basque Country UPV/EHU ``Especializacio\'on de personal investigador doctor'' program. MLLD acknowledges Coordena\c{c}\~ao de Aperfei\c{c}oamento de Pessoal de N\'ivel Superior - Brasil (CAPES) - Finance Code 001; and Conselho Nacional de Desenvolvimento Cient\'ifico e Tecnol\'ogico - Brasil (CNPq) project 142294/2018-7. GB acknowledges financial support from the National Autonomous University of Mexico (UNAM) through grant DGAPA/PAPIIT IG100319, from CONACyT through grant CB2015-252364, and from FAPESP projects 2017/02375- 2 and 2018/05392-8. M.J. Rebou\c{c}as acknowledges the support of FAPERJ under a CNE E-26/202.864/2017 grant, and  CNPq. Support by CNPq (305409/2016-6) and FAPERJ (E-26/202.841/2017) is acknowledged by DL. AB acknowledges a CNPq fellowship.
C.A.G.acknowledges support from Coordena\c{c}\~ao de Aperfei\c{c}oamento de Pessoal de N\'ivel Superior - Brasil (CAPES). EA acknowledges support from FAPESP grant FAPESP 2011/18729-1. AC acknowledges support from PNPD/CAPES. ABA acknowledges the Spanish Ministry of Economy and Com-petitiveness (MINECO) under the Severo Ochoa program SEV-2015-0548. FSK  FSK and ABA acknowledge the Spanish Ministry of Economy and Competitiveness (MINECO) under the Severo Ochoa program SEV-2015-0548,  and AYA2017-89891-P grants. FSK also thanks the  RYC2015-18693 grant. DF acknowledges  support from the Atracci\'on del Talento Cient\'ifico en Salamanca programme and the project PGC2018-096038-B-I00 by Spanish Ministerio de Ciencia, Innovaci\'on y Universidades.

SDSS - This research has made use of  SDSS, which is managed by the Astrophysical Research Consortium for the Participating Institutions of the SDSS Collaboration including the 
Brazilian Participation Group, the Carnegie Institution for Science, 
Carnegie Mellon University, the Chilean Participation Group, the French Participation Group, Harvard-Smithsonian Center for Astrophysics, Instituto de Astrof\'isica de Canarias, The Johns Hopkins University, Kavli Institute for the Physics and Mathematics of the Universe (IPMU),  
University of Tokyo, the Korean Participation Group, Lawrence Berkeley National Laboratory, Leibniz Institut f\"ur Astrophysik Potsdam (AIP), Max-Planck-Institut f\"ur Astronomie (MPIA Heidelberg), Max-Planck-Institut f\"ur Astrophysik (MPA Garching), Max-Planck-Institut f\"ur Extraterrestrische Physik (MPE), National Astronomical Observatories of China, New Mexico State University, New York University, University of Notre Dame, Observat\'orio Nacional / MCTI, The Ohio State University, Pennsylvania State University, Shanghai Astronomical Observatory, 
United Kingdom Participation Group, Universidad Nacional Aut\'onoma de M\'exico, University of Arizona, University of Colorado Boulder, University of Oxford, University of Portsmouth, 
University of Utah, University of Virginia, University of Washington, University of Wisconsin, 
Vanderbilt University, and Yale University. 
ADS - This research has made use of NASA’s Astrophysics Data System.
HEASARC - This research has made use of data and/or software provided by the High Energy Astrophysics Science Archive Research Center (HEASARC), which is a service of the Astrophysics Science Division at NASA/GSFC
This research has made use of the WEB Cosmology Calculator Wright (2006, PASP, 118, 1711)
DEEP - Funding for the DEEP2 Galaxy Redshift Survey has been provided by NSF grants AST-95-09298, AST-0071048, AST-0507428, and AST-0507483 as well as NASA LTSA grant NNG04GC89G.
This paper made use of the Hyper Suprime-Cam (HSC), which includes the astronomical communities of Japan and Taiwan, and Princeton University. The HSC instrumentation and software were developed by the National Astronomical Observatory of Japan (NAOJ), the Kavli Institute for the Physics and Mathematics of the Universe (Kavli IPMU), the University of Tokyo, the High Energy Accelerator Research Organization (KEK), the Academia Sinica Institute for Astronomy and Astrophysics in Taiwan (ASIAA), and Princeton University. Funding was contributed by the FIRST program from Japanese Cabinet Office, the Ministry of Education, Culture, Sports, Science and Technology (MEXT), the Japan Society for the Promotion of Science (JSPS), Japan Science and Technology Agency (JST), the Toray Science Foundation, NAOJ, Kavli IPMU, KEK, ASIAA, and Princeton University.

This research made use of Python (\url{www.python.org}) and several Python packages like Numpy; Astropy (\url{http://www.astropy.org}) a community-developed core Python package for Astronomy; matplotlib; IPython; Cython.

This work has made use of data from the DEEP2 Galaxy Redshift Survey, whose fundings been provided by NSF grants AST-95-09298, AST-0071048, AST-0507428, and AST-0507483 as well as NASA LTSA grant NNG04GC89G.
This work has made use of data from the European Space Agency (ESA) mission
{\it Gaia} (\url{https://www.cosmos.esa.int/gaia}), processed by the {\it Gaia}
Data Processing and Analysis Consortium (DPAC,
\url{https://www.cosmos.esa.int/web/gaia/dpac/consortium}). Funding for the DPAC
has been provided by national institutions, in particular the institutions
participating in the {\it Gaia} Multilateral Agreement.
}

\end{acknowledgements}



\bibliographystyle{aa} 
\bibliography{mini_jpas}

\appendix

\section{Log of \mjp observations}  \label{sec:log_obs}

In this appendix we provide some information on the data aquisition. More details on each tile can be found in the database in the ADQL table \texttt{minijpas.TileImage}.
In Table \ref{tab:log_obs} we indicate the number of exposures taken for each filter in each pointings (red: miniJPAS-AEGIS1, blue: miniJPAS-AEGIS2, green: miniJPAS-AEGIS3, orange: miniJPAS-AEGIS4). Different groups of filters have been observed in different observational blocks (FW$\#$), following the chosen strategy of filter changes in the \pfs\ filter wheel (see Sect.~\ref{sec:observations}). Total exposure times per pointing  are reported in the last column.

\begin{landscape}
\begin{table}[] 
\begin{center}
\caption{Log of \mjp\ observations. Number of exposures taken for each filter, for each pointing and in each observing epoch (the dates correspond to the day a new set of filter was installed on the filter wheel). The last two columns provide the total number of exposures used in the final co-added images and the total exposure time. Red: miniJPAS-AEGIS1, blue: miniJPAS-AEGIS2, green: miniJPAS-AEGIS3, orange: miniJPAS-AEGIS4.} 
\resizebox{1.3 \textwidth}{!}{
\begin{tabular}{l|c|c|c|c|c|c|c|c|c|c|c|c|c|c|c} 
\hline
\hline
\multirow{2}{*}{FILTER} & FW01 & FW02 & FW03 & FW04 & FW05 & FW06 & FW07 & FW08 & FW09 & FW10 & FW11 & FW12 & FW13  & TOTAL & TOTAL \\
 & 2018-04-27 & 2018-05-18 & 2018-06-18 & 2018-06-28 & 2018-07-11 & 2018-07-20 & 2018-08-03 & 2018-08-14 & 2018-08-24 & 2018-09-25 & 2018-10-09 & 2019-02-05 & 2019-07-11  & $N_{\textrm{img}}$ & $t_{\textrm{exp}}$(s) \\
\hline
\hline
uJAVA & & & & & & & & & & & &  {\color{red}16}, {\color{blue}8}, {\color{green}8}, {\color{orange}18} & &  {\color{red}16}, {\color{blue}8}, {\color{green}8}, {\color{orange}18} &  {\color{red}1920}, {\color{blue}960}, {\color{green}960}, {\color{orange}2160} \\
\hline
uJPAS & & & & & & & &  {\color{red}4}, {\color{blue}4}, {\color{green}4}, {\color{orange}4} & & & & & &  {\color{red}4}, {\color{blue}4}, {\color{green}4}, {\color{orange}4} &  {\color{red}480}, {\color{blue}480}, {\color{green}480}, {\color{orange}480} \\
\hline
gSDSS &  {\color{red}16}, {\color{blue}16}, {\color{green}16}, {\color{orange}16} & & & & & & & & & & & & &  {\color{red}16}, {\color{blue}16}, {\color{green}16}, {\color{orange}16} &  {\color{red}480}, {\color{blue}480}, {\color{green}480}, {\color{orange}480} \\
\hline
rSDSS &  {\color{red}6}, {\color{blue}27}, {\color{green}1}, {\color{orange}19} &  {\color{red}31} &  {\color{green}16}, {\color{orange}3} &  {\color{blue}10}, {\color{green}16}, {\color{orange}15} &  {\color{green}4} & & & & & & & & &  {\color{red}37}, {\color{blue}37}, {\color{green}37}, {\color{orange}37} &  {\color{red}1110}, {\color{blue}1110}, {\color{green}1110}, {\color{orange}1110} \\
\hline
iSDSS &  {\color{red}16}, {\color{blue}16}, {\color{green}16}, {\color{orange}16} & & & & & & & & & & & &  {\color{green}16}, {\color{orange}32} &  {\color{red}16}, {\color{blue}16}, {\color{green}32}, {\color{orange}48} &  {\color{red}480}, {\color{blue}480}, {\color{green}960}, {\color{orange}1440} \\
\hline
J0378 & & & & & & & &  {\color{red}4}, {\color{blue}8}, {\color{green}4}, {\color{orange}4} & & & & & &  {\color{red}4}, {\color{blue}8}, {\color{green}4}, {\color{orange}4} &  {\color{red}480}, {\color{blue}960}, {\color{green}480}, {\color{orange}480} \\
\hline
J0390 & &  {\color{red}11}, {\color{blue}4}, {\color{green}4}, {\color{orange}4} & & & & & & & & & &  {\color{red}16}, {\color{blue}8}, {\color{green}8}, {\color{orange}8} & &  {\color{red}27}, {\color{blue}12}, {\color{green}12}, {\color{orange}12} &  {\color{red}3240}, {\color{blue}1440}, {\color{green}1440}, {\color{orange}1440} \\
\hline
J0400 & & & &  {\color{red}9}, {\color{blue}4}, {\color{green}4}, {\color{orange}8} & & & & & & & & & &  {\color{red}9}, {\color{blue}4}, {\color{green}4}, {\color{orange}8} &  {\color{red}1080}, {\color{blue}480}, {\color{green}480}, {\color{orange}960} \\
\hline
J0410 & & & & & &  {\color{red}16}, {\color{blue}12}, {\color{green}8}, {\color{orange}4} & & & & & & & &  {\color{red}16}, {\color{blue}12}, {\color{green}8}, {\color{orange}4} &  {\color{red}1920}, {\color{blue}1440}, {\color{green}960}, {\color{orange}480} \\
\hline
J0420 & & &  {\color{red}4}, {\color{blue}4}, {\color{green}4}, {\color{orange}8} & & & & & & & &  {\color{green}4}, {\color{orange}12} & & &  {\color{red}4}, {\color{blue}4}, {\color{green}8}, {\color{orange}20} &  {\color{red}480}, {\color{blue}480}, {\color{green}960}, {\color{orange}2400} \\
\hline
J0430 & & & & & & &  {\color{red}10}, {\color{blue}4}, {\color{green}4}, {\color{orange}4} & & & & & & &  {\color{red}10}, {\color{blue}4}, {\color{green}4}, {\color{orange}4} &  {\color{red}1200}, {\color{blue}480}, {\color{green}480}, {\color{orange}480} \\
\hline
J0440 & & & & &  {\color{red}8}, {\color{blue}4}, {\color{green}4}, {\color{orange}8} & & & & & & & &  {\color{blue}4}, {\color{green}8}, {\color{orange}4} &  {\color{red}8}, {\color{blue}8}, {\color{green}12}, {\color{orange}12} &  {\color{red}960}, {\color{blue}960}, {\color{green}1440}, {\color{orange}1440} \\
\hline
J0450 & & & & & & & &  {\color{red}9}, {\color{blue}8}, {\color{green}4}, {\color{orange}8} & & & & & &  {\color{red}9}, {\color{blue}8}, {\color{green}4}, {\color{orange}8} &  {\color{red}1080}, {\color{blue}960}, {\color{green}480}, {\color{orange}960} \\
\hline
J0460 & &  {\color{red}10}, {\color{blue}4}, {\color{green}5}, {\color{orange}9} & & & & & & & & & &  {\color{red}8}, {\color{blue}8}, {\color{green}8}, {\color{orange}8} & &  {\color{red}18}, {\color{blue}12}, {\color{green}13}, {\color{orange}17} &  {\color{red}2160}, {\color{blue}1440}, {\color{green}1560}, {\color{orange}2040} \\
\hline
J0470 & & & &  {\color{red}8}, {\color{blue}4}, {\color{green}4}, {\color{orange}8} & & & & & & & & & &  {\color{red}8}, {\color{blue}4}, {\color{green}4}, {\color{orange}8} &  {\color{red}960}, {\color{blue}480}, {\color{green}480}, {\color{orange}960} \\
\hline
J0480 & & & & & &  {\color{red}12}, {\color{blue}12}, {\color{green}8}, {\color{orange}8} & & & & & & & &  {\color{red}12}, {\color{blue}12}, {\color{green}8}, {\color{orange}8} &  {\color{red}1440}, {\color{blue}1440}, {\color{green}960}, {\color{orange}960} \\
\hline
J0490 & & &  {\color{red}4}, {\color{blue}4}, {\color{green}4}, {\color{orange}8} & & & & & & & &  {\color{orange}12} & & &  {\color{red}4}, {\color{blue}4}, {\color{green}4}, {\color{orange}20} &  {\color{red}480}, {\color{blue}480}, {\color{green}480}, {\color{orange}2400} \\
\hline
J0500 & & & & & & &  {\color{red}4}, {\color{blue}4}, {\color{green}4}, {\color{orange}4} & & & & & & &  {\color{red}4}, {\color{blue}4}, {\color{green}4}, {\color{orange}4} &  {\color{red}480}, {\color{blue}480}, {\color{green}480}, {\color{orange}480} \\
\hline
J0510 & & & & &  {\color{red}8}, {\color{blue}4}, {\color{green}4}, {\color{orange}8} & & & & & & & & &  {\color{red}8}, {\color{blue}4}, {\color{green}4}, {\color{orange}8} &  {\color{red}960}, {\color{blue}480}, {\color{green}480}, {\color{orange}960} \\
\hline
J0520 & & & & & & & &  {\color{red}8}, {\color{blue}8}, {\color{green}8}, {\color{orange}8} & & & & & &  {\color{red}8}, {\color{blue}8}, {\color{green}8}, {\color{orange}8} &  {\color{red}960}, {\color{blue}960}, {\color{green}960}, {\color{orange}960} \\
\hline
J0530 & &  {\color{red}12}, {\color{blue}4}, {\color{green}4}, {\color{orange}4} & & & & & & & & & & &  {\color{blue}4}, {\color{green}8}, {\color{orange}4} &  {\color{red}12}, {\color{blue}8}, {\color{green}12}, {\color{orange}8} &  {\color{red}1440}, {\color{blue}960}, {\color{green}1440}, {\color{orange}960} \\
\hline
J0540 & & & &  {\color{red}4}, {\color{blue}4}, {\color{green}4}, {\color{orange}4} & & & & & & & & & &  {\color{red}4}, {\color{blue}4}, {\color{green}4}, {\color{orange}4} &  {\color{red}480}, {\color{blue}480}, {\color{green}480}, {\color{orange}480} \\
\hline
J0550 & & & & & &  {\color{red}12}, {\color{blue}12}, {\color{green}9}, {\color{orange}8} & & & & & & & &  {\color{red}12}, {\color{blue}12}, {\color{green}9}, {\color{orange}8} &  {\color{red}1440}, {\color{blue}1440}, {\color{green}1080}, {\color{orange}960} \\
\hline
J0560 & & &  {\color{red}8}, {\color{blue}4}, {\color{green}4}, {\color{orange}8} & & & & & & & &  {\color{orange}8} & & &  {\color{red}8}, {\color{blue}4}, {\color{green}4}, {\color{orange}16} &  {\color{red}960}, {\color{blue}480}, {\color{green}480}, {\color{orange}1920} \\
\hline
J0570 & & & & & & &  {\color{red}4}, {\color{blue}4}, {\color{green}4}, {\color{orange}4} & & & & & & &  {\color{red}4}, {\color{blue}4}, {\color{green}4}, {\color{orange}4} &  {\color{red}480}, {\color{blue}480}, {\color{green}480}, {\color{orange}480} \\
\hline
J0580 & & & & &  {\color{red}8}, {\color{blue}4}, {\color{green}4}, {\color{orange}8} & & & & & & & & &  {\color{red}8}, {\color{blue}4}, {\color{green}4}, {\color{orange}8} &  {\color{red}960}, {\color{blue}480}, {\color{green}480}, {\color{orange}960} \\
\hline
J0590 & & & & & & & &  {\color{red}4}, {\color{blue}8}, {\color{green}8}, {\color{orange}8} & & &  {\color{red}12}, {\color{green}4} & & &  {\color{red}16}, {\color{blue}8}, {\color{green}12}, {\color{orange}8} &  {\color{red}1920}, {\color{blue}960}, {\color{green}1440}, {\color{orange}960} \\
\hline
J0600 & &  {\color{red}5}, {\color{orange}4} & & & & & & & & & &  {\color{red}8}, {\color{blue}8}, {\color{green}8}, {\color{orange}8} & &  {\color{red}13}, {\color{blue}8}, {\color{green}8}, {\color{orange}12} &  {\color{red}1560}, {\color{blue}960}, {\color{green}960}, {\color{orange}1440} \\
\hline
J0610 & & & &  {\color{red}4}, {\color{blue}4}, {\color{green}4}, {\color{orange}4} & & & & & & & & & &  {\color{red}4}, {\color{blue}4}, {\color{green}4}, {\color{orange}4} &  {\color{red}480}, {\color{blue}480}, {\color{green}480}, {\color{orange}480} \\
\hline
J0620 & & & & & &  {\color{red}8}, {\color{blue}8}, {\color{green}8}, {\color{orange}12} & & & & & & & &  {\color{red}8}, {\color{blue}8}, {\color{green}8}, {\color{orange}12} &  {\color{red}960}, {\color{blue}960}, {\color{green}960}, {\color{orange}1440} \\
\hline
J0630 & & &  {\color{red}4}, {\color{blue}4}, {\color{green}4}, {\color{orange}8} & & & & & & & &  {\color{orange}4} & & &  {\color{red}4}, {\color{blue}4}, {\color{green}4}, {\color{orange}12} &  {\color{red}480}, {\color{blue}480}, {\color{green}480}, {\color{orange}1440} \\
\hline
J0640 & & & & & & &  {\color{red}8}, {\color{blue}9}, {\color{green}11}, {\color{orange}8} & & & & & & &  {\color{red}8}, {\color{blue}9}, {\color{green}11}, {\color{orange}8} &  {\color{red}960}, {\color{blue}1080}, {\color{green}1320}, {\color{orange}960} \\
\hline
J0650 & & & & &  {\color{red}8}, {\color{blue}4}, {\color{green}4}, {\color{orange}8} & & & & & & & & &  {\color{red}8}, {\color{blue}4}, {\color{green}4}, {\color{orange}8} &  {\color{red}960}, {\color{blue}480}, {\color{green}480}, {\color{orange}960} \\
\hline
J0660 &  {\color{red}8}, {\color{blue}12}, {\color{green}4}, {\color{orange}5} & & & & & & & & & & &  {\color{red}8}, {\color{blue}8}, {\color{green}8}, {\color{orange}8} & &  {\color{red}16}, {\color{blue}20}, {\color{green}12}, {\color{orange}13} &  {\color{red}1920}, {\color{blue}2400}, {\color{green}1440}, {\color{orange}1560} \\
\hline
J0670 & &  {\color{red}12}, {\color{blue}4}, {\color{green}4}, {\color{orange}6} & & & & & & & & & &  {\color{red}8}, {\color{blue}8}, {\color{green}4}, {\color{orange}4} & &  {\color{red}20}, {\color{blue}12}, {\color{green}8}, {\color{orange}10} &  {\color{red}2400}, {\color{blue}1440}, {\color{green}960}, {\color{orange}1200} \\
\hline
J0680 & & & &  {\color{red}4}, {\color{blue}4}, {\color{green}4}, {\color{orange}4} & & & & & & & & &  {\color{blue}4}, {\color{green}8}, {\color{orange}4} &  {\color{red}4}, {\color{blue}8}, {\color{green}12}, {\color{orange}8} &  {\color{red}480}, {\color{blue}960}, {\color{green}1440}, {\color{orange}960} \\
\hline
J0690 & & & & & &  {\color{red}8}, {\color{blue}8}, {\color{green}8}, {\color{orange}9} & & & & & & &  {\color{blue}4}, {\color{green}8}, {\color{orange}4} &  {\color{red}8}, {\color{blue}12}, {\color{green}16}, {\color{orange}13} &  {\color{red}960}, {\color{blue}1440}, {\color{green}1920}, {\color{orange}1560} \\
\hline
J0700 & & &  {\color{red}4}, {\color{blue}4}, {\color{green}4}, {\color{orange}8} & & & & & & & & & & &  {\color{red}4}, {\color{blue}4}, {\color{green}4}, {\color{orange}8} &  {\color{red}480}, {\color{blue}480}, {\color{green}480}, {\color{orange}960} \\
\hline
J0710 & & & & & & &  {\color{red}4}, {\color{blue}4}, {\color{green}4}, {\color{orange}4} & & & & & & &  {\color{red}4}, {\color{blue}4}, {\color{green}4}, {\color{orange}4} &  {\color{red}480}, {\color{blue}480}, {\color{green}480}, {\color{orange}480} \\
\hline
J0720 & & & & &  {\color{red}8}, {\color{blue}4}, {\color{green}4}, {\color{orange}8} & & & & & & & & &  {\color{red}8}, {\color{blue}4}, {\color{green}4}, {\color{orange}8} &  {\color{red}960}, {\color{blue}480}, {\color{green}480}, {\color{orange}960} \\
\hline
J0730 & & & & & & & &  {\color{red}4}, {\color{blue}8}, {\color{green}7}, {\color{orange}8} & & &  {\color{red}8} & & &  {\color{red}12}, {\color{blue}8}, {\color{green}7}, {\color{orange}8} &  {\color{red}1440}, {\color{blue}960}, {\color{green}840}, {\color{orange}960} \\
\hline
J0740 & &  {\color{red}12}, {\color{blue}4}, {\color{green}4}, {\color{orange}4} & & & & & & & & & & &  {\color{blue}4}, {\color{green}8}, {\color{orange}4} &  {\color{red}12}, {\color{blue}8}, {\color{green}12}, {\color{orange}8} &  {\color{red}1440}, {\color{blue}960}, {\color{green}1440}, {\color{orange}960} \\
\hline
J0750 & & & &  {\color{red}4}, {\color{blue}4}, {\color{green}4}, {\color{orange}4} & & & & & & & & & &  {\color{red}4}, {\color{blue}4}, {\color{green}4}, {\color{orange}4} &  {\color{red}480}, {\color{blue}480}, {\color{green}480}, {\color{orange}480} \\
\hline
J0760 & & & & & &  {\color{red}9}, {\color{blue}4}, {\color{green}8}, {\color{orange}9} & & & & & & & &  {\color{red}9}, {\color{blue}4}, {\color{green}8}, {\color{orange}9} &  {\color{red}1080}, {\color{blue}480}, {\color{green}960}, {\color{orange}1080} \\
\hline
J0770 & & &  {\color{red}4}, {\color{blue}4}, {\color{green}4}, {\color{orange}8} & & & & & & & & & & &  {\color{red}4}, {\color{blue}4}, {\color{green}4}, {\color{orange}8} &  {\color{red}480}, {\color{blue}480}, {\color{green}480}, {\color{orange}960} \\
\hline
J0780 & & & & & & &  {\color{red}8}, {\color{blue}4}, {\color{green}4}, {\color{orange}4} & & & & & & &  {\color{red}8}, {\color{blue}4}, {\color{green}4}, {\color{orange}4} &  {\color{red}960}, {\color{blue}480}, {\color{green}480}, {\color{orange}480} \\
\hline
J0790 & & & & &  {\color{red}8}, {\color{blue}4}, {\color{green}4}, {\color{orange}8} & & & & & & & & &  {\color{red}8}, {\color{blue}4}, {\color{green}4}, {\color{orange}8} &  {\color{red}960}, {\color{blue}480}, {\color{green}480}, {\color{orange}960} \\
\hline
J0800 & & & & & & & & &  {\color{red}8}, {\color{blue}18}, {\color{green}8}, {\color{orange}8} & & & & &  {\color{red}8}, {\color{blue}18}, {\color{green}8}, {\color{orange}8} &  {\color{red}960}, {\color{blue}2160}, {\color{green}960}, {\color{orange}960} \\
\hline
J0810 & & & & & & & & &  {\color{red}8}, {\color{blue}8}, {\color{green}8}, {\color{orange}8} & & & & &  {\color{red}8}, {\color{blue}8}, {\color{green}8}, {\color{orange}8} &  {\color{red}960}, {\color{blue}960}, {\color{green}960}, {\color{orange}960} \\
\hline
J0820 & & & & & & & & &  {\color{red}8}, {\color{blue}8}, {\color{green}8}, {\color{orange}8} & & & & &  {\color{red}8}, {\color{blue}8}, {\color{green}8}, {\color{orange}8} &  {\color{red}960}, {\color{blue}960}, {\color{green}960}, {\color{orange}960} \\
\hline
J0830 & & & & & & & & &  {\color{red}16}, {\color{blue}8}, {\color{green}16}, {\color{orange}8} & & & & &  {\color{red}16}, {\color{blue}8}, {\color{green}16}, {\color{orange}8} &  {\color{red}1920}, {\color{blue}960}, {\color{green}1920}, {\color{orange}960} \\
\hline
J0840 & & & & & & & & &  {\color{red}16}, {\color{blue}8}, {\color{green}8}, {\color{orange}8} & & & & &  {\color{red}16}, {\color{blue}8}, {\color{green}8}, {\color{orange}8} &  {\color{red}1920}, {\color{blue}960}, {\color{green}960}, {\color{orange}960} \\
\hline
J0850 & & & & & & & & &  {\color{red}13}, {\color{blue}8}, {\color{green}8}, {\color{orange}8} & & & & &  {\color{red}13}, {\color{blue}8}, {\color{green}8}, {\color{orange}8} &  {\color{red}1560}, {\color{blue}960}, {\color{green}960}, {\color{orange}960} \\
\hline
J0860 &  {\color{red}12}, {\color{blue}12}, {\color{green}4}, {\color{orange}4} & & & & & & & & & & & & &  {\color{red}12}, {\color{blue}12}, {\color{green}4}, {\color{orange}4} &  {\color{red}1440}, {\color{blue}1440}, {\color{green}480}, {\color{orange}480} \\
\hline
J0870 & & & & & & & & & &  {\color{red}12}, {\color{blue}16}, {\color{green}20}, {\color{orange}8} & & & &  {\color{red}12}, {\color{blue}16}, {\color{green}20}, {\color{orange}8} &  {\color{red}1440}, {\color{blue}1920}, {\color{green}2400}, {\color{orange}960} \\
\hline
J0880 & & & & & & & & & &  {\color{red}12}, {\color{blue}13}, {\color{green}12}, {\color{orange}8} & & & &  {\color{red}12}, {\color{blue}13}, {\color{green}12}, {\color{orange}8} &  {\color{red}1440}, {\color{blue}1560}, {\color{green}1440}, {\color{orange}960} \\
\hline
J0890 & & & & & & & & & &  {\color{red}12}, {\color{blue}8}, {\color{green}12}, {\color{orange}8} & & & &  {\color{red}12}, {\color{blue}8}, {\color{green}12}, {\color{orange}8} &  {\color{red}1440}, {\color{blue}960}, {\color{green}1440}, {\color{orange}960} \\
\hline
J0900 & & & & & & & & & &  {\color{red}12}, {\color{blue}8}, {\color{green}12}, {\color{orange}8} & & & &  {\color{red}12}, {\color{blue}8}, {\color{green}12}, {\color{orange}8} &  {\color{red}1440}, {\color{blue}960}, {\color{green}1440}, {\color{orange}960} \\
\hline
J0910 & & & & & & & & & &  {\color{red}4}, {\color{blue}8}, {\color{green}4}, {\color{orange}8} & & & &  {\color{red}4}, {\color{blue}8}, {\color{green}4}, {\color{orange}8} &  {\color{red}480}, {\color{blue}960}, {\color{green}480}, {\color{orange}960} \\
\hline
J1007 & & & & & & & & & &  {\color{red}4}, {\color{blue}7}, {\color{green}12}, {\color{orange}8} & & & &  {\color{red}4}, {\color{blue}7}, {\color{green}12}, {\color{orange}8} &  {\color{red}480}, {\color{blue}840}, {\color{green}1440}, {\color{orange}960} \\
\hline
\hline
\end{tabular}
 \label{tab:log_obs}
} 
\end{center}
\end{table}
\end{landscape}

\section{Background correction} \label{sec:app_bgk}

Extending the discussion from Sect.~\ref{sec:single_frames}, here we
provide more technical details on the background 
treatment of the \mjp\ images. In particular, we describe a background
correction derived from the illumination correction and a ``superbackground'' correction, as we call it.

\subsection{Background correction derived from illumination correction}

The illumination correction is becoming a standard step in the image reduction chain when the acquisition system involves large FoVs and/or includes lenses like field correctors. Such telescopes are prone to suffer from illumination issues. Their origin is explained in \citet{andersen1995}. Nowadays, projects involving that type of telescopes take that effect into account (e.g. DES, \citet{bernstein2017} and SDSS, \citet{betoule2013}).
The approach adopted for the illumination correction in the OAJ reduction pipeline is based on \citet{manfroid1995}.
It requires the acquisition of a set of specific observations. In the case of \mjp\, these  were planned
during \mjp\ observing campaign. From these observations a master
illumination correction image (ICOR, hereafter) was constructed for
each filter and this was applied to the images just after the flat
field correction\footnote{The procedure is equivalent to the
  correction of the actual flat field.}. However, images corrected by
the illumination correction showed a background pattern not visible 
after the preceding flat field correction.

Before continuing with the description of the procedure aiming at removing this apparent
worsening of the images after the illumination correction, let us 
explain its origin.

Due to the optics of the system, a flat illumination arriving to the
telescope can result in a non uniform distribution of light in the
detector. This affects the scientific images in two ways. First, they show
a non uniform background pattern which is an additive component that
should be subtracted. Second, it affects the flat field
correction. If not taken into account, the non uniform distribution of
light in the flat field images is assumed as a variation of efficiency
(which is not). This has two effects on the scientific images after
being divided by the master flat. First, it introduces a 2D variation
of the photometry of the objects (i.e. a 2D variation of the
photometric zero point). This is the effect corrected by the ICOR. The
second effect is to apparently remove the previous background pattern
of the scientific images. However, this background should be corrected
by subtraction and not by division and, therefore, this apparent
background removal is actually wrong. And, in fact, any additive
background present in the raw image, should be present in the images
after flat field correction. That is the reason why after the
illumination correction, which corrects the flat field itself, the
original additive background pattern present in the raw image
reappears in the scientific image.

Then, knowing that there is a background component of the images
which has the same spatial distribution as the ICOR and that it is an
additive component, a new background correction is computed having 
the 2D distribution of the ICOR and an average value of the average
background of the target image. This background correction is
subtracted just after applying ICOR resulting in an reduced image with
flatter background  and flatter zero point.

Finally, to check that this background subtraction did not affect 
the actual photometry of the objects of the images, we processed the
individual images used for constructing an ICOR, controlling  that the
differences in the photometry of individual objects observed in
different images were compatible with zero.

\subsection{Superbackground correction}

After the background subtraction from the ICOR, many single frames
still display background patterns whose spatial distribution made them
difficult to be properly managed in a massive automatic way during the
source detection process, i.e. with \sext.  These patterns have two
main characteristics that make the correction complex: presence of
strong background gradients and strong temporal variability of the
patterns, either in intensity or spatial distribution\footnote{It is
  due to this temporal variation that they are not removed by the
  background coming from the ICOR which only removes the stable
  illumination common to flat field and scientific images.}.

Under the assumptions that these patterns are instrumental, and
therefore do not depend on the sky position (at least for close enough
pointings), and that they are stable in time scales of few minutes, we
can consider that images taken close in time (with the same filter)
present the same background pattern. Therefore, a median combination
of scientific images with celestial objects in different CCD positions
should keep the background pattern while removing the real
astrophysical objects. This is a procedure analogue to the
construction of master fringing images or superflat images, for this
reason we dubbed these background images as ``superbackground images''
(SBGK, hereafter).

The observational strategy of \mjp\ in which the single
exposures of the same pointing and same filter were taken with small
dithering and one after the other, introduced one problem and provided
an advantage. First, to not subtract light of real objects when
constructing the SBGK, it is necessary to avoid to have celestial objects
falling in the same parts of the images combined to
construct the SBGK. Due to the temporal variation of the patterns
(with detected changes on scales of several minutes), it was not
possible to combine images of the same filter and different pointings,
since they already showed changes in the background. Therefore, we
were left with the only possibility of having to use the images of the
same pointing and, due to the small dithering pattern, the chances of
having areas of the image covered by actual objects were high, in
particular for larger diffuse objects. To minimize this possibility,
before combining the images, the objects detected in a preliminary
reduction without SBKG correction were masked with masks a bit larger
than the actual detected size of the object to take into account
extended diffuse light. This masking procedure was fine-tuned until it
was checked that there were no residuals of extended galaxies in the
SBGK (for bright saturated stars we allowed for light of their halos
to be in the SBKG). In the specific case of \mjp\ data, being a
cosmological field devoid of large galaxies (i.e., galaxies occupying a large
area of the CCD), the possible oversubtraction of background due to
this procedure was not a big issue, although for the studies of faint
extended studies this has to be kept in mind.

In summary, the SBKG construction consisted, first, in identifying the
images to be combined. That was done taking advantage that exposures
of the same pointing and same filter were done consecutively and,
therefore, we could consider that in the interval between the first
and last exposure the background pattern was stable (although in few
ocassions we detected that this wasn't the case and some residuals
were left). Then, objects were masked and images
median-combined. Finally, the resulting image was smoothed using
\sext. The last step was to subtract the SBKG to the individual
scientific frames.

Again, in order to verify that the SBGK correction did not affect the
photometry of the objects, we simulated the construction of an ICOR (a
procedure that by construction shows changes in photometry of object
across the CCD) applying all the corrections to the images to be used,
finding that the resulting map was compatible with no correction.

\section{\texttt{SExtractor}'s parameters} \label{sec:sext_params}

Source detection and basic parameters of the objects in the images
have been done using \sext\ \citet{bertin1996}. Details on
how \sext\ works can be found either in \citet{bertin1996} and in the \sext\
User's Manual. In Table~\ref{tab:sext_params} we provide the
list of parameters that have been used to obtain the photometric
catalogues that are the base of this data release.
We only list those parameters that are common to all the images. Other parameters, such as  \texttt{CATALOG\_NAME}, \texttt{CHECKIMAGE\_NAME}, are obviously specific to each image and are not included in the table. 
{\tt 
\begin{table*}[htb]
 \caption{List of \sext's parameters used to construct the \mjp\ source catalogues.}
  \centering
{\tt
  \begin{tabular}{|lr|}
 \hline
 {\bf \sext parameter} & {\bf value} \\
 \hline
    ANALYSIS\_THRESH & 2.0 \\
    BACKPHOTO\_THICK & 24 \\
    BACKPHOTO\_TYPE & LOCAL \\
    BACK\_FILTERSIZE & 3 \\
    BACK\_FILTTHRESH & 0.0 \\
    BACK\_SIZE & 512 \\
    BACK\_TYPE & AUTO \\
    BACK\_VALUE & 0.0,0.0 \\
    CLEAN & Y \\
    CLEAN\_PARAM & 1.0 \\
    DEBLEND\_MINCONT & 0.005 \\
    DEBLEND\_NTHRESH & 32 \\
    DETECT\_MINAREA & 5 \\
    DETECT\_THRESH & 0.9 (dual-mode) / 2 (single-mode)\\
    DETECT\_TYPE & CCD \\
    FILTER & Y \\
    FILTER\_NAME & gauss\_3.0\_5x5.conv \\
    GAIN\_KEY & GAIN \\
    MASK\_TYPE & CORRECT \\
    PHOT\_AUTOAPERS & 0.0,0.0 \\
    PHOT\_AUTOPARAMS & 2.5,3.5 \\
    PHOT\_FLUXFRAC & 0.5 \\
    PHOT\_PETROPARAMS & 2.0,3.5 \\
    PIXEL\_SCALE & 0.2267 \\
    STARNNW\_NAME & default.nnw \\
    THRESH\_TYPE & RELATIVE \\
    VERBOSE\_TYPE & NORMAL \\
    WEIGHT\_GAIN & Y \\
    WEIGHT\_THRESH & 0,0 \\
    WEIGHT\_TYPE & MAP\_WEIGHT,MAP\_WEIGHT \\
    \hline
  \end{tabular}
}
 
  \label{tab:sext_params}
\end{table*}
}

\end{document}